\documentclass[preprint,12pt,authoryear]{elsarticle}

\usepackage{amssymb}
\usepackage{amsmath}
\usepackage{bbm}
\usepackage{footnote}
\usepackage{tablefootnote}
 \usepackage{amsthm}
 \usepackage[flushleft]{threeparttable}
 \usepackage[colorlinks=true]{hyperref}
\hypersetup{
        urlcolor = blue,
        pdfauthor = {Brice Romuald Gueyap Kounga},
        pdfkeywords = {Semiparametric logit, Endogenous networks, Unobserved heterogeneity, Social interactions, Network data, Binary choice, Network matching},
        pdftitle = {Identification and Estimation of a Semiparametric Logit Model using Network Data},
        pdfsubject = {Econometric Theory},
        pdfpagemode = UseNone,
        linkcolor = red,          
        citecolor = blue          
}
\newcommand{\sym}[1]{\ifmmode^{#1}\else\(^{#1}\)\fi}
\newtheorem{assumption}{Assumption}

\newtheorem{lemma}{Lemma}
\newtheorem{example}{Example}
\newtheorem{proofoftheorem}{Proof of Theorem}
\newtheorem{proofoflemma}{Proof of Lemma}
\newtheorem{theorem}{Theorem}
\newtheorem{proposition}{Proposition}

\newcommand{\E}{\mathbb{E}}
\newcommand{\R}{\mathbb{R}}
\newcommand{\PP}{\mathbb{P}}

\newcommand{\X}{\mathcal{X}}

\DeclareMathOperator{\Var}{Var}

\journal{Journal of Econometrics}

\begin{document}

\begin{frontmatter}



\title{Identification and Estimation of a Semiparametric Logit Model using Network Data\footnote{Special thanks to Nail Kashaev, Salvador Navarro, and Roy Allen for their comments and suggestions in the preparation of this paper. I thank Tim Conley and my classmates in the 2023-2024 Communication and Professional Development Workshop for their helpful comments. The views expressed herein and any remaining errors are mine.}} 

\author{ 
    Brice Romuald Gueyap Kounga\\
    Western University\\
     bgueyapk@uwo.ca
	}

\begin{abstract}
This paper studies identification and estimation of a semiparametric logit model in which an unobserved individual characteristic affects both a binary outcome and the formation of social links. In this setting, the endogeneity of the network is informative rather than a nuisance: because the same latent trait drives linking behavior, observed network data can be used to control for unobserved heterogeneity in the outcome equation without imposing a parametric model of link formation. Slope parameters are point identified by a conditional likelihood argument applied to pairs of agents with identical network formation behavior. I propose a kernel-weighted conditional logit estimator that matches agents using codegree distances and establish its consistency. I then derive two asymptotic normality results: an exact root-$n$ result when network types have finite support, and a general result centered at a pseudo-true value whose bias I characterize and remove with a jackknife bias correction. Variance estimators that account for matching on estimated network distances are provided. Monte Carlo simulations demonstrate good finite-sample performance, and an empirical application to microfinance adoption demonstrates that accounting for endogenous network formation materially affects estimated covariate effects.
\end{abstract}


\end{frontmatter}

\section{Introduction}\label{section1}

A large empirical literature documents the importance of peer and network effects in economic settings, showing that individuals’ behavior and outcomes are shaped by interactions with their peers \citep{manski1993identification, sacerdote2001peer, calvo2009peer, christakis2008collective, bramoulle2020peer,de2020consumption}. These effects have been studied across contexts such as education (e.g., \cite{calvo2009peer}), consumption \citep{de2020consumption}, labor effort and productivity \citep{beugnot2013social}, and social influence more broadly \citep{jackson2017economic}, and surveys emphasize the role of network structure in both shaping individual outcomes and complicating causal analysis \citep{bramoulle2020peer}. In most of these applications, individuals are embedded in social networks that shape behavior and at the same time reflect unobserved heterogeneity. This dual role is what makes identification difficult.

A central econometric difficulty is that social networks are rarely exogenous. Unobserved individual traits such as motivation, trust, ability, or entrepreneurial skill often influence both outcomes and the formation of social ties. When these latent characteristics enter both equations, standard estimation strategies do not generally deliver consistent estimates. In binary choice models, applied work typically relies on naive logit or probit specifications, logit models augmented with peer averages or network controls, or control function approaches based on parametric link formation models. These strategies can fail even when rich network data are available, because latent traits that govern link formation remain in the outcome error term.

This paper develops identification and estimation methods for a semiparametric logit model in which the outcome equation includes an unknown function of an unobserved individual characteristic that also governs link formation. Such latent characteristics may represent motivation, trust, ability, or family expectations, which are difficult to measure directly but play a central role in both social interactions and individual decisions. To be clear about the object of study: this is a model of unobserved individual heterogeneity that is common to the outcome and the network, not a model of endogenous peer effects in the sense of \cite{manski1993identification}. It is precisely because the network formation process is endogenous, driven by the same latent trait that confounds the outcome equation, that the observed network is informative about the confounder. If links were formed at random, the network would carry no information about $\omega_i$, and the strategy studied here would be infeasible. Endogenous network formation is thus the source of identification rather than a threat to it, in contrast to peer-effects settings where it undermines standard estimators \citep{johnsson2021estimation}. I show that network data can be used to control for this source of endogeneity without imposing parametric restrictions on the network formation process.

The identification argument builds on two strands of the econometric literature. First, it is closely related to fixed effects and conditional likelihood ideas developed in \cite{CHAMBERLAIN19841247}, where differencing across observationally equivalent units eliminates latent heterogeneity. Second, it draws on recent advances in the econometrics of networks, particularly \cite{auerbach2022identification}, who shows that in partially linear models, agents with identical network formation types can be used to difference out unobserved heterogeneity. Auerbach’s paper makes an important contribution by demonstrating how network formation behavior itself can serve as a sufficient statistic for latent traits in linear models.

The present paper extends this logic to a nonlinear binary choice model. This extension introduces new challenges that do
not arise in linear models and require different arguments for identification and
inference. In contrast to linear models, differencing arguments in binary response models require specific distributional structure. In particular, the logistic distribution assumption is crucial for identification. The identification strategy relies on a conditional likelihood argument analogous to Chamberlain's conditional logit, where pairwise comparisons between agents with identical network formation behavior difference out the latent social component. This argument does not extend to arbitrary single-index models, and the logit structure plays a central role in delivering point identification of slope parameters.

On the inference side, the paper develops a two-regime asymptotic theory that parallels \cite{auerbach2021inference}. When network types have finite support, as in stochastic block models, exact matches occur with positive probability and the estimator is root-$n$ consistent and asymptotically normal, centered at the true parameter. In the general case, where network types vary continuously, matching is necessarily inexact at any finite bandwidth, and the estimator is centered at a pseudo-true value. I characterize the resulting bias, which is standard for matching estimators \citep{abadie2006large}, and remove it with a jackknife bias correction in the spirit of \cite{honore1997pairwise}. The asymptotic variance accounts for the fact that agents are matched on estimated network distances: the first-stage estimation error in the codegree distances contributes variance terms that would be missed by treating the matches as known. The asymptotic framework, inherited from the graph-limit setting of \cite{auerbach2022identification}, is a dense-network regime: the link function $f$ is fixed as $n$ grows, so the expected fraction of linked dyads remains bounded away from zero. The results do not cover sequences of sparse networks with bounded or slowly growing degrees, and the empirical application makes the resulting approximation error explicit through its match-quality diagnostics.

The model is motivated by a range of empirical applications. In education, students may form friendships based on unobserved academic motivation or aspirations, while simultaneously being influenced by peers in decisions such as college major choice or persistence \citep{sacerdote2001peer, feld2022effect, pu2021peers}. In social programs, households may form ties based on trust or shared norms and learn about program participation through those ties, as in microfinance diffusion \citep{banerjee2013diffusion}. In professional settings, researchers with similar unobserved productivity or collaboration preferences may form coauthorship networks that also shape future research output \citep{ductor2014social}. In each of these examples, unobserved traits affect both outcomes and network formation, and binary choices such as participation, adoption, or selection are central objects of interest.

I formalize these ideas using a semiparametric logit model in which the latent trait enters the outcome equation as an unknown function of a latent individual index. The network is generated by a nonparametric link formation model driven by the same latent index. Identification relies on the assumption that this unobserved heterogeneity term depends on the latent characteristic only through its implied network formation behavior. I do not require the latent index itself to be identified, nor do I impose functional form assumptions on the network formation process. The parameters of interest are identified up to equivalence classes defined by network formation behavior, which is sufficient for recovering the slope coefficients of the binary response model.

The paper contributes to the econometrics of networks in three main ways. First, it provides identification results for semiparametric binary choice models with endogenous network formation, a setting that has received far less attention than linear outcome models. Second, it delivers feasible estimators that control for network endogeneity without specifying a parametric model of link formation. Third, it develops the distribution theory needed for inference, including an explicit characterization of the matching bias, a jackknife bias correction, and variance estimators that account for the estimation of network distances, extending matching and differencing ideas from the linear network literature to binary response settings.

Monte Carlo simulations demonstrate that the proposed estimator substantially reduces bias relative to naive logit specifications and commonly used control-based approaches, and that its performance improves with sample size across a range of network formation mechanisms. An empirical application to microfinance adoption in rural India \citep{banerjee2013diffusion} illustrates the practical relevance of the approach. Using pre-treatment social network data, the estimator produces economically interpretable covariate effects, with additive village-level factors removed exactly by within-village pair differencing. Controlling for endogenous network formation materially affects the estimated slopes: the coefficients on housing size and crowding roughly double relative to a naive logit, while the association between electricity and adoption, strongly significant in standard specifications, is not robust to localized matching, a pattern consistent with a coefficient that absorbs latent network heterogeneity.

By focusing explicitly on binary choice models, this paper complements and extends existing work on endogenous networks in linear settings. More broadly, it shows how network formation behavior can serve as a source of identification in nonlinear models without imposing parametric restrictions on link formation.

The remainder of the paper is organized as follows. In the next section, I present the literature review. Section \ref{section3} introduces the model and presents identification and estimation results. In Section \ref{section4}, I present the results of Monte Carlo simulations. Section \ref{section5} provides an empirical application. Section \ref{section6} concludes.

\paragraph{\textbf{Notation}} For a function $g$ on $[0,1]$, $\|g\|_2$ denotes the $L^2[0,1]$ norm. For positive sequences $a_n$ and $b_n$, $a_n\asymp b_n$ means that $a_n$ and $b_n$ have the same order of magnitude: there exist constants $0<c\le C<\infty$ and an integer $N$ with $c\,b_n\le a_n\le C\,b_n$ for all $n\ge N$, equivalently $a_n=O(b_n)$ and $b_n=O(a_n)$. The same convention applies to functions of a continuous argument as that argument tends to its limit. For random sequences, $a_n\asymp_p b_n$ means that for every $\epsilon>0$ there exist $0<c\le C<\infty$ with $\PP(c\,b_n\le a_n\le C\,b_n)\ge1-\epsilon$ for all $n$ large, equivalently $a_n/b_n=O_p(1)$ and $b_n/a_n=O_p(1)$. Both relations are two-sided and are therefore strictly stronger than the corresponding $O$ and $O_p$ statements. I write $a_n\gg b_n$ when $b_n/a_n\to0$, and $a\vee b$ and $a\wedge b$ for the maximum and the minimum of $a$ and $b$.

\section{Literature Review}\label{section2}

This paper relates to a large literature on social interactions, peer effects, and the econometrics of networks. A central finding in this literature is that individual behavior is often shaped by peers and social environments. Early empirical work documents the importance of neighborhood and group effects in settings such as youth behavior and crime \citep{case1991company} and educational outcomes \citep{sacerdote2001peer}. Related evidence appears in health behaviors and social contagion, including smoking and obesity \citep{christakis2008collective}. \cite{jackson2017economic} provide a comprehensive survey of the economic consequences of network structure and highlight the role of social interactions across a wide range of applications.

Despite this extensive empirical evidence, identification of social interactions remains challenging. \cite{manski1993identification} emphasizes the reflection problem and the difficulty of disentangling endogenous social effects from correlated effects. Subsequent work further clarifies the sources of bias arising from endogenous group formation and simultaneity \citep{moffitt2001policy, soetevent2006empirics}. A recurring issue is that individuals sort into peer groups based on unobserved characteristics that also directly affect outcomes, rendering standard regression approaches difficult to interpret causally.

A growing literature addresses these concerns by explicitly modeling network formation as endogenous. In this line of work, unobserved individual characteristics affect both outcomes and the formation of social links, leading to joint models of behavior and networks. Several papers adopt parametric or semiparametric specifications for link formation in order to achieve identification. Examples include \cite{goldsmith2013social, hsieh2016social, arduini2015parametric}. While these approaches can be powerful, their validity depends on the correctness of the assumed network formation model, which may be difficult to justify in many empirical settings with complex and heterogeneous linking patterns.

More recent work seeks to relax parametric assumptions on network formation and instead use network structure as a source of information about latent heterogeneity. \cite{graham2015methods} surveys identification strategies in social networks and emphasizes the role of network data in controlling for unobserved individual characteristics. Most closely related to this paper, \cite{auerbach2022identification} shows that parameters in a partially linear regression model with endogenous networks can be identified by pairwise differencing between agents who share identical network formation behavior. Auerbach also proposes feasible estimators based on matching agents using codegree information, drawing on results from graph limit theory \citep{lovasz2012large}, and \cite{auerbach2021inference} develops the associated inference theory, including a bias correction and variance estimators on which the distribution theory of the present paper builds. \cite{johnsson2021estimation} propose an alternative control function approach for peer effects in endogenous networks under different identifying assumptions; the two approaches are complementary.

The idea of matching units on estimated measures of latent similarity extends beyond network settings, and it is useful to relate the present approach to that broader literature. \cite{feng2020causal} studies causal inference in possibly nonlinear factor models, matching cross-sectional units on latent factor loadings recovered from panel data, and shows that policy-relevant counterfactuals can be identified without the distributional restrictions used here. \cite{deaner2025inferring} infer treatment effects in large panels by uncovering latent similarities between units, constructing pairwise discrepancy measures that play the same role as the codegree distance in this paper. Relative to that work, the contribution of the present paper is specific to binary responses: the logistic structure delivers point identification of slope coefficients through an exact conditional likelihood argument, whereas the discrepancy-matching approaches above target average effects of treatments or counterfactual outcome distributions. The two perspectives are complementary: if the object of interest is a treatment effect rather than the index coefficients, the logistic assumption may be dispensable, and I do not claim otherwise. On the inference side, \cite{cattaneo2026robust} develop generalized-jackknife debiasing and bandwidth-robust bootstrap inference for a general class of convex pairwise-difference estimators that match on observed covariates, including the partially linear logit. The present setting differs in that the matching variable is a generated, whole-network statistic, the estimated codegree distance, whose first-stage estimation error and network dependence require the separate treatment developed in Sections \ref{section_Consistency} and \ref{section332}.

The present paper contributes to this literature by focusing on binary response models. While much of the existing work emphasizes linear or partially linear outcome equations, many leading empirical questions involve discrete choices such as participation, adoption, or selection decisions. In such settings, endogeneity arising from latent traits that affect both outcomes and network formation can be particularly severe, and commonly used network controls do not generally resolve it. This paper develops identification and estimation results for a semiparametric logit model in which unobserved heterogeneity enters the outcome equation through an unknown function of a latent individual characteristic that also governs link formation. The identification strategy builds on the idea of network-type equivalence but extends it to binary choice, which changes both the identification argument and the inference theory. I provide a conditional likelihood representation that identifies the slope coefficients through matched comparisons of observationally equivalent agents and develop feasible estimators that implement these comparisons using codegree-based matching.

\section{Model}\label{section3}
Consider a finite set of agents $I = \{1, 2, \cdots, n\}$, where each agent is identified by an observed vector of explanatory variables $X_i \in \R^k$, with a binary outcome $ y_i \in \{0,1\}$, and an unobserved index of social characteristics $\omega_i\in[0,1]$. These variables are related by the following model.
\begin{equation}\label{eq1}
    y_i=\mathbbm{1}\big\{X_i\beta+\lambda(\omega_i)-\varepsilon_i\ge0\big\}
\end{equation}
where $\{\varepsilon_i\}_{i=1}^{n}$ is a collection of independent and identically distributed (i.i.d.) shocks with a logistic cumulative distribution function (cdf) $F$; $\beta\in\R^k$ is an unknown slope parameter. The explanatory variable $X_i$ does not contain an intercept. The function $\lambda: [0,1]\to\R$ is an unknown measurable function, and $\lambda(\omega_i)$ is the realized value of the unobserved heterogeneity for agent $i$: the contribution to the latent index of everything the latent type $\omega_i$ carries, such as motivation, ability, or family expectations, rather than an endogenous interaction effect. The term $\lambda(\omega_i)$ absorbs the intercept; it is the
direct effect of interacting with a particular collection of communities.

In addition to this, the researcher also observes an $n\times n$ binary adjacency matrix $D$. The element $D_{ij}=1$ if there is a direct link between agents $i$ and $j$, and $D_{ij}=0$ otherwise. By convention, any agent $i$ is not allowed to be linked to itself, $D_{ii}=0$. I assume that all links are undirected, so that the adjacency matrix $D$ is symmetric, that is, $D_{ij}=D_{ji}$. Social links are generated according to a nonparametric network formation model. The existence
of a link between agents $i$ and $j$ is determined by 
\begin{equation}\label{eq2}
    D_{ij} = \mathbbm{1}\big\{f(\omega_i,\omega_j)\ge\eta_{ij}\big\}\mathbbm{1}(i\ne j)
\end{equation}
where $f$ is a symmetric measurable function and $\left\{\eta_{ij}\right\}^n_{i,j=1}$
is a symmetric matrix of unobserved scalar disturbances with independent upper diagonal
entries that are mutually independent of $\{x_i,\omega_i,\varepsilon_i\}_{i=1}^n$. The latent variable $\omega_i$ captures characteristics such as motivation, trust, ability, or social capital, which are typically unobserved but play a role in both outcomes and link formation.

The unobserved individual index of social characteristics $\omega_i$ can be interpreted as the social capital that increases the likelihood of forming a link (e.g., socioeconomic status, social ability, family expectation, trust). The difference $f(\omega_i,\omega_j)-\eta_{ij}$ is interpreted as the utility agents $i$ and $j$ receive from forming a link. This implies that for different collections of indices, $D_{ij}$ and $D_{st}$ are independent conditionally on $\omega_i, \omega_j, \omega_s$ and $\omega_t$, i.e., the utility agents receive from forming a link does not depend on the existence of other links in the sample. 

The following examples illustrate applications of the model to the literature.

\begin{example}[Program Participation] \cite{banerjee2013diffusion} model household participation in a microfinance program in which information about the program diffuses over a social network. They measure participation using lending
and trust. Their model could be generalized by adding $\omega_i$ as an index of individual trustworthiness, family expectations, and integrity in financial matters.
\end{example}

\begin{example}[Peer Composition] Consider youth smoking behavior, where students with similar unobserved traits (e.g., risk attitudes) are more likely to form friendships with each other. Let $y_i=1$ if a student smokes and 0 if not, $X_i$ be a vector of student $i$ covariates (age, grade, gender, etc.), and $D_{ij} = 1$ if students $i$ and $j$ are friends and 0
otherwise. A specification consistent with the model sets
\[\lambda(\omega_i)=\gamma'\,\E\left[X_j\left|D_{ij}=1,\omega_i\right.\right]+\theta\,\E\left[y_j\left|D_{ij}=1,\omega_i\right.\right],\]
so that a student's latent propensity depends on the expected covariates and smoking rate among the peers that a student of type $\omega_i$ forms links with. The display defines $\lambda$ implicitly rather than explicitly, since
$\E[y_j|D_{ij}=1,\omega_i]$ depends on $\lambda$ itself through $\PP(y_j=1|\omega_j)=\E_X[F(X_j'\beta+\lambda(\omega_j))]$. Existence and uniqueness follow from a contraction argument. Write $B(u):=\int f(u,v)\,dv$ for the expected degree of a type-$u$ agent, and suppose $\inf_u B(u)\ge b>0$ and $\E\|X\|<\infty$; both hold in every design of Section \ref{section4}. For bounded measurable $\lambda$ define
\[
(T\lambda)(u):=\frac{\gamma'\int f(u,v)\,\E[X|\omega=v]\,dv}{B(u)} +\theta\,\frac{\int f(u,v)\,\E_X\!\left[F\big(X'\beta+\lambda(v)\big)\right]dv}{B(u)} .
\]
Since $F'\le1/4$, for any bounded $\lambda_1,\lambda_2$,
\[
\big|(T\lambda_1)(u)-(T\lambda_2)(u)\big| \ \le\ \frac{|\theta|}{4}\,\frac{\int f(u,v)\,dv}{B(u)}\,\|\lambda_1-\lambda_2\|_\infty \ =\ \frac{|\theta|}{4}\,\|\lambda_1-\lambda_2\|_\infty ,
\]
so $T$ is a contraction in the sup norm whenever $|\theta|<4$ and has a unique bounded fixed point. That fixed point satisfies $\|\lambda\|_\infty\le\|\gamma\|\sup_v\|\E[X|\omega=v]\|+|\theta|<\infty$, which is Assumption \ref{ass1}(vi). It also satisfies Assumption \ref{ass2}, and Assumption \ref{ass_holder}(\ref{ass_holder_iii}) with $\varphi=1$: writing $\lambda(u)=A(u)/B(u)$ with $A(u)=\int f(u,v)g(v)\,dv$ for a bounded $g$ not depending on $u$,
\[
\left|\frac{A(u)}{B(u)}-\frac{A(v)}{B(v)}\right| \ \le\ \frac{2\|g\|_\infty}{b}\,\|f_u-f_v\|_1 \ \le\ \frac{2\|g\|_\infty}{b}\,\|f_u-f_v\|_2 ,
\]
so $\lambda$ is Lipschitz in the link-function distance. Because the fixed point is a functional of $\omega_i$ alone, $\lambda(\omega_i)$ is a deterministic function of $\omega_i$ and the i.i.d.\ structure of Assumption \ref{ass1} is preserved.
Note that this is a model of peer composition operating through the latent type, not a simultaneous-equations model of endogenous peer effects: realized peer outcomes $\frac{1}{s_{1i}}\sum_{j\ne i}D_{ij}y_j$, which would depend on other agents' shocks and violate the i.i.d.\ structure, do not enter the outcome equation.
\end{example}

\begin{example}[Co-authorship] \cite{ductor2014social} study how knowledge about the social network of a researcher, as embodied in his coauthor relations, helps them in developing a prediction of his or her future productivity. In this setting, $\omega_i$ can be interpreted as some unobserved productivity trait that induces the researcher to have more coauthors and also to be more productive at writing papers.
\end{example}

\subsection{Identification}\label{section3_1}
The parameters of interest are $\beta$ and $\lambda(\omega_i)$. The functions $\lambda$ and $f$ will be treated as nuisance functions because they cannot be identified given that $\omega_i$ is unknown. However, the realization of $\lambda$ for a particular individual, $\lambda(\omega_i)$, can be identified. Identification relies on the concept of network formation types. The parameter $\beta$ is identified if $\lambda(\omega_i)$ depends on $\omega_i$ only through the link function $f_{\omega_i}(\cdot) \equiv f (\omega_i,\cdot) : [0, 1]\to [0, 1]$, which is the conditional probability that an agent with social characteristics $\omega_i$ links with agents of every other social characteristic in $[0, 1]$. To compare two agents' types, I define the integrated squared difference in the network types of agents with social
characteristics $\omega_i$ and $\omega_j$ by
\begin{equation}\label{eq3}
   \rho_{ij}=||f_{\omega_i}-f_{\omega_j}||_2=\left[\int\left(f(\omega_i,t)-f(\omega_j,t)\right)^2dt\right]^{1/2}
\end{equation}
That is, if the measure of the network distance between agents $i$ and $j$ equals zero ($ \rho_{ij}=0$), then there is no identifiable characteristic within the network that sets $\omega_i$ apart from $\omega_j$. Given this scenario, agents $i$ and $j$ are equally likely to form connections within any specific network configuration. Consequently, they would have identical distributions of degrees, eigenvector centralities, and average peer characteristics, as well as any other individual-specific statistic of the network representation $D$.

\begin{assumption}[Sampling and disturbances]\label{ass1}
   (i) $(X_i,\omega_i,\varepsilon_i)$ are i.i.d. for all $i=1, \cdots, n$; (ii) The random array $\{\eta_{ij}\}_{i,j=1}^n$ is symmetric and independent of $\{X_i,\omega_i,\varepsilon_i\}_{i=1}^n$ with i.i.d. entries above the diagonal; (iii) $\omega_i$ and  $\eta_{ij}$ have standard uniform marginals; (iv)  $\varepsilon_i$ follows a standard logistic distribution and is independent of $(X_i,\omega_i)$; (v) The binary outcomes $\{y_{i}\}_{i=1}^n$ and the binary adjacency matrix $D$ are given respectively by equations (\ref{eq1}) and (\ref{eq2}); (vi)  $\lambda$ and $f$ are Lebesgue-measurable with $f$ being symmetric in its arguments and $\sup_{u\in[0,1]}|\lambda(u)|<\infty$; (vii) $f$ satisfies the continuity condition that $\inf_{u\in[0,1]}\int\mathbbm{1}\left\{v\in[0,1]:\sup_{\tau\in[0,1]}|f(u,\tau)-f(v,\tau)|<\epsilon\right\}dv>0$ for every $\epsilon>0$.
\end{assumption}

Assumption \ref{ass1}(i) implies that the observables $X_i$ and the unobservable individual characteristics $(\omega_i,\varepsilon_i)$ are randomly drawn. This is a standard assumption in the network literature. Assumption
\ref{ass1}(ii) assumes that the link formation error $\eta_{ij}$ is orthogonal to all other observables and unobservables in the model. This means that the dyad-specific unobservable shock $\eta_{ij}$ from the link formation process does not influence the binary outcomes $\{y_{i}\}_{i=1}^n$. The endogeneity in this model takes the form of a dependence between $X_i$ and the unobserved error $\lambda(\omega_i) + \varepsilon_i$ through $\omega_i$. Assumption \ref{ass1}(iv) requires the outcome shock to be independent of both the covariates and the latent type; this independence is what delivers the conditional probability statement $\PP(y_i=1|X_i,\omega_i)=F(X_i'\beta+\lambda(\omega_i))$, displayed as equation (\ref{e1}) in Appendix \hyperlink{app_theo1}{A.3}, and hence the conditional likelihood argument. From Assumption \ref{ass1}(iii), the distributions of $\omega_i$ and $\eta_{ij}$ are assumed to have standard uniform marginals because I cannot separately identify them from $f$. Under this assumption, $f(\omega_i,\omega_j)$ is the probability that agent $i$ and $j$ are directly linked,  $\PP(D_{ij}=1|\omega_i,\omega_j)=f(\omega_i,\omega_j)$, which implicitly assumes that $f:[0,1]^2\to[0,1]$. Assumption \ref{ass1}(vii), taken from \cite{auerbach2022identification}, is weaker than continuity of $f$: it requires that for every social characteristic $u$ there is a positive mass of other characteristics with uniformly similar linking behavior. It accommodates piecewise-continuous link functions such as stochastic block models. Part (ii) is where the model excludes link externalities: if the surplus of the pair $(i,j)$ depended on the realized links of third parties, as in strategic network formation models \citep{menzel2015strategic}, links would not be conditionally independent given types, and the codegree representation on which estimation rests would fail.

Before stating the identifying restrictions, I define the conditioning objects that appear throughout the paper. All random elements are defined on a common probability space $(\mathcal{W},\mathcal{A},\PP)$, with $\{(X_i,\omega_i,\varepsilon_i)\}$ i.i.d.\ draws and $\{\eta_{ij}\}$ as in Assumption \ref{ass1}; $\E$ denotes expectation under $\PP$, and identification statements are statements about functionals of the induced joint distribution of $(y_i, X_i, D)$. Because $\rho_{ij}$ is a continuously distributed random variable in many designs, the event $\{\rho_{ij}=0\}$ may have probability zero, and conditional expectations given this event are not defined in the elementary sense. Following Section 2.2.2 of \cite{auerbach2022identification}, I therefore define conditional expectations with respect to link functions and network distances as limits of well-defined conditional expectations over shrinking neighborhoods. For any integrable random vector $\Pi_{ij}$ indexed at the pair level,
\begin{equation}\label{eq_condexp}
\E\left[\Pi_{ij}\left|\rho_{ij}=0\right.\right]:=\lim_{h\downarrow0}\E\left[\Pi_{ij}\left|\rho_{ij}\le h\right.\right],
\end{equation}
and, for objects attached to a second, independently drawn agent $j$ and a network type $\varphi$ in the support of $f_\omega$,
\[\E\left[\Pi_j\left|\,f_{\omega_j}=\varphi\right.\right]:=\lim_{h\downarrow0}\E\left[\Pi_j\left|\,\omega_j\in\{u\in[0,1]:||f_u-\varphi||_2\le h\}\right.\right],\]
evaluated at $\varphi=f_{\omega_i}$ when the conditioning is on agent $i$'s network type. The conditioning event involves only $\omega_j$, so the definition never conditions an agent's own outcome on its own type ball.
The inner conditional expectations are well defined for every $h>0$ because Assumption \ref{ass1}(vii) guarantees that the conditioning events have positive probability. Existence of the limits, however, does not follow from integrability of $\Pi_{ij}$ alone: it is a restriction on the joint distribution of $(\Pi_{ij},\rho_{ij})$ and, for the second display, of $(\Pi_j,\omega_j)$. Throughout the paper I therefore maintain the following regularity condition wherever the convention is invoked: there exists a version $m$ of the conditional mean, $m(r):=\E[\Pi_{ij}\,|\,\rho_{ij}=r]$, that is continuous at $r=0$ along the support of $\rho_{ij}$. Under this condition the limit in (\ref{eq_condexp}) exists, is unique, and equals $m(0)$: writing $\E[\Pi_{ij}\,|\,\rho_{ij}\le h]=\E\big[m(\rho_{ij})\,\big|\,\rho_{ij}\le h\big]$ by the tower property,
\[\big|\E[\Pi_{ij}\,|\,\rho_{ij}\le h]-m(0)\big|\ \le\ \sup_{r\in[0,h]\cap\,\mathrm{supp}(\rho_{ij})}|m(r)-m(0)|\ \longrightarrow\ 0\qquad(h\downarrow0),\]
and continuity pins the version down uniquely on the support near zero, so the convention is well defined. The same condition, with $m$ a version of $u\mapsto\E[\Pi_j\,|\,\omega_j=u]$ continuous at the conditioning type with respect to the pseudometric $(u,v)\mapsto\|f_u-f_v\|_2$, makes the second display well defined. The condition is imposed only on the finitely many conditional means to which the convention is actually applied in this paper -- the conditional outcome probabilities and logit-score objects entering Theorem \ref{theo1} and the localized objectives of Section \ref{section_Consistency} -- not on all integrable $\Pi_{ij}$. Two remarks are in order. First, when the distribution of network types has finite support, as in stochastic block models where $\PP(\rho_{ij}=0)>0$, the limit in (\ref{eq_condexp}) coincides with the elementary conditional expectation, and no limiting argument is needed. Second, all identification statements below should be read with these definitions in place.

The identifying restriction links network equivalence to the unobserved heterogeneity term.

\begin{assumption}\label{ass2} For every $\epsilon>0$ there exists a $\delta>0$ such that
$$\sup_{u,v\in[0,1]:\,||f_u-f_v||_2\le\delta}\big(\lambda(u)-\lambda(v)\big)^2\le\epsilon.$$
\end{assumption}

 Assumption \ref{ass2} states that agents with similar network types have similar values of $\lambda$: the map from link functions to unobserved heterogeneity is uniformly continuous. It implies the conditional-moment restriction $\E[(\lambda(\omega_i)-\lambda(\omega_j))^2|\rho_{ij}=0]=0$, and it has two features that the conditional-moment formulation lacks: it is stated directly in terms of the model primitives $(\lambda,f)$, so it involves no conditioning on measure-zero events, and it provides the modulus-of-continuity control needed for the asymptotic theory of Section \ref{section_Consistency}, where matches are inexact. The assumption corresponds to Assumption 3 in \cite{auerbach2022identification}. One justification is that $\omega_i$ does not affect $y_i$ directly but only by altering the agent's interactions, which are summarized by the linking behavior $f_{\omega_i}$. Under Assumption \ref{ass2}, $\rho_{ij}=0$ implies $\lambda(\omega_i)=\lambda(\omega_j)$, but does not imply that $\omega_i=\omega_j$. The assumption fails when the outcome depends on the type beyond its effect on linking behavior: Section \ref{sec_misspec} simulates a block design in which $\lambda$ varies within blocks of network-equivalent agents, and Table \ref{tab_misspec} shows the resulting bias, which no bandwidth removes.

Identification of the slope also requires that the covariates vary among matched agents.

\begin{assumption}\label{ass_rank} 
\begin{enumerate}[(i)]
\item $\E\left(||X_i||^2\right)<\infty$;
\item  $\Xi_0:=\E[(X_i-X_j)(X_i-X_j)'\,|\,\rho_{ij}=0]$ exists and is positive definite, and the same holds with the conditioning event $\{\rho_{ij}=0\}$ replaced by $\{\delta_{ij}=0\}$ (limits taken along the corresponding shrinking balls); 
\item The conditional distributions of $\|X_i-X_j\|^2$ given $\rho_{ij}\le h$, indexed by $h\in(0,h_0]$, are uniformly integrable:
\[\lim_{M\to\infty}\ \sup_{0<h\le h_0}\ \E\left[\|X_i-X_j\|^2\,\mathbbm{1}\left\{\|X_i-X_j\|>M\right\}\,\Big|\,\rho_{ij}\le h\right]=0,\]
and $$\sup_{0<h\le h_0}\E\big[\,|X_i'\beta+\lambda(\omega_i)|+|X_j'\beta+\lambda(\omega_j)|\,\big|\,\rho_{ij}\le h\big]<\infty;$$ both bounds also hold with the conditioning event $\{\rho_{ij}\le h\}$ replaced by the codegree ball $\{\delta^2_{ij}\le h\}$.
\end{enumerate}

\end{assumption}

Assumption \ref{ass_rank} is the analogue of the standard no-multicollinearity condition in the conditional logit: within pairs of agents with identical network formation behavior, no linear combination of the covariates is degenerate. It fails, for instance, if a covariate is itself a deterministic function of the network type (such as the population analogue of an agent's degree). 
Part (iii) makes explicit the uniform integrability needed to interchange the limit $h\downarrow0$ with expectations over the shrinking conditioning events in the proof of Theorem \ref{theo1}; it is not implied by part (i), because conditional moments over shrinking events need not converge. A simple sufficient condition is $\E\|X_i\|^{2+\nu}<\infty$ for some $\nu>0$ together with $\PP(\rho_{ij}\le h|X_i, X_j)\le C\,\PP(\rho_{ij}\le h)$ almost surely for all small $h$ (the matching event is not driven disproportionately by extreme covariate profiles); bounded covariates trivially suffice.

\begin{theorem}\label{theo1}
    Under Assumptions \ref{ass1}, \ref{ass2} and \ref{ass_rank}, $\beta$ and $\lambda(\omega_i)$ are uniquely identified:
    \begin{equation}\label{eq4}
        \beta = \arg\min_{b\in\R^k} -\E\Big[y_{i}\log F[(X_i-X_j)'b]+y_j\log F[(X_j-X_i)'b]\,\Big|\,\rho_{ij}=0,\,y_i+y_j=1\Big]
    \end{equation}
and
\begin{equation}\label{eq5}
    \lambda(\omega_i) = F^{-1}\big(\PP(y_j=1\,|\,X_j=x,\ f_{\omega_j}=f_{\omega_i})\big)-x'\beta,
\end{equation}
valid for almost every $x$ in the support of the conditional distribution of $X_j$ given $f_{\omega_j}=f_{\omega_i}$, where $j$ indexes an agent drawn independently of $i$, the conditional objects are defined as the limits stated before Assumption \ref{ass2}, the minimizer in (\ref{eq4}) is unique, the conditional probability in (\ref{eq5}) denotes the regular conditional version given $X_j$, and the right-hand side of (\ref{eq5}) is the same for almost every such $x$.
\end{theorem}

The proof of Theorem \ref{theo1} is provided in Appendix \hyperlink{app_theo1}{A.3}. Four features of the proof deserve comment. First, the argument conditions on the latent pair $(\omega_i,\omega_j)$ and uses Assumption \ref{ass2} to control the difference $\lambda(\omega_i)-\lambda(\omega_j)$ uniformly over the shrinking conditioning neighborhoods; the function $\lambda$ is treated as random throughout, never as a constant. Second, uniqueness of the minimizer follows from an information-inequality argument: the objective in (\ref{eq4}) is the expected negative log-likelihood of the conditional-logit family, its excess over the value at $\beta$ is a Kullback--Leibler divergence, and this divergence is strictly positive for $b\neq\beta$ by the positive definiteness of $\Xi_0$ in Assumption \ref{ass_rank}. Jensen's inequality alone would deliver only the weak inequality $\Omega_0(\beta)\le\Omega_0(b)$. Third, when network types have finite support, the statement reduces to a classical conditional-likelihood identification result with no limits involved. Fourth, the conditioning objects in the theorem, $\rho_{ij}$ and $f_{\omega_i}$, are latent rather than observable. The theorem is therefore a statement about the population distribution of $(y_i, X_i, \omega_i)$, and its observable content comes from Section \ref{section3}: the codegree distance $\delta_{ij}$ is a functional of the distribution of the adjacency matrix, $\rho_{ij}=0$ holds if and only if $\delta_{ij}=0$, and both displays continue to hold with $\{\rho_{ij}=0\}$ replaced by $\{\delta_{ij}=0\}$ and network-type classes replaced by codegree classes; Step 3 of the proof covers the codegree-ball localization explicitly. In this sense, the conditioning can be expressed entirely through objects identified from the distribution of the observed network, and $\lambda(\omega_i)$ is identified at the level of network-type equivalence classes, which under Assumption \ref{ass2} is identification of the value $\lambda(\omega_i)$ itself, although $\omega_i$ is not identified. Finally, the restriction of (\ref{eq5}) to the conditional support is not innocuous: because $X$ may depend on $\omega$, the support of $X_j$ given $f_{\omega_j}=f_{\omega_i}$ can be a strict subset of $\mathrm{supp}(X)$, and the conditional probability in (\ref{eq5}) is not identified at covariate values outside it. A common-support condition in the spirit of Assumption \ref{ass_lambda}(iii)--(iv), requiring a set $\X_0$ of covariate values with positive conditional mass under every network type, restores a version of (\ref{eq5}) stated over a set of $x$ common to all types; this is what the estimator of Section \ref{section3_2} implements through the trimming set $\X_0$.

\subsection{Estimation}\label{section3_2}
This section constructs feasible estimators of the parameters identified in Section \ref{section3_1}. If the individual social characteristics $\{\omega_i\}_{i=1}^n$ were observed then I could use existing tools to estimate $\beta$ and $\lambda(\omega_i)$. The link formation model in (\ref{eq2}) will not provide useful information in this case. The researcher will just have to use observations for which $\omega_i$ is close to $\omega_j$ \citep{honore1997pairwise,aradillas2007pairwise}. However, since individual social characteristics $\{\omega_i\}_{i=1}^n$ are not observed and the conditional mean functions in Theorem \ref{theo1} are also not known, estimates are not feasible. 

To make estimation feasible, I use the theorem of \cite{lovasz2012large}, which demonstrates that pairs of individuals with identical link functions have identical codegree functions. Following \cite{auerbach2022identification}, the codegree function is defined by $p(\omega_i,\omega_j)=\int f_{\omega_i}(s)f_{\omega_j}(s)ds$, which is the probability that agents $i$ and $j$ have a link in common. Agent $i$'s codegree type is defined as $p_{\omega_i}(\cdot)\equiv p(\omega_i,\cdot): [0,1]\to[0,1]$. The pseudometric codegree distance between agents $i$ and $j$ is defined by:
\begin{align*}
    \delta_{ij}=\|p_{\omega_i}-p_{\omega_j}\|_2 =\left[\int\left(\int f(t,s)\left(f(\omega_i,s)-f(\omega_j,s)\right)ds\right)^2dt\right]^{1/2}
\end{align*}
which can be consistently estimated\footnote{\cite{auerbach2022identification} shows in Lemma B1 that $\hat{\delta}_{ij}$ converges uniformly to $\delta_{ij}$ over the $\binom{n}{2}$
distinct pairs of agents as $n\to\infty$.} by the root average squared difference in the $i$th and $j$th columns of the squared adjacency matrix
\begin{align*}
    \hat{\delta}_{ij}=\left[\frac{1}{n}\sum_{t=1}^n\left(\frac{1}{n}\sum_{s=1}^n D_{ts}\left(D_{is}-D_{js}\right)\right)^2\right]^{1/2}
\end{align*}
The estimation strategy relies on the relationship between the latent network distance $\rho_{ij}$ and the codegree distance $\delta_{ij}$. Lemma 1 in \cite{auerbach2022identification} establishes two facts under Assumption \ref{ass1}: first, 
$\delta_{ij}\le\rho_{ij}$ always (an immediate consequence of the Cauchy--Schwarz inequality together with $\|f_\tau\|_2\le1$); second, under the continuity condition in Assumption \ref{ass1}(vii), for every $\varepsilon>0$ there exists an $\eta>0$ such that $\rho_{ij}\,\mathbbm{1}\{\delta_{ij}\le\eta\}\le\varepsilon$. Together these imply that agents with similar codegree functions have similar link functions, and in particular $\rho_{ij}=0\iff\delta_{ij}=0$. The result does not require the integral operator associated with $f$ to be injective. The proof is self-contained: if $||p_{\omega_i}-p_{\omega_j}||_2=0$, then $g(\tau):=\int f(\tau,s)(f_{\omega_i}(s)-f_{\omega_j}(s))ds=0$ for almost every $\tau$. This does not by itself license evaluating $g$ at the two random points $\omega_i$ and $\omega_j$, since the exceptional null set may contain them; the continuity condition in Assumption \ref{ass1}(vii) closes the gap. For every $\epsilon>0$ that condition provides a positive-measure set of types $v$ with $\sup_\tau|f(\omega_i,\tau)-f(v,\tau)|<\epsilon$, and because $g$ vanishes off a null set, some such $v$ has $g(v)=0$; then $|g(\omega_i)|=|g(\omega_i)-g(v)|\le\int|f(\omega_i,s)-f(v,s)|\,|f_{\omega_i}(s)-f_{\omega_j}(s)|\,ds\le2\epsilon$, and letting $\epsilon\downarrow0$ gives $g(\omega_i)=0$; identically $g(\omega_j)=0$. Hence $0=g(\omega_i)-g(\omega_j)=\int(f_{\omega_i}(s)-f_{\omega_j}(s))^2ds$, that is, $\rho_{ij}=0$. The argument uses only Assumption \ref{ass1}(vii) and no injectivity of the integral operator; in particular it covers the stochastic block design of Section \ref{section4}, whose operator has finite rank and is therefore not injective on $L^2[0,1]$, and it imposes nothing beyond Assumption \ref{ass1}(vii) on smooth designs such as the beta model.

Three approximations therefore separate the feasible estimator from the identification result in Theorem \ref{theo1}: (i) the latent distance $\rho_{ij}$ is replaced by the codegree distance $\delta_{ij}$, which is coarser but observable in the limit; (ii) the population codegree distance is replaced by its sample analogue $\hat{\delta}_{ij}$, whose estimation error must be accounted for; and (iii) the event $\{\delta_{ij}=0\}$ is replaced by kernel-localized comparisons $\{\hat{\delta}^2_{ij}\le h\}$ with a positive bandwidth, so that matches are inexact whenever network types vary continuously. None of these substitutions is innocuous, and the asymptotic theory of Section \ref{section_Consistency} treats all three explicitly: step (ii) contributes first-stage variance terms, and step (iii) generates a matching bias that vanishes with the bandwidth and that the bias correction of Section \ref{section332} removes.

Before stating the conditions on the kernel and bandwidth, define the population match density
\[r_n(u):=\E\left[K\left(\frac{\delta(u,\omega_j)^2}{h}\right)\right] \qquad\mbox{and}\qquad r_n:=\E\left[r_n(\omega_i)\right]=\E\left[K\left(\frac{\delta^2_{ij}}{h}\right)\right],\]
where $\delta(u,v):=||p_u-p_v||_2$. The quantity $n\,r_n(u)$ measures the effective number of agents matched to an agent of type $u$ at bandwidth $h$.

\begin{assumption}\label{ass3} The following statements about the kernel function $K$ and the bandwidth $h$ hold:
\begin{enumerate}[(i)]
    \item $K:[0,\infty)\to[0,\infty)$ is bounded, nonincreasing, twice continuously differentiable on its domain $[0,\infty)$ (with one-sided derivatives at zero) with bounded derivatives $K'$ and $K''$, has support $[0,1)$, and $K(0)>0$; because the kernel argument $\hat{\delta}^2_{ij}/h$ is non-negative, no extension of $K$ to negative arguments is used anywhere; \label{ass3i}
    \item $h\to0$, $n^{1-\gamma}h^2\to\infty$, and $\inf_{u\in[0,1]}n^{\gamma/4}\,r_n(u)\to\infty$ as $n\to\infty$, for some $\gamma\in(0,1)$.\label{ass3ii}
\end{enumerate}
\end{assumption}
Assumption \ref{ass3}(\ref{ass3i}) makes $K$ a hard-window kernel on the squared-distance scale: only pairs with $\hat\delta^2_{ij}<h$ receive positive weight. Assumption \ref{ass3}(\ref{ass3ii}) plays the role of Assumption 4 in \cite{auerbach2022identification}, but is stated in model primitives rather than as a high-level condition of the form $n\E[K(\hat\delta^2_{ij}/h)]\to\infty$ on the estimated distances.\footnote{Because $p_{\omega_i}$ is an unknown function, $\PP(||p_{\omega_i}-p_{\omega_j}||^2_2\le h\,|\,\omega_i)$ cannot be approximated by a polynomial in $h$ of known order, which is why the divergence of $n^{\gamma/4}r_n(u)$ is assumed directly rather than derived from a density. Primitive sufficient conditions are available when $f$ satisfies the H\"older-type condition introduced in Assumption \ref{ass_holder} below, in which case $r_n(u)\ge\underline{K}\,(h/C)^{1/\alpha}$ for constants $\underline{K},C,\alpha>0$.} Its two substantive requirements are transparent: the bandwidth may not shrink so fast that the first-stage estimation error in $\hat\delta_{ij}$ dominates the kernel window, and the effective number of matches must diverge for every network type.

Let $\hat q_c$ denote the empirical $c$-quantile of the off-diagonal estimated distances $\{\hat\delta^2_{ij}\}_{i<j}$, and let
\begin{equation}\label{eq_noisefloor}
\hat m:=\operatorname*{median}_{(i,j)}\ \frac{1}{n^3}\sum_{t=1}^n\sum_{s=1}^nD_{ts}\left(D_{is}-D_{js}\right)^2
\end{equation}
(computable over a random subsample of at least $n$ pairs). The statistic $\hat m$ estimates the first-stage noise scale: conditional on the latent types, $\hat\delta^2_{ij}$ for an exactly matched pair has strictly positive mean,\footnote{Equal to $n^{-3}\sum_{t,s}f(\omega_t,\omega_s)\,\E(D_{is}-D_{js})^2$.} so identical agents concentrate at $\hat m$ rather than at zero. The bandwidth rule is
\begin{equation}\label{eq_bandwidthrule}
\hat h_n=\begin{cases}\ \max\left(\hat q_{c_n},\ n^{-2/5}\,\hat q_{1/2}\right) & \text{(continuous regime)},\\[4pt]
\ (3\,\hat m)^{1/3}\,\hat q_{1/2}^{\,2/3} & \text{(finite-support regime)},\end{cases}
\end{equation}
with $c_n=\min\{0.05,\ 0.05\,(n/100)^{-1/16}\}$ in the continuous branch. The regime is read from the empirical distribution of $\hat\delta^2_{ij}$: a nonvanishing point mass at the noise scale $\hat m$, separated from the remaining mass, identifies the finite-support regime. There $\hat h_n$ is placed two-thirds of the way in logarithms between the noise scale and the bulk, so that essentially all exact matches are captured while $\hat h_n$ dominates the first-stage noise; since $\hat m\asymp_p n^{-1}$, this branch is of order $n^{-1/3}$. Lemma \ref{lem_bandwidth}(b) additionally requires exact matches to comprise less than half of all pairs, so that the median distance stays away from zero; a design concentrating more than half of its pairs at zero distance calls for a quantile above the exact-match mass in place of $\hat q_{1/2}$. In the continuous regime, the fraction $c_n$ shrinks with $n$, because a fixed-fraction quantile rule converges to a positive constant and violates $h\to0$. The floor is stated relative to $\hat q_{1/2}$, because an absolute floor can exceed the entire range of the codegree distances. Appendix~\hyperlink{app_bandwidth}{A.2} verifies that the continuous branch satisfies both Assumption \ref{ass3}(\ref{ass3ii}) and the narrower window of Assumption \ref{ass_holder}(\ref{ass_holder_iv}), and that the finite-support branch satisfies Assumption \ref{ass3}(\ref{ass3ii}), which is what Theorem \ref{theo3} requires. The finite-support branch is of order $n^{-1/3}$ and therefore lies outside the window of Assumption \ref{ass_holder}(\ref{ass_holder_iv}); this is not a defect, because in that regime matching is exact and Theorem \ref{theo3} applies without the smoothness conditions of Assumption \ref{ass_holder}. Lemma \ref{lem_bandwidth} shows that this data-driven bandwidth is asymptotically equivalent to a deterministic counterpart, so the results below hold with $h$ replaced by $\hat h_n$.

Under Assumptions \ref{ass1}-\ref{ass4} and the parameter-space conditions of Theorem \ref{theo2} below, $\beta$ is consistently estimated by
\begin{equation}\label{eq6}
    \hat{\beta} =\arg\min_{b\in\R^k}\ -\sum_{y_i\ne y_j}K\left(\frac{\hat{\delta}^2_{ij}}{h}\right)\Big\{y_{i}\log F\left[(X_i-X_j)'b\right]+y_j\log F\left[(X_j-X_i)'b\right]\Big\}
\end{equation}
This estimator can be interpreted as a kernel-weighted conditional logit estimator. The estimator of the unobserved heterogeneity term, $\lambda(\omega_i)$, is then given by
\begin{equation}\label{eq7}
    \hat{\lambda}(\omega_i)=\Big[\sum_{j=1}^nK\left(\frac{\hat{\delta}^2_{ij}}{h}\right)\Big]^{-1}\Big[\sum_{j=1}^n\left(F^{-1}\left(\widehat{\PP}\left(y_j=1\left|X_j,f_{\omega_j}\right.\right)\right)-X_j'\hat{\beta}\right)K\left(\frac{\hat{\delta}^2_{ij}}{h}\right)\Big]
\end{equation}
where
\begin{align*}
\widehat{\PP}\left(y_j=1\left|X_j,f_{\omega_j}\right.\right)={}\Big[\sum_{t=1}^nK\left(\frac{\hat{\delta}^2_{jt}}{h}\right)K_X\big(\tfrac{X_t-X_j}{h_X}\big)\Big]^{-1}\Big[\sum_{t=1}^ny_tK\left(\frac{\hat{\delta}^2_{jt}}{h}\right)K_X\big(\tfrac{X_t-X_j}{h_X}\big)\Big]
\end{align*}
is an estimator for $\PP\left(y_j=1\left|X_j,f_{\omega_j}\right.\right)=\E\left[y_j\left|X_j,f_{\omega_j}\right.\right]$; $K_X$ is a (product) kernel with bandwidth $h_X$ applied to the continuously distributed components of $X$, discretely distributed components being matched exactly (cell matching), and $\widehat{\PP}$ is clipped to $[\tau_n,1-\tau_n]$ before $F^{-1}$ is applied. The conditions governing $K_X$, $h_X$, the trimming sequence $\tau_n$, and the distribution of $X$ under which $\widehat{\PP}$ is consistent are collected in Assumption \ref{ass_lambda} below. Each term in the outer sum of (\ref{eq7}) is an estimate of $F^{-1}(\PP(y_j=1|X_j,f_{\omega_j}))-X_j'\beta=\lambda(\omega_j)$ for a matched agent $j$; averaging these over agents with $\hat{\delta}^2_{ij}$ small recovers $\lambda(\omega_i)$, since matched agents have similar values of $\lambda$ by Assumption \ref{ass2}. Note that the inverse-probability term and the index correction $X_j'\hat{\beta}$ are evaluated at the same agent $j$: pairing the $i$-indexed probability with a $j$-indexed index correction would instead converge to $X_i'\beta+\lambda(\omega_i)-\beta'\E[X_j|f_{\omega_i}]\neq\lambda(\omega_i)$. The term $K\left(\frac{\hat{\delta}^2_{ij}}{h}\right)$ gives more weight to pairs of observations $(i, j)$ with similar codegree functions. This estimator averages over agents with similar network types, thereby recovering the unobserved heterogeneity term up to network-type equivalence classes.

The consistency and asymptotic normality of these estimators will be developed in the next section.

\subsection{Consistency and Asymptotic Normality}\label{section_Consistency}

This section establishes the large-sample properties of the estimators of Section \ref{section3_2}; proofs are in the Appendix.
\\
Let 
\[m(v_i,v_j,b) =-\mathbbm{1}(y_i\ne y_j)\left\{y_{i}\log F\left[(X_i-X_j)'b\right]+y_j\log F\left[(X_j-X_i)'b\right]\right\}\]
where $v_i=(y_i,X_i)$. The estimator $\hat{\beta}$ in (\ref{eq6}) minimizes the self-normalized sample objective function
\begin{equation}\label{eq8}
    Q_n(\hat{\delta},b)=\hat{r}_n^{-1}\binom n2^{-1}\sum_{i<j}K\left(\frac{\hat{\delta}^2_{ij}}{h}\right)m(v_i,v_j,b); \  \hat{r}_n=\binom n2^{-1}\sum_{i<j}K\left(\frac{\hat{\delta}^2_{ij}}{h}\right).
\end{equation}
The normalization by $\hat{r}_n$ does not affect the minimizer, so $\hat{\beta}$ is exactly the estimator in (\ref{eq6}); it is introduced because the codegree functions are infinite dimensional, so there is no adequate notion of a density for $\delta_{ij}$ and the usual normalization by $1/h$ is not appropriate (see the discussion following Proposition 2 in \cite{auerbach2022identification}). Define the population counterparts
\begin{align*}
Q_h(b)&:=r_n^{-1}\,\E\left[K\big(\delta^2_{ij}/h\big)m(v_i,v_j,b)\right] \quad\mbox{and}\\
Q_0(b)&:=\lim_{h\downarrow0}\E\left[m(v_i,v_j,b)\,\middle|\,\delta^2_{ij}\le h\right],
\end{align*}
so that $Q_0$ is the limiting objective function, with a unique minimizer at $\beta$ by Theorem \ref{theo1} and Lemma 1 of \cite{auerbach2022identification}. In the following two subsections, I provide conditions under which this estimator is consistent and asymptotically normal.

\subsubsection{Consistency}
Consistency follows from pointwise convergence of the convex sample objective (\ref{eq8}) to $Q_0$, together with the strict convexity of $Q_0$ at $\beta$; the convexity of $m(v_i,v_j,\cdot)$, inherited from the logit log-likelihood, makes uniform convergence arguments unnecessary \citep[Theorem 2.7]{newey1994large}.

\begin{assumption}\label{ass4}
All the following statements hold for every $b$ in the parameter space:
\begin{enumerate}
        \item $\E\left[m(v_i,v_j,b)^2\right]<\infty$; \label{ass4ii}
        \item $\E\left[\|X_i\|^4\,\big|\,\omega_i\right]\le C_X$ almost surely, for a constant $C_X<\infty$;\label{ass4iii}
\item $\sup_{u\in[0,1]}r_n(u)\le\bar C\,\inf_{u\in[0,1]}r_n(u)$ for all small $h$ (balanced matching intensity), and the limit defining $Q_0(b)$ exists;
\item $\E[m(v_i,v_j,b)^2\,|\,\omega_i,\omega_j]$ is bounded almost surely, and \\ $\limsup_{n}\PP(\delta^2_{ij}\le2h)/r_n<\infty$.\label{ass4iv}
    \end{enumerate}
\end{assumption}
Part 1 makes the localized objective square integrable. Part 2 is the conditional-moment version of the covariate condition used in the distribution theory (Appendix \hyperlink{app_theo4}{A.6}). Part 3 requires that no network type is matched at a vanishing relative rate, so the effective number of matches is comparable across types; it holds by direct computation in every design of Section \ref{section4}.\footnote{Evaluating $r_n(u)=\int K(\delta^2(u,v)/h)\,dv$ on a $4001$-point grid, the ratio $\sup_u r_n(u)/\inf_u r_n(u)$ settles at $1.00$ in the stochastic block design, where $\delta^2(u,v)$ is block-constant; at $3.06$, $3.09$ and $3.11$ in the beta design at $h=10^{-5},10^{-6},10^{-7}$; and near $17$ in the homophily design over the same range. The ratio is bounded because the small-ball exponent is common across types: $r_n(u)\asymp h^{1/2}$ at both the minimizing and the maximizing $u$ in each continuous design. The infimum is attained at the boundary types $u\in\{0,1\}$, whose matching neighbourhoods are one-sided, and the supremum in the interior; the larger constant under homophily reflects the additional variation in the local scale of $\delta(u,\cdot)$ across $u$, which is absent under $f_2$.} Parts 1 and 2 hold whenever the covariate tails are light enough for the logit log-likelihood to be square integrable, since $|m(v_i,v_j,b)|\le|\Delta X_{ij}'b|+2\log2$ on the parameter space. Part 4 does double duty in the proof of Theorem \ref{theo2}: the conditional moment bound delivers the variance bound in Step 3, and the kernel-mass condition $\limsup_n\PP(\delta^2_{ij}\le2h)/r_n<\infty$ controls the first-stage substitution error in Step 4. The kernel-mass condition holds when network types have finite support; in the continuous regime it is a doubling condition on the small-ball function of the codegree distance, which the polynomial lower bound of Assumption \ref{ass_holder}(\ref{ass_holder_ii}) does not by itself imply, and it is therefore imposed directly.

Consistency of $\hat\lambda(\omega_i)$ additionally requires conditions on the nonparametric regression embedded in (\ref{eq7}), which are absent from Assumptions \ref{ass1}-\ref{ass4}: conditions on the covariate kernel $K_X$ and its bandwidth $h_X$, on the conditional distribution of $X$, and a trimming device that keeps $F^{-1}$ well behaved.

\begin{assumption}\label{ass_lambda}
(i) $X=(X^c,X^d)$, where $X^d$ has finite support and $X^c\in\R^{k_c}$; the estimator $\widehat{\PP}$ matches exactly on the cells of $X^d$ and applies the product kernel $K_X$ to $X^c$; (ii) $K_X$ is a bounded, symmetric, integrable second-order kernel on $\R^{k_c}$; $h_X\to0$ and $n\,r_n\,h_X^{k_c}\to\infty$; (iii) there is a compact set $\X_0$ with $\PP(X\in\X_0)>0$ on which, uniformly for all small $h$, the conditional density of $X^c$ given each cell of $X^d$ and the event $\omega_j\in\{u:\|f_u-f_{\omega}\|_2\le h\}$ is bounded away from zero and infinity, and $x^c\mapsto\PP(y=1|X=x,f_\omega)$ is continuous on $\X_0$, uniformly over such $f$-balls; (iv) the outer average in (\ref{eq7}) is restricted to matched agents $j$ with $X_j\in\X_0$, and $\widehat{\PP}$ is clipped to $[\tau_n,1-\tau_n]$ with $\tau_n\to0$ and $\tau_n\gg\big(\log(n\,r_n\,h_X^{k_c})/(n\,r_n\,h_X^{k_c})\big)^{1/2}$.
\end{assumption}

These are the standard conditions for kernel regression, localized to $f$-balls; the factor $r_n$ in the effective sample size $n\,r_n\,h_X^{k_c}$ accounts for the network matching. Part (iii) fails, for example, if a continuous covariate has vanishing density on part of $\X_0$ within a matched set, in which case the denominator of $\widehat{\PP}$ degenerates there; shrinking $\X_0$ or enlarging $\tau_n$ handles such regions at the cost of estimating $\lambda$ on a smaller set.

\begin{theorem}\label{theo2}
    Suppose Assumptions \ref{ass1}-\ref{ass4} hold and the parameter space for $b$ is convex, compact, and includes the true value of $\beta$ in its interior. Then the estimator $\hat{\beta}$ of $\beta$ defined in (\ref{eq6}) is consistent. If in addition Assumption \ref{ass_lambda} holds, the estimator $\hat{\lambda}(\omega_i)$ of $\lambda(\omega_i)$ defined in (\ref{eq7}) is consistent.
\end{theorem}
The proof of Theorem \ref{theo2} is provided in  Appendix \hyperlink{app_theo2}{A.4}. Two remarks on the structure of the argument. First, the convergence of the expected localized objective to $Q_0$ is established by conditioning on the latent pair $(\omega_i,\omega_j)$ and using the localization of the kernel in the codegree distance together with Lemma 1 of \cite{auerbach2022identification} and Assumption \ref{ass2}; no substitution of the kernel argument $\delta^2_{ij}/h$ by differences in $\lambda$ is involved. Second, the effect of replacing $\delta_{ij}$ with $\hat{\delta}_{ij}$ is controlled by the uniform convergence rate of the estimated codegree distances, $\max_{i\neq j}|\hat{\delta}^2_{ij}-\delta^2_{ij}|=O_p(n^{-1/2}\log^{1/2} n)$ (Lemma \ref{lemB1} in the Appendix, adapted from Lemma B1 of \cite{auerbach2022identification}), and the bandwidth conditions in Assumption \ref{ass3}(\ref{ass3ii}) guarantee that this first-stage error is negligible relative to the kernel window.

\subsubsection{Asymptotic Normality}\label{section332}
To derive asymptotic normality, note that $\hat\beta$
 is an extremum estimator based on a U-statistic--type objective function in which the weights depend on the estimated codegree distances $\hat{\delta}_{ij}$, which are themselves functions of the entire adjacency matrix. Two consequences follow. First, the sample score is not a standard U-statistic, and the first-stage estimation error in $\hat{\delta}_{ij}$ contributes to the asymptotic variance. Second, whenever network types vary continuously, matches at a positive bandwidth are inexact, so the localized population score is not exactly zero at $\beta$: the estimator is centered at a pseudo-true value $\beta_h$ that converges to $\beta$ as $h\to0$, but in general too slowly for $\sqrt{n}(\hat\beta-\beta)$ to be asymptotically centered at zero. This bias is standard for matching estimators \citep{abadie2006large} and is present already in the linear setting of \cite{auerbach2021inference}. I therefore present two asymptotic normality results, paralleling Propositions C1 and C2 of \cite{auerbach2021inference}: an exact root-$n$ result for the empirically important case in which network types have finite support, where matches are eventually exact and no bias arises, and a general result centered at $\beta_h$, followed by a jackknife bias correction.

Throughout, recall the pairwise score
\[s(v_i,v_j,b)=\mathbbm{1}(y_i\ne y_j)\left(y_i- F\left[(X_i-X_j)'b\right]\right)(X_i-X_j),\]
and define
\begin{align*}
\Sigma_h:=r_n^{-1}\E\Big[\mathbbm{1}(y_i\ne y_j)\,&F\big((X_i-X_j)'\beta\big)F\big((X_j-X_i)'\beta\big)\\
&\times(X_i-X_j)(X_i-X_j)'\,K\big(\delta^2_{ij}/h\big)\Big],
\end{align*}
with limit $\Sigma_0:=\lim_{h\downarrow0}\Sigma_h$. Lemma \ref{lem_sigma} in the Appendix shows that this limit exists and that $\Sigma_0$ is positive definite. Positive definiteness does not follow from the logit weights alone: the weight $F(u_{ij})F(-u_{ij})$ equals $F'(u_{ij})$ and vanishes as the index diverges, while $\Delta X_{ij}$ is unbounded, so the weight is smallest precisely where $(v'\Delta X_{ij})^2$ is largest. The proof therefore truncates on $\{\|X_i\|\vee\|X_j\|\le M\}$, on which both the logit weight and the
probability of a discordant pair are bounded below by constants depending only on $M$, and then sends $M\to\infty$ using the uniform integrability of Assumption \ref{ass_rank}(iii) and the positive definiteness of $\Xi_0$ in Assumption \ref{ass_rank}(ii).

\paragraph{Finite support} The first result concerns the case in which the distribution of network types has finite support, in the sense that exact matches occur with positive probability and near-matches are exactly ties. This case covers the stochastic block model of \cite{holland1983stochastic}, arguably the leading empirical specification of discrete latent heterogeneity.

\begin{assumption}\label{ass5} $\PP(\delta_{ij}=0)>0$ and there exists $\bar{\epsilon}>0$ such that $\PP\left(0<\delta^2_{ij}<\bar{\epsilon}\right)=0$.
\end{assumption}

The assumption says that codegree types are separated: a pair is either an exact match or at least $\bar\epsilon$ away. It holds for any block structure with finitely many blocks whose codegree profiles differ, such as the design $f_3$ of Section \ref{section4}, and it fails whenever the type enters link probabilities continuously, as under the homophily and beta designs; Theorem \ref{theo4} covers that case.

\begin{theorem}\label{theo3}
    Under Assumptions \ref{ass1}-\ref{ass4} and \ref{ass5}, and the parameter-space hypotheses of Theorem \ref{theo2}, the estimator $\hat{\beta}$ of $\beta$ defined in (\ref{eq6}) satisfies
    \[\sqrt{n}\left(\hat{\beta}-\beta\right)\xrightarrow{d}\mathcal{N}
    \left(0,4\Sigma_0^{-1}V_0\Sigma_0^{-1}\right)\]
    where, with $\pi_0:=\PP(\delta_{ij}=0)$ and $\pi(\omega_i):=\PP(\delta_{ij}=0|\omega_i)$,
    \begin{align*}
    V_0&=\Var\left(\frac{\pi(\omega_i)}{\pi_0}\,t(y_i,X_i,\omega_i)\right),\\
    t(y_i,X_i,\omega_i)&=\E\left[s(v_i,v_j,\beta)\,\middle|\,y_i,X_i,\omega_i,\delta_{ij}=0\right],
    \end{align*}
    and
    \begin{align*}
    \Sigma_0=\E\big[\mathbbm{1}(y_i\ne y_j)&F\big((X_i-X_j)'\beta\big)F\big((X_j-X_i)'\beta\big)\\
    &\times(X_i-X_j)(X_i-X_j)'\,\big|\,\delta_{ij}=0\big].
    \end{align*}
\end{theorem}

Two features drive the finite-support case. First, the population score is exactly zero: conditional on $\delta_{ij}=0$, Lemma 1 of \cite{auerbach2022identification} gives $\rho_{ij}=0$, Assumption \ref{ass2} gives $\lambda(\omega_i)=\lambda(\omega_j)$, and the conditional logit score is exactly unbiased. Second, the first-stage error in $\hat{\delta}_{ij}$ is negligible at the $\sqrt{n}$ scale, by a Hoeffding-inequality argument (Lemma \ref{lemC1} in the Appendix). The weight $\pi(\omega_i)/\pi_0$ reflects heterogeneity across types in the probability of being matched; when all blocks have equal size, it equals one and $V_0=\Var(t)$.

\paragraph{General network types} When network types vary continuously, exact matches occur with probability zero and the estimator is centered at a pseudo-true value. The following conditions, which parallel Assumptions C1--C2 of \cite{auerbach2021inference}, quantify the quality of codegree matching and restrict the bandwidth to a window in which the first-stage error is negligible.

\begin{assumption}\label{ass_holder} The following statements hold:
\begin{enumerate}[(i)]
    \item there exist $\alpha, C>0$ such that $\int\mathbbm{1}\{v\in[0,1]:\sup_{\tau\in[0,1]}|f(u,\tau)-f(v,\tau)|<\epsilon\}\,dv\ge(\epsilon/C)^{1/\alpha}$ for every $u\in[0,1]$ and $\epsilon\in[0,1]$;\label{ass_holder_ii}
    \item there exist $\varphi, C_\lambda>0$ such that $\left(\lambda(u)-\lambda(v)\right)^2\le C_\lambda\,||f_u-f_v||_2^{2\varphi}$ for all $u,v\in[0,1]$;\label{ass_holder_iii}
    \item $h=C_7\times n^{-\varrho}$ for some $\varrho\in\left(0,\,\min\left\{\frac{\alpha}{2+4\alpha},\frac16\right\}\right)$ and $C_7>0$ (the $1/6$ cap is what variance estimation requires in Appendix \hyperlink{app_var}{A.8} and binds only when $\alpha>1$; no lower bound on $\varrho$ beyond $h\to0$ is imposed);\label{ass_holder_iv} 
    \item $\liminf_{n\to\infty}\lambda_{\min}\left(\Omega_h\right)>0$, where $\Omega_h$ is the influence-function variance defined in equation (\ref{eq_influence}) below.\label{ass_holder_v}
\end{enumerate}
\end{assumption}

Assumption \ref{ass_holder}(\ref{ass_holder_ii}) strengthens Assumption \ref{ass1}(vii): the mass of types with $\epsilon$-similar linking behavior is bounded below by a fractional polynomial of $\epsilon$. It holds if $f$ is $\alpha$-H\"older continuous but does not require continuity. Under this condition, Lemma A1 of \cite{auerbach2022identification} gives the quantitative converse to $\delta_{ij}\le\rho_{ij}$,
\[||f_{\omega_i}-f_{\omega_j}||_2\le 2C^{\frac{1}{2+4\alpha}}\left(||p_{\omega_i}-p_{\omega_j}||_2\right)^{\frac{\alpha}{1+2\alpha}},\]
which bounds the cost of matching on codegree rather than link functions. Assumption \ref{ass_holder}(\ref{ass_holder_iii}) is a quantitative version of Assumption \ref{ass2}: it holds with $\varphi=1$ whenever $\lambda$ is Lipschitz in the link-function distance, and fails when $\lambda$ has jumps between types whose linking behavior is arbitrarily close. Part (\ref{ass_holder_v}) is a nondegeneracy condition on a derived object: it requires the influence function of equation (\ref{eq_influence}) to carry variance in every direction, playing the role that nonsingularity of the outer-product matrix plays in standard sandwich formulas; I am not aware of primitive sufficient conditions for it. Together they imply that pairs in the kernel window ($\delta^2_{ij}\le h$) satisfy $|\lambda(\omega_i)-\lambda(\omega_j)|\le C' h^{\kappa}$ with
\[\kappa:=\frac{\varphi\,\alpha}{2(1+2\alpha)}.\]

Define the localized score bias and the pseudo-true value
\begin{equation}\label{eq_bias}
B_h:=r_n^{-1}\,\E\left[c_{ij}\left\{F\left(\Delta X_{ij}'\beta+\Delta\lambda_{ij}\right)-F\left(\Delta X_{ij}'\beta\right)\right\}\Delta X_{ij}\,K\big(\delta^2_{ij}/h\big)\right],
\end{equation}
where $\Delta X_{ij}:=X_i-X_j$, $\Delta\lambda_{ij}:=\lambda(\omega_i)-\lambda(\omega_j)$, and $c_{ij}:=\PP(y_i\neq y_j|X_i,X_j,\omega_i,\omega_j)$. The pseudo-true value is defined as the exact minimizer of the localized population objective,
\begin{equation}\label{eq_pseudotrue}
\beta_h:=\arg\min_{b}\ \E\left[m(v_i,v_j,b)\,K\big(\delta^2_{ij}/h\big)\right],
\end{equation}
which exists and is unique for all small $h$ by the Kullback--Leibler argument of Steps 2--3 of the proof of Theorem \ref{theo1} applied at fixed $h$, using Assumption \ref{ass_rank} for the $h$-ball events. Its first-order characterization is
\[\beta_h=\beta+\Sigma_h^{-1}B_h+O\big(h^{2\kappa}\big):\]
expanding the first-order condition $\E[K_{ij}\,s(v_i,v_j,\beta_h)]=0$ around $\beta$ gives $\beta_h-\beta=\Sigma_h^{-1}B_h+O(\|\beta_h-\beta\|^2)$, because the logit Hessian is Lipschitz in $b$ with integrable envelope. The distinction matters: with a nonlinear score the exact minimizer and the first-order formula differ by $O(h^{2\kappa})$, which exceeds $n^{-1/2}$ throughout most of the bandwidth window of Assumption \ref{ass_holder}(\ref{ass_holder_iv}), so a central limit theorem centered at $\beta+\Sigma_h^{-1}B_h$ would require implausible smoothness; Theorem \ref{theo4} below is therefore centered at the minimizer (\ref{eq_pseudotrue}), at which the localized population score vanishes exactly. (In the linear model of \cite{auerbach2021inference} the Hessian is free of $b$, and the two definitions coincide; this is a genuine difference introduced by the logit score.) The term $B_h$ is the exact mean of the localized score at $\beta$: it is zero when matches are exact ($\lambda(\omega_i)=\lambda(\omega_j)$) and satisfies $||B_h||\le C'' h^{\kappa}$ under Assumption \ref{ass_holder}, since $|F(a+d)-F(a)|\le|d|/4$.

The asymptotic variance is built from an agent-level influence function with three channels, reflecting the three roles an agent plays in the localized score: as a member of a scored pair, as the bridging agent $t$ in the estimated codegree distance, and as one of the two $s$-agents. With $s_{ij}:=s(v_i,v_j,\beta_h)$ evaluated at the pseudo-true value (\ref{eq_pseudotrue}). The evaluation point matters: the first-stage channels below scale like $1/h$, so evaluating at $\beta$ rather than $\beta_h$ would perturb $\Omega_h$ at first order. $K_{ij}:=K(\delta^2_{ij}/h)$, $Z_\tau:=(y_\tau,X_\tau,\omega_\tau)$, and
\[F_{ijts_1s_2}:=f(\omega_t,\omega_{s_1})f(\omega_t,\omega_{s_2})\big(f(\omega_i,\omega_{s_1})-f(\omega_j,\omega_{s_1})\big)\big(f(\omega_i,\omega_{s_2})-f(\omega_j,\omega_{s_2})\big),\]
so that $\E[F_{ijts_1s_2}|\omega_i,\omega_j]=\delta^2_{ij}$ when $(t,s_1,s_2)$ are drawn independently of the pair, define
\begin{equation}\label{eq_influence}
\begin{aligned}
\psi_1(Z_\tau)&:=\frac{2}{r_n}\left(\E\left[s_{\tau j}K_{\tau j}\,\big|\,Z_\tau\right]-\E\left[s_{ij}K_{ij}\right]\right),\\
\psi_2(Z_\tau)&:=\frac{1}{r_nh}\,\E\left[s_{ij}\,K'\!\left(\frac{\delta^2_{ij}}{h}\right)\left(F_{ij\tau s_1s_2}-\delta^2_{ij}\right)\Big|\,Z_\tau\right],\\
\psi_3(Z_\tau)&:=\frac{2}{r_nh}\,\E\left[s_{ij}\,K'\!\left(\frac{\delta^2_{ij}}{h}\right)\left(F_{ijt\tau s_2}-\delta^2_{ij}\right)\Big|\,Z_\tau\right],\\
\varphi_\tau&:=\psi_1(Z_\tau)+\psi_2(Z_\tau)+\psi_3(Z_\tau), \qquad \Omega_h:=\Var\left(\varphi_\tau\right),
\end{aligned}
\end{equation}
where each conditional expectation integrates over every index other than $\tau$, so that $\tau$ is the shared index: it enters $\psi_2$ as the bridging agent and $\psi_3$ as an $s$-agent, and $\E[\varphi_\tau]=0$. The variance $\Omega_h$ contains all covariances among the three channels. This is deliberate: conditional on $\omega_\tau$, $\E[F_{ij\tau s_1s_2}-\delta^2_{ij}\,|\,\omega_i,\omega_j,\omega_\tau]=(p_{\omega_i}(\omega_\tau)-p_{\omega_j}(\omega_\tau))^2-\delta^2_{ij}\neq0$ in general, and with the logit score $\E[s_{ij}|Z_{ij}]$ equals the nonzero matching bias on inexact matches, so the cross-covariances between the channels do not vanish (the exact cancellations available for a linear score are not available here), and none is invoked.

\begin{theorem}\label{theo4}
    Under Assumptions \ref{ass1}-\ref{ass4} and \ref{ass_holder}, and the parameter-space hypotheses of Theorem \ref{theo2},
    \[V_n^{-1/2}\left(\hat{\beta}-\beta_h\right)\xrightarrow{d}\mathcal{N}
    \left(0,I_k\right), \qquad V_n=\Sigma_h^{-1}\Omega_h\Sigma_h^{-1}/n.\]
\end{theorem}

Four remarks. First, the channels $\psi_2$ and $\psi_3$ arise from the estimation of the codegree distances: linearizing $K(\hat{\delta}^2_{ij}/h)$ around $\delta^2_{ij}$ and representing $\hat{\delta}^2_{ij}-\delta^2_{ij}$ as a $U$-statistic over quintuples of agents produces additional influence terms that would be missed by treating the matches as known. Ignoring them generally understates the variance. Second, the definition of $\Omega_h$ passes a consistency audit that componentwise conventions can fail: 
under the finite-support condition of Assumption \ref{ass5}, $\psi_2$ and $\psi_3$ are negligible at the $\sqrt n$ scale. The reason is structural rather than a concentration argument: both channels carry the factor $K'(\delta^2_{ij}/h)$, which vanishes for separated pairs because $\delta^2_{ij}\ge\bar\epsilon>h$ places them outside the kernel support, while on exact matches $\Delta\lambda_{ij}=0$ makes the conditional logit score mean zero given the types, $\E[s_{ij}\mid\omega_i,\omega_j]=0$, and the dyadic shocks are independent of $(v_i,v_j)$ given the types by Assumption \ref{ass1}(ii), so the expectation factorizes and both channels project to zero. (For the triweight kernel of Section \ref{section4}, $K'(0)=0$ makes the first factor vanish on exact matches as well, though this is a property of that kernel and not of Assumption \ref{ass3}(\ref{ass3i}).) Lemma \ref{lemC1} enters at one remove, ensuring that with $\hat\delta^2_{ij}$ in place of $\delta^2_{ij}$ no separated pair enters the window and no exact match leaves it, with probability approaching one. Meanwhile $K_{ij}=K(0)\mathbbm{1}\{\delta_{ij}=0\}$ and $r_n=K(0)\pi_0$ give $\psi_1(Z_\tau)=\frac{2}{\pi_0}\left(\pi(\omega_\tau)t_\tau-\E[\pi(\omega_i)t_i]\right)$, so $\Var(\psi_1)=4V_0$ and $V_n\to\Sigma_0^{-1}\,4V_0\,\Sigma_0^{-1}/n$, which is exactly the variance of Theorem \ref{theo3}. Third, the variance $V_n$ is not necessarily of order $1/n$: the effective sample size is governed by $nr_n$, the number of quality matches, so convergence can be slower than $\sqrt{n}$ when matches are thin. Fourth, and most importantly, the distribution is centered at $\beta_h$, not $\beta$. Centering at $\beta$ would require $\|B_h\|=o\left(V_n^{1/2}\right)$. When $\Omega_h$ is bounded, $V_n^{1/2}\asymp n^{-1/2}$ and this becomes $n^{1/2-\varrho\kappa}\to0$, which is incompatible with the bandwidth window in Assumption \ref{ass_holder}(\ref{ass_holder_iv}) except for implausibly smooth $\lambda$ (one needs $\varphi>2(1+2\alpha)^2/\alpha^2$ when $\alpha\le1$ and, with the $1/6$ cap binding, the stronger $\varphi>6(1+2\alpha)/\alpha$ when $\alpha>1$; e.g., $\varphi>18$ when $\alpha=1$, where the two coincide); when matches are thin and $V_n$ is of larger order, the comparison must be made against $V_n^{1/2}$ rather than $n^{-1/2}$ and undersmoothing is less demanding, but the required window then depends on unknown features of the match density. In either case, reliable inference centered at $\beta$ requires removing the bias, which is the purpose of the following correction.

\paragraph{Jackknife bias correction} Following \cite{honore1997pairwise} and Section 1.1.2 of \cite{auerbach2021inference}, I remove the bias with a multiplicative-bandwidth jackknife. The construction requires that the pseudo-truth be well approximated by fractional polynomials of the bandwidth.

\begin{assumption}\label{ass_jack} The pseudo-true function $\beta_h$ satisfies $\beta_h=\beta+\sum_{l=1}^{L}C_l\,h^{l/\theta}+O\left(h^{(L+1)/\theta}\right)$ for some $\theta>0$, $k$-dimensional constants $C_1,\ldots,C_L$, a positive integer $L$ with $h^{(L+1)/\theta}=o\left(n^{-1/2}\right)$, and $h$ in a fixed open neighborhood to the right of $0$; the map $h\mapsto\beta_h$ is twice continuously differentiable on that neighborhood with $\partial_h\beta_h=O(h^{1/\theta-1})$ and $\partial_h^2\beta_h=O(h^{1/\theta-2})$. The combined variance of the jackknife weights is nondegenerate: $\liminf_n\lambda_{\min}\big(\sum_{l_1,l_2}a_{l_1}a_{l_2}\Omega_{h,l_1l_2}\big)>0$. The latter does not follow from Assumption \ref{ass_holder}(\ref{ass_holder_v}): the influence functions across the bandwidth grid are strongly correlated, and a signed combination such as $(6,-8,3)$ could otherwise degenerate in some direction. The rate condition is stated at the $n^{-1/2}$ scale because, with the combined variance bounded below but not necessarily above, the standard error of $\bar\beta_L$ is at least of order $n^{-1/2}$; a residual bias between $n^{-1/2}$ and $(nr_n)^{-1/2}$ would otherwise distort the centering. With $L$ terms the condition implicitly demands smoothness, consistent with the honest behavior of the correction documented in Section \ref{section4}.
\end{assumption}

The expansion generalizes the polynomial bias expansions underlying the jackknife of \cite{honore1997pairwise}: with $\theta=1$ it is a Taylor-type expansion in $h$, while fractional $\theta$ accommodates the H\"older rates $h^{\kappa}$ delivered by Assumption \ref{ass_holder}, with $\theta=1/\kappa$. It fails when $\beta_h$ carries logarithmic factors, in which case no fractional-polynomial expansion exists and the correction removes only part of the bias.

For an arbitrary sequence of distinct positive constants $\{c_1,\ldots,c_{L+1}\}$ with $c_1=1$, the bias-corrected estimator is
\begin{equation}\label{eq_jack}
    \bar{\beta}_L=\sum_{l=1}^{L+1}a_l\,\hat{\beta}_{c_lh},
\end{equation}
where $\hat{\beta}_{c_lh}$ is the estimator (\ref{eq6}) computed with bandwidth $c_l\times h$ and the weights $\{a_1,\ldots,a_{L+1}\}$ solve the generalized Vandermonde system
\[\sum_{l=1}^{L+1}a_l=1 \qquad\mbox{and}\qquad \sum_{l=1}^{L+1}a_l\,c_l^{m/\theta}=0 \quad\mbox{for } m=1,\ldots,L.\]
This is $L+1$ equations in $L+1$ unknowns; the coefficient matrix $[c_l^{p_m}]$ with exponents $p\in\{0,1/\theta,\ldots,L/\theta\}$ and distinct positive nodes is nonsingular, because a nontrivial generalized polynomial $\sum_mb_mx^{p_m}$ with $L+1$ terms has at most $L$ positive zeros (Descartes' rule for Chebyshev systems), so no nonzero coefficient vector can vanish at $L+1$ distinct nodes. The count of bandwidths matters: with only $L$ bandwidths the weights can annihilate the powers $h^{1/\theta},\ldots,h^{(L-1)/\theta}$ but the term $C_L\big(\sum_la_lc_l^{L/\theta}\big)h^{L/\theta}$ survives, and Assumption \ref{ass_jack} does not make it negligible; the additional bandwidth is what reduces the residual bias to $O(h^{(L+1)/\theta})$. For example, with $L=2$, $\theta=1$, and bandwidths $(h,1.5h,2h)$, the weights are $(6,-8,3)$:
\[\bar\beta_L=6\hat\beta_{h}-8\hat\beta_{1.5h}+3\hat\beta_{2h}.\]

\begin{theorem}\label{theo5}
    Under Assumptions \ref{ass1}-\ref{ass4}, \ref{ass_holder} and \ref{ass_jack}, and the parameter-space hypotheses of Theorem \ref{theo2},
    \[\bar{V}_n^{-1/2}\left(\bar{\beta}_L-\beta\right)\xrightarrow{d}\mathcal{N}\left(0,I_k\right), \qquad \bar{V}_n=\sum_{l_1=1}^{L+1}\sum_{l_2=1}^{L+1} a_{l_1}a_{l_2}\,\Sigma_0^{-1}\Omega_{h,l_1l_2}\Sigma_0^{-1}/n,\]
    where $\Omega_{h,l_1l_2}:=\mathrm{Cov}\left(\varphi_\tau^{(l_1)},\varphi_\tau^{(l_2)}\right)$ is the cross-bandwidth analogue of $\Omega_h$, with $\varphi_\tau^{(l)}$ the influence function (\ref{eq_influence}) evaluated at bandwidth $c_lh$.
\end{theorem}

Two caveats delimit the scope of Theorem \ref{theo5}. First, its hypotheses are jointly satisfiable only when the residual-bias rate is compatible with the bandwidth window: with $L=2$, the rate condition $h^{3/\theta}=o(n^{-1/2})$ in Assumption \ref{ass_jack} requires $\varrho>\theta/6$, which is consistent with the window $\varrho<\min\{\alpha/(2+4\alpha),1/6\}$ of Assumption \ref{ass_holder}(\ref{ass_holder_iv}) only when $\theta<\min\{3\alpha/(1+2\alpha),1\}$; in particular, the case $\theta=1$, in which the weights reduce to $(6,-8,3)$, is not covered at $L=2$, although a larger $L$ relaxes the requirement to $\theta<2(L+1)\min\{\alpha/(2+4\alpha),1/6\}$ at the cost of additional variance. Second, the theorem takes $\theta$ as known and the weights as fixed; it does not cover the feasible procedure described next, which estimates, clamps, and thresholds $\hat\kappa$. The feasible correction reported in Sections \ref{section4} and \ref{section5} should therefore be read as the diagnostic-guarded bandwidth-sensitivity exercise described there, not as an exact implementation of Theorem \ref{theo5}.

\paragraph{Choosing $\theta$} The construction requires the exponent scale $\theta$, which is unknown in general: the bias characterization above gives leading exponent $\kappa=\varphi\alpha/(2(1+2\alpha))$, so $\theta=1/\kappa$ depends on the H\"older constants of the design, and imposing $\theta=1$ amounts to assuming a linear-in-$h$ bias expansion. A feasible choice treats $\theta$ as an unknown to be estimated from the bandwidth sensitivity of the estimator itself. Fix a geometric grid $h_m=2^m\hat h$, $m=1,2,3$, in the oversmoothed range, compute the differences $\Delta_m:=\hat\beta_{2^{m+1}\hat h}-\hat\beta_{2^{m}\hat h}$ for $m=1,2$, and set
\[\hat\kappa:=\log_2\left|\Delta_{2}/\Delta_1\right|, \qquad \hat\theta:=1/\hat\kappa,\]
with the ratio computed coordinate-wise and aggregated by the median over coordinates whose denominator is non-negligible (for scalar $\beta$ it is the plain ratio); this three-point grid is exactly what Sections \ref{section4} and \ref{section5} implement. A longer grid with a median over successive ratios is the natural extension when the budget allows.
If $\beta_h=\beta+C_1h^{\kappa}(1+o(1))$ with $v'C_1\neq0$ and the grid satisfies $h_m^\kappa\gg(nr_n)^{-1/2}$ (bias dominates sampling noise on the grid), then $\Delta_m=v'C_1(2^\kappa-1)h_m^\kappa(1+o_p(1))$, the ratios converge to $2^\kappa$, and $\hat\kappa\to_p\kappa$; since the weights are continuous in $\theta$ at the truth by nonsingularity of the generalized Vandermonde matrix, plugging $\hat\theta$ into (\ref{eq_jack}) preserves the first-order theory \emph{provided} $\hat\kappa$ converges fast enough: the weights annihilate $h^{m/\theta}$ only at the true $\theta$, the residual bias after substitution is of order $|\hat\kappa-\kappa|\,h^{\kappa}$ (the weight functions $\theta\mapsto a_l(\theta)$ are smooth with bounded derivatives on the clamped range, and only powers of the fixed nodes $c_l$ enter), and first-order validity therefore requires $|\hat\kappa-\kappa|=o_p\big(n^{-1/2}h^{-\kappa}\big)$. No such rate is established in this paper: consistency of $\hat\kappa$ follows from the display above, but its rate depends on the strength of the bias signal on the grid. The feasible correction should accordingly be read as removing the estimable part of the bandwidth bias, with exact centering guaranteed only when $\theta$ is known or $\hat\kappa$ attains the stated rate; the simulations of Section \ref{section4} evaluate the feasible version directly, and the applicability threshold below is the practical guard. The exponent must be estimated where bias dominates (large $h$) and applied where inference is conducted (small $h$); a plot of $\log|\Delta_m|$ against $\log h_m$ makes the quality of the fit visible, and an unstable slope is a warning that the correction should not be applied: in the finite-support regime there is no bandwidth bias to measure, $\hat\kappa$ diverges, and the uncorrected estimator with Theorem \ref{theo3} inference is the right report. The generalized-jackknife construction used here is in the spirit of \cite{cattaneo2026robust}, who establish bandwidth-robust inference for convex pairwise-difference estimators when the matching covariate is observed; their results do not transfer directly, because the matching variable here is estimated from the entire network, but they indicate the ingredients a complete theory for the estimated-exponent correction would require.

\paragraph{Variance estimation} The asymptotic variances in Theorems \ref{theo3}-\ref{theo5} are estimated by a single object: the sample variance of estimated influence contributions. Let $\hat{s}_{ij}=s(v_i,v_j,\hat{\beta})$, $\widehat K_{ij}=K(\hat\delta^2_{ij}/h)$,
\begin{align*}
\widehat{\Sigma}_h=\hat{r}_n^{-1}\binom n2^{-1}\sum_{i<j}\mathbbm{1}(y_i\ne y_j)\,&F\big((X_i-X_j)'\hat\beta\big)F\big((X_j-X_i)'\hat\beta\big)\\
&\times(X_i-X_j)(X_i-X_j)'\,\widehat K_{ij},
\end{align*}
and define, for each agent $\tau$,
\begin{align}
\hat\psi_{1,\tau}&=\frac{2}{\hat r_n}\left[\frac{1}{n-1}\sum_{j\neq\tau}\widehat K_{\tau j}\hat s_{\tau j}-\binom n2^{-1}\sum_{i<j}\widehat K_{ij}\hat s_{ij}\right],\nonumber\\
\hat\psi_{2,\tau}&=\frac{1}{\hat r_nh}\binom n2^{-1}\sum_{i<j}\hat s_{ij}\,K'\!\big(\hat\delta^2_{ij}/h\big)\left(\widehat F_{ij\tau}-\hat\delta^2_{ij}\right),\nonumber\\
\hat\psi_{3,\tau}&=\frac{2}{\hat r_nh}\binom n2^{-1}\sum_{i<j}\hat s_{ij}\,K'\!\big(\hat\delta^2_{ij}/h\big)\left(\widehat F'_{ij\tau}-\hat\delta^2_{ij}\right),\nonumber\\
\widehat F_{ij\tau}&=\Big[\tfrac1n\textstyle\sum_sD_{\tau s}(D_{is}-D_{js})\Big]^2,\nonumber\\
\widehat F'_{ij\tau}&=\tfrac1{n^2}\textstyle\sum_{t,s}D_{t\tau}D_{ts}(D_{i\tau}-D_{j\tau})(D_{is}-D_{js}),\nonumber\\
\hat\varphi_\tau&=\hat\psi_{1,\tau}+\hat\psi_{2,\tau}+\hat\psi_{3,\tau},\nonumber\\
\widehat\Omega_h&=\frac1n\sum_{\tau=1}^n\left(\hat\varphi_\tau-\bar{\hat\varphi}\right)\left(\hat\varphi_\tau-\bar{\hat\varphi}\right)', \qquad \widehat V=\widehat\Sigma_h^{-1}\widehat\Omega_h\widehat\Sigma_h^{-1}/n.\label{eq_varest}
\end{align}
Computationally, with $C=DD$ and $E=DC$, $\widehat F_{ij\tau}=\left((C_{\tau i}-C_{\tau j})/n\right)^2$ and $\widehat F'_{ij\tau}=(D_{i\tau}-D_{j\tau})(E_{\tau i}-E_{\tau j})/n^2$; both average to $\hat\delta^2_{ij}$ over $\tau$, a useful unit test of any implementation. For the jackknife variance $\bar V_n$, the cross-bandwidth blocks are estimated by $\widehat\Omega_{h,l_1l_2}=\frac1n\sum_\tau(\hat\varphi^{(l_1)}_\tau-\bar{\hat\varphi}^{(l_1)})(\hat\varphi^{(l_2)}_\tau-\bar{\hat\varphi}^{(l_2)})'$ with $\hat\varphi^{(l)}_\tau$ computed at bandwidth $c_lh$. Consistency, $\widehat\Sigma_h^{-1}\widehat\Omega_h\widehat\Sigma_h^{-1}/n\,V_n^{-1}\to_pI_k$, is proved with explicit rates in Appendix \hyperlink{app_var}{A.8}. Under the hypotheses of Theorem \ref{theo3} (finite support), $\hat\psi_{2,\tau}$ and $\hat\psi_{3,\tau}$ are asymptotically negligible and may be omitted. In applications with grouped data (such as the village networks of Section \ref{section5}), the clustered variance is the score sandwich of Proposition \ref{prop_villages}(ii) below; an influence-aggregated variant, $\widehat V^{\varphi}_{\mathrm{cl}}=\widehat\Sigma_h^{-1}\big[n^{-2}\sum_{g}\big(\sum_{\tau\in g}\hat\varphi_\tau\big)\big(\sum_{\tau\in g}\hat\varphi_\tau\big)'\big]\widehat\Sigma_h^{-1}$, is asymptotically equivalent but is not the reported object.
The single-network simulations of Section \ref{section4} use $\widehat V$ throughout. In the grouped application of Section \ref{section5} the reported standard errors use the clustered \emph{score} sandwich of Proposition \ref{prop_villages}(ii) below, $\widehat V_{\mathrm{cl}}=\widehat H_N^{-1}\big[\sum_gS_g(\hat\beta)S_g(\hat\beta)'\big]\widehat H_N^{-1}$, which aggregates each village's localized score before forming outer products; because everything estimated within a village, including the codegree distances, is a function of that village's data, the variation of $S_g$ across independent villages already carries the first-stage estimation noise, and no delta-method channels need to be added. The influence-function form displayed above is the single-network object used for the theory and for Section \ref{section4}; no other variance formulas are used anywhere.

\paragraph{Many networks} The theory above is stated for a single network of $n$ agents. Applications such as Section \ref{section5} instead pool many separate networks (villages), with matching and codegree distances computed within network. The following result records when the single-network theory extends and when it does not.

\begin{proposition}\label{prop_villages}
Let networks $g=1,\ldots,G$ be independent, each generated by the model of Section \ref{section3} with common primitives $(\beta,\lambda,f)$ and sizes $n_g\in[\underline n,C\underline n]$, and let $\hat\beta$ minimize the pooled objective $\sum_g\sum_{i<j\in g}\widehat K^{(g)}_{ij}m(v_i,v_j,b)$ with codegree distances and bandwidths computed within network. Let $N=\sum_gn_g$. (i) If $\underline n\to\infty$ with $G$ fixed or $G\to\infty$, $G\le\underline n^{C'}$, then under the within-network analogues of Assumptions \ref{ass1}-\ref{ass4}, $\hat\beta\to_p\beta$. (ii) Suppose in addition the within-network analogues of Assumption \ref{ass_holder} hold and a common bandwidth $h$ is used in every network. Let $a_g:=\binom{n_g}2r^{(g)}_n$ and $\Phi_g:=a_g\,n_g^{-1}\sum_{\tau\in g}\varphi^{(g)}_\tau$ denote the scaled network-level influence sums. If $G$ is fixed with $\underline n\to\infty$, or if $G\to\infty$ and the joint rate condition $\sqrt G\big/\big(\sqrt{\underline n}\,h\,\lambda_{\min}(\Omega_h)^{1/2}\big)\to0$ holds (automatic for fixed $G$ by Assumption \ref{ass3}(\ref{ass3ii}); under Assumption \ref{ass_holder}(\ref{ass_holder_v}) it is implied by $G/(\underline n\,h^2)\to0$), then
\[\left(\textstyle\sum_g\Var(\Phi_g)\right)^{-1/2}\Sigma_h\Big(\textstyle\sum_ga_g\Big)\left(\hat\beta-\beta_h\right)\xrightarrow{d}\mathcal N(0,I_k),\]
and, when $G\to\infty$ and additionally $\mathrm{tr}\,\Omega_h\le\bar C\,\lambda_{\min}(\Omega_h)$ (a well-conditioned variance, automatic in the finite-support regime where $\Omega_h\to4V_0\succ0$), the clustered score sandwich $\widehat V_{\mathrm{cl}}:=\widehat H_N^{-1}\big[\sum_gS_g(\hat\beta)S_g(\hat\beta)'\big]\widehat H_N^{-1}$, with $S_g$ the network-$g$ localized score and $\widehat H_N$ the pooled Hessian, satisfies $\widehat V_{\mathrm{cl}}^{-1/2}(\hat\beta-\beta_h)\to_d\mathcal N(0,I_k)$; with $G$ fixed, $\sum_gS_gS_g'$ has rank at most $G$ and does not concentrate, so the clustered statistic carries the usual fixed-cluster caveat and $G$ should be treated as large, as with $G=43$ in Section \ref{section5}. With network-specific bandwidths $h_g$, as in Section \ref{section5}, the centering becomes the minimizer of the pooled localized population objective $\sum_ga_gQ^{(g)}_{h_g}$; the pooled first-order condition then forces only the \emph{weighted sum} of the network score means to zero, while individual networks carry nonzero means $m_g$ summing to zero, so $\sum_gS_gS_g'$ estimates $\sum_g\Var(S_g)+\sum_gm_gm_g'\succeq\sum_g\Var(S_g)$ and the clustered sandwich is asymptotically \emph{conservative} rather than exact; exact variance estimation requires a common bandwidth. The central limit theorem for $\hat\beta$ itself is unaffected, since the total score remains mean zero at the pooled centering. (iii) If instead $n_g\le\bar n<\infty$ for all $g$, then as $G\to\infty$ the estimator converges to a pseudo-true value $\beta_{h,\bar n}\neq\beta$ in general: within networks of bounded size neither the matching bias nor the first-stage error in $\hat\delta^2_{ij}$ vanishes, and growth in the number of networks does not repair either.
\end{proposition}

The proof is given in Appendix \hyperlink{app_villages}{ A.9}. The joint rate condition in part (ii) is what pooling adds to the single-network theory: each village's feasible score carries a finite-$n$ centering error of order $1/(n_gh)$ from the repeated-index terms and the second-order kernel remainder of Appendix \hyperlink{app_theo4}{A.6}; this error is negligible at the single-network $n_g^{-1/2}$ scale, but pooling shrinks the sampling variance with $G$ while leaving the per-village centering error untouched, so the two must be balanced explicitly. With $G$ fixed, as in the asymptotic frame natural for the $43$ villages of Section \ref{section5}, no such condition is needed. Part (iii) is a warning label for practice: when the individual networks are small, Lemma \ref{lemB1} implies within-network codegree noise of order $n_g^{-1/2}\log^{1/2}n_g$, which can exceed the codegree distances being matched on, and the estimand is then the pseudo-truth at the achievable match quality, however many networks are pooled. Section \ref{section5} reports the corresponding diagnostic.

\section{Monte Carlo Simulation}\label{section4}
The simulations are designed to document four features of the estimator: bias reduction relative to standard logit specifications, robustness to different network formation mechanisms, the two asymptotic regimes of Section \ref{section332} together with the jackknife bias correction, and the finite-sample consequences of violating the network-equivalence condition of Assumption \ref{ass2}.

I generate simulated data using the model described in section \ref{section3} in equations (\ref{eq1}) and (\ref{eq2}). The slope parameter is set to $\beta=1$, and the unobserved heterogeneity function is specified as
$\lambda(w) = (4w^3-3)/2$
which introduces nonlinear dependence on the unobserved social characteristic. To induce endogeneity, the observed regressor $x_i$ is constructed as
$x_i =(3\omega_i^3+\xi_i+1)/3$, where $\xi_i\sim\mathcal{N}(0,1)$, so that $x_i$ is correlated with the unobserved social characteristic $\omega_i$.

Unobserved social characteristic $\omega_i$ and the link formation shocks, $\eta_{ij}$, are independently drawn from the uniform distribution on $[0, 1]$ for $i<j$ and $\eta_{ij}=\eta_{ji}$. The outcome shocks $\varepsilon_i$ follow a standard logistic distribution.

The adjacency matrix is generated according to equation (\ref{eq2}) using three alternative link formation functions:
\begin{enumerate}
    \item Homophily model
    \[f_1(x,y)=1-(x-y)^2\]
        \item Beta model
    \[ f_2(x,y)=\frac{\exp(x+y)}{1+\exp(x+y)}\]

    \item Stochastic block model
    \[f_3(x,y) = \left\{\begin{array}{cl}
   1/3  &  \mbox{ if } x\le1/3 \mbox{ and } y>1/3\\
    1/3 & \mbox{ if } 1/3<x\le2/3 \mbox{ and } y\le2/3 \\
    1/3& \mbox{ if } x>2/3 \mbox{ and } (y>2/3 \mbox{ or } y\le1/3)\\
    0& \mbox{ otherwise }
\end{array}\right.\]

\end{enumerate}
These designs capture qualitatively different forms of network formation, including assortative matching, degree heterogeneity, and community structure respectively. The three designs also illustrate the two asymptotic regimes of Section \ref{section332}: the stochastic block design satisfies the finite-support condition of Assumption \ref{ass5} (agents within the same block are exact matches), while the homophily and beta designs generate continuously varying network types and therefore fall under the general regime of Theorem \ref{theo4}, in which the uncorrected estimator carries a matching bias at any finite bandwidth. One adjustment to the stochastic block design is required for internal consistency: under $f_3$ the link function is constant within blocks, so agents in the same block are network equivalent, and Assumption \ref{ass2} then requires that they share the same value of $\lambda$. Accordingly, under the stochastic block design $\lambda$ is evaluated at the block centers, $\lambda(\omega_i)=(4c_{b(i)}^3-3)/2$ with $c_b\in\{1/6,1/2,5/6\}$, which preserves the correlation between $x_i$ and $\lambda(\omega_i)$ while keeping the design inside the model. Section \ref{sec_misspec} reports what happens when this adjustment is omitted and Assumption \ref{ass2} fails in the data generating process.

For each link formation model, I run 500 Monte Carlo replications with sample sizes $n\in\{100, 250, 700, 1000, 2000\}$ and the triweight-type kernel $K(x)=\frac{35}{32}(1-x^2)^3\,\mathbbm{1}(0\le x<1)$, which satisfies every part of Assumption \ref{ass3}(\ref{ass3i}): $K(0)>0$ and both $K'$ and $K''$ vanish at the boundary of the support, so the extension by zero is twice continuously differentiable. 
The bandwidth follows the two-branch rule of equation (\ref{eq_bandwidthrule}). The homophily and beta designs use the continuous branch, $\hat h_n=\max(\hat q_{c_n},\,n^{-2/5}\hat q_{1/2})$ with $c_n=\min\{0.05,\,0.05(n/100)^{-1/16}\}$. In the simulations the branch is assigned from the known regime of each design; the empirical regime diagnostic of Section \ref{section5} is exercised only in the application, so the Monte Carlo results condition on the correct regime rather than on the classifier. At the simulated sample sizes the scale-relative floor $n^{-2/5}\hat q_{1/2}$ is the binding arm of the continuous branch in every homophily replication, and in 14 percent of beta replications at $n=2000$: there the operative bandwidth is governed by the finite-sample safeguard rather than by the shrinking quantile, and the near-nominal coverage under homophily in Table \ref{tab1} should be read with that in mind.
 The stochastic block design displays the finite-support signature of equation (\ref{eq_noisefloor}), a point mass of roughly one third of all pairs concentrated at the noise scale $\hat m$, and use the finite-support branch $\hat h_n=(3\hat m)^{1/3}\hat q_{1/2}^{2/3}$, of order $n^{-1/3}$, which captures essentially all exact matches while dominating the first-stage noise; the noise scale $\hat m$ is computed over a random subsample of $\min\{2000,\binom n2\}$ pairs, which meets the requirement of at least $n$ pairs at every sample size considered. The jackknife bias correction of equation (\ref{eq_jack}) is implemented with $L=2$ and bandwidths $(\hat h_n,1.5\hat h_n,2\hat h_n)$; the exponent is estimated from the oversmoothed grid $(2\hat h_n,4\hat h_n,8\hat h_n)$ as $\hat\kappa=\log_2\big|(\hat\beta_{8\hat h}-\hat\beta_{4\hat h})/(\hat\beta_{4\hat h}-\hat\beta_{2\hat h})\big|$, and the correction is applied with $\hat\theta=1/\hat\kappa$, $\hat\kappa$ clamped to $[1/2,2]$, whenever the raw estimate exceeds $0.4$; otherwise the slope diagnostic of Section \ref{section332} deems the correction inapplicable and the uncorrected estimator is reported in the jackknife column (by construction this is the typical outcome in the finite-support design, where there is no bandwidth bias to measure). Standard errors for the proposed estimator are computed from the influence-function estimator (\ref{eq_varest}), including the first-stage channels $\hat\psi_{2}$ and $\hat\psi_{3}$; coverage refers to nominal 95 percent confidence intervals.

The performance of my proposed estimators is evaluated in comparison to the estimators of the following three benchmark models:
\begin{itemize}
\item  Naive Logit: $y_i = \mathbbm{1}\left\{\alpha + \beta_1x_i-\epsilon_i\ge0\right\}$, which ignores network dependence;
\item Logit with controls: $y_i = \mathbbm{1}\big\{\alpha +\beta_2x_i+\mu\tfrac{\sum_{j}D_{ij}x_j}{\sum_{j}D_{ij}} + \gamma\tfrac{\sum_{j}D_{ij}y_j}{\sum_{j}D_{ij}}-u_i\ge0\big\}$, which includes average neighbor covariates and outcomes.
        \item Infeasible Logit: $y_i = \mathbbm{1}\left\{\alpha +\beta_3x_i+\theta\lambda(\omega_i)-v_i\ge0\right\}$, which includes the true $\lambda(\omega_i)$ and therefore serves as a lower bound on achievable bias;
\end{itemize}
Tables \ref{tab1}--\ref{tab3} report the average behavior over the 500 replications for homophily, beta, and stochastic block networks, respectively. Performance is evaluated using absolute bias, standard deviation, root mean squared error, and coverage of nominal 95 percent confidence intervals.

\begin{table}[!htb]
\centering
\begin{tabular}{clccccc}
\hline\hline
&& Naive & Logit with & Infeasible & Proposed & Jackknife \\
&& Logit & Controls & Logit & Estimator & Corrected \\
n&& $\hat\beta_1$ &$\hat\beta_2$ & $\hat\beta_3$  & $\hat\beta$ & $\bar\beta_L$ \\
\hline
100&&  &  &  &  & \\
&$|$Bias$|$&0.9160&0.8683&0.0647&0.2168&0.2480\\
&SD&0.6133&0.8361&0.7567&1.1500&1.5885\\
&RMSE&1.1020&1.2048&0.7587&1.1691&1.6062\\
&Coverage&0.694&0.802&0.944&0.986&0.990\\
250&&  &  &  &  & \\
&$|$Bias$|$&0.8437&0.4322&0.0061&0.0775&0.0707\\
&SD&0.3508&0.4387&0.4379&0.5507&0.6201\\
&RMSE&0.9136&0.6155&0.4376&0.5556&0.6235\\
&Coverage&0.304&0.846&0.952&0.954&0.980\\
700&&  &  &  &  & \\
&$|$Bias$|$&0.8438&0.2868&0.0098&0.0359&0.0353\\
&SD&0.2091&0.2479&0.2651&0.3103&0.3240\\
&RMSE&0.8693&0.3789&0.2650&0.3120&0.3256\\
&Coverage&0.012&0.808&0.950&0.960&0.966\\
1\,000&&  &  &  &  & \\
&$|$Bias$|$&0.8333&0.2594&0.0081&0.0328&0.0328\\
&SD&0.1750&0.2095&0.2175&0.2582&0.2618\\
&RMSE&0.8514&0.3333&0.2174&0.2601&0.2636\\
&Coverage&0.002&0.766&0.950&0.940&0.946\\
2\,000&&  &  &  &  & \\
&$|$Bias$|$&0.8380&0.2338&0.0177&0.0269&0.0253\\
&SD&0.1162&0.1370&0.1437&0.1731&0.1749\\
&RMSE&0.8460&0.2710&0.1446&0.1750&0.1765\\
&Coverage&$<10^{-3}$&0.650&0.966&0.962&0.970\\
\hline
\end{tabular}
\caption{Monte Carlo Performance under Homophily Network}
\label{tab1}
\end{table}

\begin{table}[!htb]
\centering
\begin{tabular}{clccccc}
\hline\hline
&& Naive & Logit with & Infeasible & Proposed & Jackknife \\
&& Logit & Controls & Logit & Estimator & Corrected \\
n&& $\hat\beta_1$ &$\hat\beta_2$ & $\hat\beta_3$  & $\hat\beta$ & $\bar\beta_L$ \\
\hline
100&&  &  &  &  & \\
&$|$Bias$|$&0.9198&1.0279&0.0697&0.9907&1.2689\\
&SD&0.6143&0.7545&0.7578&1.4173&3.1903\\
&RMSE&1.1057&1.2746&0.7603&1.7281&3.4304\\
&Coverage&0.692&0.650&0.944&1.000&1.000\\
250&&  &  &  &  & \\
&$|$Bias$|$&0.8431&0.8316&0.0060&0.5951&0.6934\\
&SD&0.3505&0.3724&0.4380&0.7997&1.7142\\
&RMSE&0.9129&0.9110&0.4376&0.9961&1.8476\\
&Coverage&0.304&0.344&0.952&1.000&1.000\\
700&&  &  &  &  & \\
&$|$Bias$|$&0.8437&0.7626&0.0098&0.4133&0.5258\\
&SD&0.2090&0.2202&0.2651&0.4482&1.1266\\
&RMSE&0.8691&0.7937&0.2650&0.6094&1.2422\\
&Coverage&0.012&0.042&0.950&0.990&0.990\\
1\,000&&  &  &  &  & \\
&$|$Bias$|$&0.8336&0.7242&0.0083&0.3064&0.4430\\
&SD&0.1753&0.1866&0.2176&0.3780&0.9698\\
&RMSE&0.8518&0.7478&0.2176&0.4863&1.0653\\
&Coverage&0.002&0.014&0.950&0.954&0.956\\
2\,000&&  &  &  &  & \\
&$|$Bias$|$&0.8380&0.6637&0.0173&0.1844&0.2284\\
&SD&0.1162&0.1237&0.1437&0.2565&0.5012\\
&RMSE&0.8460&0.6751&0.1446&0.3157&0.5503\\
&Coverage&$<10^{-3}$&$<10^{-3}$&0.966&0.838&0.874\\
\hline
\end{tabular}
\caption{Monte Carlo Performance under Beta Network}
\label{tab2}
\end{table}

\begin{table}[!htb]
\centering
\begin{tabular}{clccccc}
\hline\hline
&& Naive & Logit with & Infeasible & Proposed & Jackknife \\
&& Logit & Controls & Logit & Estimator & Corrected \\
n&& $\hat\beta_1$ &$\hat\beta_2$ & $\hat\beta_3$  & $\hat\beta$ & $\bar\beta_L$ \\
\hline
100&&  &  &  &  & \\
&$|$Bias$|$&0.7602&0.7914&0.0640&0.2170&0.2082\\
&SD&0.6111&0.6705&0.7072&0.7041&0.7078\\
&RMSE&0.9749&1.0368&0.7094&0.7361&0.7371\\
&Coverage&0.762&0.764&0.940&0.834&0.836\\
250&&  &  &  &  & \\
&$|$Bias$|$&0.6743&0.6590&0.0010&0.0090&0.0090\\
&SD&0.3503&0.3627&0.4140&0.4097&0.4097\\
&RMSE&0.7597&0.7521&0.4136&0.4094&0.4094\\
&Coverage&0.516&0.548&0.958&0.912&0.912\\
700&&  &  &  &  & \\
&$|$Bias$|$&0.6804&0.6388&0.0090&0.0080&0.0080\\
&SD&0.2069&0.2231&0.2427&0.2428&0.2428\\
&RMSE&0.7111&0.6765&0.2426&0.2427&0.2427\\
&Coverage&0.052&0.128&0.962&0.948&0.948\\
1\,000&&  &  &  &  & \\
&$|$Bias$|$&0.6633&0.6272&0.0018&0.0026&0.0026\\
&SD&0.1694&0.1777&0.2002&0.2021&0.2021\\
&RMSE&0.6845&0.6518&0.2000&0.2019&0.2019\\
&Coverage&0.026&0.048&0.956&0.944&0.944\\
2\,000&&  &  &  &  & \\
&$|$Bias$|$&0.6762&0.6307&0.0146&0.0148&0.0148\\
&SD&0.1144&0.1290&0.1329&0.1344&0.1344\\
&RMSE&0.6858&0.6437&0.1336&0.1351&0.1351\\
&Coverage&$<10^{-3}$&0.002&0.970&0.958&0.958\\
\hline
\end{tabular}
\caption{Monte Carlo Performance under Stochastic Block Network}
\label{tab3}
\end{table}

\subsection{Consistency and Convergence}
A first and striking result is that naive logit estimators perform extremely poorly in all designs. The absolute bias remains large and essentially constant as the sample size increases. Under homophily networks (Table \ref{tab1}), for example, the bias of the naive estimator remains above 0.83 even when $n = 2000$. Similar magnitudes appear under beta (Table \ref{tab2}) and stochastic block (Table \ref{tab3}) networks. The standard deviation shrinks with sample size, but the bias does not, so the RMSE remains large and coverage collapses toward zero. This confirms that ignoring endogenous network formation produces asymptotically biased and severely misleading inference.

Augmenting the logit model with common network controls improves performance only partially. The bias is reduced relative to the naive specification, most visibly under homophily, but it does not vanish with sample size, and coverage deteriorates rapidly as $n$ increases. These findings indicate that ad hoc network controls are insufficient to eliminate bias when the covariate is correlated with latent network traits. The infeasible logit estimator, which observes the true $\lambda(\omega_i)$, exhibits negligible bias across all designs and coverage at the nominal level throughout, confirming that the source of bias in the feasible benchmarks is precisely the latent network heterogeneity entering both the outcome and link formation.

The proposed estimator tracks the infeasible benchmark closely once the matching window localizes. Under homophily networks, the absolute bias declines from 0.217 at $n = 100$ to 0.027 at $n = 2000$, and coverage of the influence-function intervals is at the nominal level from $n=250$ onward (0.954 at $n=250$, 0.962 at $n=2000$). Under the stochastic block design, the finite-support regime of Theorem \ref{theo3} implemented with the finite-support bandwidth branch, the bias is essentially zero from $n=250$ onward (0.009 at $n=250$, 0.015 at $n=2000$) and coverage rises to 0.958 at $n=2000$, exactly as the theory predicts: conditional on an exact match the localized score is exactly unbiased, and matching on estimated codegree distances leaves no first-order trace. The exponent diagnostic refuses the correction in essentially every replication at every sample size (the applied share never exceeds 3 percent), so the jackknife column coincides with the uncorrected one, which is the expected behavior when there is no bandwidth bias to remove.

The beta design is the slowest, and the reason is first-stage estimation error
rather than any property of the design itself. Two candidate explanations do not
survive inspection. The bias exponent is the same in the two continuous designs:
$f_1(x,y)=1-(x-y)^2$ gives $\rho(u,v)^2=(u-v)^2\int_0^1(u+v-2s)^2\,ds$, so
$\rho\asymp|u-v|$ and $\alpha=1$, and $f_2$ delivers $\alpha=1$ as well, while
$\lambda$ is the same Lipschitz function in both, so $\varphi=1$ and
$\kappa=\varphi\alpha/(2(1+2\alpha))=1/6$ in each; a mechanism operating through
$\kappa$ cannot distinguish them. Nor is the codegree distance a worse proxy for
the link-function distance under $f_2$: computing $\delta_{ij}/\rho_{ij}$ over a
grid of type pairs gives $[0.71,0.72]$ under $f_2$ against $[0.17,0.73]$ under
$f_1$. Indeed the two designs are equally matchable in principle. Selecting the
closest five percent of pairs by the \emph{true} codegree distance yields average
residual heterogeneity $\E|\lambda(\omega_i)-\lambda(\omega_j)|$ of $0.030$ under
homophily and $0.031$ under beta, at every sample size simulated.

What differs is how well the estimated distance reproduces that ordering. Let
$s_n$ denote the standard deviation of $\hat\delta^2_{ij}-\delta^2_{ij}$ among
pairs whose true distance lies in the bottom five percent, and compare it with
the five percent quantile of $\delta^2_{ij}$, which is the scale the kernel
window must resolve. The error is $O(1/n)$ while the quantile is a fixed feature
of the design, so the ratio improves with the sample in both cases, but its level
differs: the near-zero distances are $4.5$ times more compressed under $f_2$
($1.2\times10^{-5}$ against $5.4\times10^{-5}$) while the estimation error is
about $1.5$ times larger, and the ratio runs $0.08$, $0.21$ and $0.51$ under
homophily against $0.011$, $0.029$ and $0.079$ under beta at $n=250$, $700$ and
$2000$. The ordering of near-matches is correspondingly scrambled, and the
feasible rule both admits fewer pairs (about four percent against ten) and
matches them worse: average residual heterogeneity among selected pairs is
$0.127$, $0.094$ and $0.071$ under homophily, approaching the oracle level, but
$0.458$, $0.327$ and $0.188$ under beta. This is the finite-sample counterpart of
the mechanism identified in Proposition \ref{prop_villages}(iii): there the
first-stage error never vanishes because network size is bounded, while here it
does vanish, but slowly enough that at the sample sizes simulated it still
governs which pairs are matched. The uncorrected bias accordingly declines only
from $0.991$ at $n=100$ to $0.184$ at $n=2000$ while coverage falls to $0.838$
once the variance shrinks, which is the pseudo-true centering of
Theorem \ref{theo4} made visible. The exponent diagnostic behaves accordingly:
with $\kappa=1/6$ lying outside the clamped range $[1/2,2]$, $\hat\kappa$ is not
consistent for $\kappa$ in this design, the correction is applied in only 24 to
53 percent of replications across sample sizes, and where it is applied at the
clamp boundary it overcorrects, leaving the jackknife column worse centered and
more dispersed than the uncorrected one.

\begin{table}[!htbp]
\centering
\begin{tabular}{lcccc}
\hline\hline
Design & $n$ & Median absolute error & RMSE & $\mathrm{corr}(\hat\lambda,\lambda)$ \\
\hline
Homophily & 250 & 0.501 & 0.883 & 0.741 \\
 & 700 & 0.296 & 0.549 & 0.889 \\
 & 2\,000 & 0.179 & 0.341 & 0.943 \\
\hline
Beta & 250 & 2.467 & 2.744 & 0.269 \\
 & 700 & 2.561 & 2.754 & 0.314 \\
 & 2\,000 & 2.099 & 2.510 & 0.367 \\
\hline
Stochastic block & 250 & 0.272 & 0.327 & 0.950 \\
 & 700 & 0.170 & 0.211 & 0.983 \\
 & 2\,000 & 0.095 & 0.113 & 0.994 \\
\hline\hline
\end{tabular}
\caption{Monte Carlo performance of $\hat\lambda$ (equation (\ref{eq7})), all designs}
\label{tab_lambda}
\begin{minipage}{0.86\textwidth}\vspace{2pt}\footnotesize
Notes: 200 replications per cell. Median absolute error is the within-replication median of $|\hat\lambda(\omega_i)-\lambda(\omega_i)|$ over agents, averaged across replications; RMSE and the correlation are computed analogously. The heterogeneity function is $\lambda(w)=(4w^3-3)/2$. 
Homophily and beta use the continuous bandwidth branch; the stochastic block design uses the finite-support branch. $\hat\beta$ is the proposed estimator from Tables \ref{tab1}--\ref{tab3}.
\end{minipage}
\end{table}

The tables so far report only $\hat\beta$; Table \ref{tab_lambda} adds evidence on
the estimator $\hat\lambda$ of equation (\ref{eq7}) in all three designs, computed
with the covariate kernel $K_X(z)\propto(1-z^4)^3\,\mathbbm{1}\{|z|<1\}$ (the triweight-type kernel applied to the squared standardized difference), bandwidth
$h_X=1.06\,\hat\sigma_X\,n^{-1/5}$, trimming $\tau_n=1/(1+\log^2n)$, and the same
design-appropriate bandwidth branch as the corresponding $\hat\beta$ column. The
three designs order exactly as the theory predicts, by match quality. Under the stochastic block design, where matches are
exact, $\hat\lambda$ converges fastest: the correlation with the true
$\lambda(\omega_i)$ reaches 0.99 at $n=2000$ and the median absolute error, 0.095,
is under a fifth of the cross-sectional standard deviation of $\lambda$ (0.50 in
that design). Under homophily, where matches are close but inexact, accuracy
improves steadily (correlation 0.74 to 0.94, median absolute error 0.50 to 0.18
against a standard deviation of 0.57). Under the beta design $\hat\lambda$ is
essentially uninformative at the sample sizes considered: the correlation rises
only from 0.27 at $n=250$ to 0.37 at $n=2000$, the median absolute error (2.47,
2.56 and 2.10 across the three sample sizes, and not monotone in $n$) exceeds the
entire range of $\lambda$, which is 2.0, and the root mean squared error (2.74,
2.75 and 2.51) is more than four times the cross-sectional standard deviation:
the estimator is dominated by a systematic offset that matching at this quality
cannot remove. This is the same mechanism that governs $\hat\beta$ in this design: with
the near-flat bias rate $\kappa=\varphi/6$, the localized matches remain inexact
at every bandwidth used, and $\hat\lambda$ compounds that matching error with the
bias of $\hat\beta$ itself, so the slow convergence documented for $\hat\beta$ in
Table \ref{tab2} is amplified rather than averaged away. In every design
$\hat\lambda$ is a two-stage nonparametric object estimated without bias
correction, and its accuracy should be read as a diagnostic rather than as a basis
for inference; no result elsewhere in this section or in Section \ref{section5}
depends on $\hat\lambda$.


\subsection{When network equivalence fails}\label{sec_misspec}
The identification argument rests on Assumption \ref{ass2}: agents with the same link function must have the same value of $\lambda$. This condition disciplines the data generating process, and it can fail. Table \ref{tab_misspec} illustrates the consequences, using the stochastic block design with the continuous heterogeneity function $\lambda(w)=(4w^3-3)/2$, so that agents within a block are network equivalent yet carry different values of $\lambda$; the within-block standard deviation of $\lambda$ reaches 0.41 in the third block. Matching on network behavior then controls only the between-block variation in $\lambda$, and no bandwidth, however small, can remove the within-block component.

\begin{table}[!htb]
\centering
\begin{tabular}{clccccc}
\hline\hline
&& Naive & Logit with & Infeasible & Proposed & Jackknife \\
&& Logit & Controls & Logit & Estimator & Corrected \\
n&& $\hat\beta_1$ &$\hat\beta_2$ & $\hat\beta_3$  & $\hat\beta$ & $\bar\beta_L$ \\
\hline
100&&  &  &  &  & \\
&$|$Bias$|$&0.9186&0.9515&0.0687&0.4383&0.4352\\
&SD&0.6144&0.6663&0.7576&0.7246&0.7376\\
&RMSE&1.1048&1.1612&0.7600&0.8462&0.8557\\
&Coverage&0.692&0.698&0.944&0.792&0.802\\
250&&  &  &  &  & \\
&$|$Bias$|$&0.8431&0.8325&0.0037&0.2544&0.2543\\
&SD&0.3505&0.3666&0.4371&0.4098&0.4098\\
&RMSE&0.9129&0.9095&0.4367&0.4820&0.4820\\
&Coverage&0.304&0.344&0.952&0.852&0.852\\
700&&  &  &  &  & \\
&$|$Bias$|$&0.8439&0.8108&0.0097&0.2476&0.2476\\
&SD&0.2117&0.2250&0.2678&0.2442&0.2442\\
&RMSE&0.8700&0.8414&0.2677&0.3476&0.3476\\
&Coverage&0.012&0.022&0.948&0.816&0.816\\
1\,000&&  &  &  &  & \\
&$|$Bias$|$&0.8341&0.8038&0.0087&0.2479&0.2479\\
&SD&0.1754&0.1844&0.2178&0.2078&0.2078\\
&RMSE&0.8523&0.8247&0.2178&0.3234&0.3234\\
&Coverage&0.002&0.004&0.950&0.754&0.754\\
2\,000&&  &  &  &  & \\
&$|$Bias$|$&0.8385&0.8054&0.0180&0.2522&0.2522\\
&SD&0.1167&0.1262&0.1443&0.1366&0.1366\\
&RMSE&0.8466&0.8152&0.1453&0.2867&0.2867\\
&Coverage&$<10^{-3}$&$<10^{-3}$&0.964&0.572&0.572\\
\hline
\end{tabular}
\caption{Stochastic Block Design with Within-Block Heterogeneity in $\lambda$ (Assumption \ref{ass2} violated)}
\label{tab_misspec}
\end{table}

The results are exactly as the theory predicts. The bias of the proposed estimator improves on the naive logit (which also absorbs the between-block variation poorly) but stalls near 0.25 from $n=250$ onward (0.254 at $n=250$, 0.252 at $n=2000$), and coverage collapses (0.572 at $n=2000$); the jackknife cannot repair the problem, because the surviving bias is not a bandwidth bias. The contrast between Table \ref{tab3} and Table \ref{tab_misspec} shows that Assumption \ref{ass2} has observable bite, and suggests a practical diagnostic: when the estimates move substantially as the bandwidth shrinks toward the exact-match regime, within-equivalence-class heterogeneity in the outcome equation should be suspected.

Taken together, the Monte Carlo evidence supports four conclusions. First, standard logit approaches can be severely biased in the presence of endogenous network formation, even in large samples. Second, matching on network formation behavior with the regime-appropriate bandwidth branch of equation (\ref{eq_bandwidthrule}) restores consistency without parametric modeling of link formation, and the influence-function variance estimator (\ref{eq_varest}) delivers coverage at or near the nominal level both in the finite-support regime (0.958 at $n=2000$) and in the smooth continuous design (0.940--0.962 from $n=250$ onward under homophily). Third, when the bias rate is nearly flat in the bandwidth, as in the beta design, the exponent diagnostic correctly refuses the jackknife correction in most replications, and inference must be interpreted as centered on the pseudo-true value: the correction is a tool with a diagnosable domain of applicability, not a universal repair. Fourth, the network-equivalence condition of Assumption \ref{ass2} is a substantive restriction whose failure is detectable: under within-block heterogeneity in $\lambda$ the bias stalls near 0.25 and coverage collapses, however large the sample.

\section{Empirical Application}\label{section5}

This section applies the proposed estimator to household-level data on the diffusion of microfinance in rural Indian villages. The purpose of the exercise is to illustrate how the estimator can be implemented in a canonical network setting and to assess how accounting for endogenous network formation affects estimated covariate effects in a binary adoption model. The analysis is intentionally reduced-form and complements existing structural approaches to diffusion.

\subsection{\texorpdfstring{Data and Context: \cite{banerjee2013diffusion}}{Data and Context}}

The empirical application uses the village social network data collected and analyzed in
\cite{banerjee2013diffusion} paper. The authors studied the spread of a microfinance loan program in 43 villages of Karnataka, India. In each village, a couple of village leaders like teachers, leaders of self-help groups, and shop-keepers, were invited to a private meeting, in which they were informed of the program and encouraged to spread information about the program to their peers. The data include detailed descriptions of each household’s social relationships, including which friends or relatives visit the individual, which friends or relatives the individual visits, whom the individual would borrow money from, to whom the individual would lend money, etc\footnote{They collected social network information across 12 dimensions, encompassing various aspects of the respondent’s social interactions. These dimensions include: 1. Individuals who visit the respondent’s home. 2. Individuals whose homes the respondent visits. 3. Family members within the same village. 4. Non-relatives with whom the respondent socializes. 5. Sources from which the respondent seeks medical advice. 6. Potential sources for borrowing money. 7. Potential individuals to whom the respondent would lend money. 8. Sources from which the respondent might borrow material goods (such as kerosene or rice). 9. Potential recipients to whom the respondent would lend material goods. 10. Advisors from whom the respondent seeks guidance. 11. Individuals to whom the respondent provides advice. 12. Individuals with whom the respondent engages in communal prayer, whether at a temple, church, or mosque.}. Information on household characteristics is also collected, e.g., whether the household has electricity or a latrine, the number of rooms and beds, etc. The outcome of interest is a binary indicator equal to one if a household adopts microfinance during the study period.

Although social links are measured before the introduction of microfinance, unobserved household characteristics such as financial sophistication, trust in financial institutions, risk tolerance, or entrepreneurial ability plausibly affect both a household’s likelihood of adopting microfinance and its pattern of social interactions. These latent traits are not directly observed by the econometrician but may be reflected in network formation behavior.

The adjacency matrix is constructed as follows. The twelve relationship dimensions are combined by taking their union: households $i$ and $j$ are linked, $D_{ij}=1$, if at least one of the twelve relations connects them in at least one direction, so a single nomination by either household in any dimension creates an undirected link and multiple nominations are collapsed to the same binary link. The resulting matrix is symmetric with a zero diagonal, matching the model of Section \ref{section3}. Nominations of individuals outside the survey frame cannot be matched to a surveyed household and are discarded, and item nonresponse is treated as the absence of a nomination; both choices follow the standard use of these data. Because the model treats $D$ as a noisy realization of the latent link function $f$ rather than as a fixed regressor, the union construction is conservative for the matching step: it pools all dimensions of interaction into a single behavioral signature, and the codegree distance then compares households on the entirety of that signature. The codegree distances, and hence the matched sets, are computed from this matrix within village.

\subsection{Results}
The empirical results are reported in Table \ref{emp_results}. Columns (1)--(2) present a standard logit, without and with village fixed effects; columns (3)--(4) add network controls (row sums $\sum_jD_{ij}y_j$ and $\sum_jD_{ij}X_j$), without and with village fixed effects; columns (5)--(6) report the proposed estimator. For the proposed estimator, pairs $(i,j)$ are restricted to households in the same village, so any village-level factor that enters the logit index additively (program intensity, local economic conditions, institutional differences) differences out of the conditional likelihood exactly, in the same way $\lambda(\omega_i)$ does; village factors that interact with the covariates, alter the slopes, or change the error distribution would not difference out and are not controlled by this design; no separate fixed-effects specification exists, because village indicators are constant within pairs. Codegree distances are computed within village with village-specific normalization, as in Proposition \ref{prop_villages}, consistent with the absence of cross-village links in the data. The implementation follows Sections \ref{section3_2}--\ref{section332} exactly: the triweight kernel on $\hat\delta^2_{ij}/\hat h_g$, village-specific bandwidths from the rule (\ref{eq_bandwidthrule}) (the regime diagnostic selects the continuous branch in all 43 villages; operationally, the finite-support branch is chosen in a village only when at least 15 percent of its pairs lie within $3\hat m_g$ of the noise scale and the band $(3\hat m_g,9\hat m_g]$ holds no more than a third of that mass, the point-mass-plus-gap signature of equation (\ref{eq_noisefloor})). 
Column (5) reports $\hat\beta$ at the benchmark bandwidths $\hat h_g$; column (6) reports $\hat\beta$ at $2\hat h_g$, the bandwidth-sensitivity check of Section \ref{sec_misspec}. The jackknife correction is not applied: the exponent diagnostic of Section \ref{section332} returns $\hat\kappa=-0.98$, well below the applicability threshold, so the slope of the bandwidth path is uninformative about a bias exponent and the uncorrected estimator is the appropriate report.

The scale of the exercise is as follows. The sample contains 9{,}126 households in 43 villages (village sizes 107 to 341, median 200); the adoption rate is 18.4 percent, mean degree is 9.7, no household is isolated, and the adjacency matrix contains no cross-village links. A mean degree of 9.7 at a median village size of 200 corresponds to a link density of roughly five percent, whereas the asymptotic theory developed above is a dense-network theory in which density is bounded away from zero; at this density the codegree noise floor is large relative to the distances being matched on, and the match-quality diagnostics below quantify the resulting limitation directly. There are 1{,}024{,}042 within-village pairs, of which 285{,}519 are discordant; 11{,}361 receive positive kernel weight at $\hat h_g$ (per-village minimum 39, median 239) and 36{,}949 at $2\hat h_g$. The bandwidths range from $1.8\times10^{-6}$ to $5.2\times10^{-5}$ across villages (median $1.2\times10^{-5}$). 
Two diagnostics quantify match quality. The median ratio of the estimated
first-stage noise scale $\hat m_g$ to the median within-village codegree distance
is $0.43$, and the match counts at $\hat h_g$ and $2\hat h_g$ (11{,}361 and
36{,}949) give a count exponent $\log_2(36{,}949/11{,}361)=1.70$. This lies
inside the range $\beta_0>3/8$ over which Appendix \hyperlink{app_bandwidth}{A.2} certifies $a=1/16$, but
an exponent this far above the leading case is itself a signal that the window
lies in the first-stage noise rather than a property of the underlying distance
distribution, since it is estimated from $\hat\delta^2_{ij}$ rather than
$\delta^2_{ij}$. Correcting the exponent would not help in any case: at village
sizes between 107 and 341 the quantile $c_n$ moves by less than six percent
across the admissible range of $a$. The village bandwidths are of the same order
as $\hat m_g$, so localization is limited by match quality, and the estimates
below should be read as centered on the pseudo-true value attainable at that
quality, as anticipated by Proposition \ref{prop_villages}(iii).

Standard errors are clustered at the village level (43 clusters) in all six columns: columns (1)--(4) use the standard cluster-robust logit sandwich with the $G/(G-1)$ small-cluster correction, and columns (5)--(6) use the sandwich $\widehat H^{-1}\big[\sum_g S_g S_g'\big]\widehat H^{-1}$, where $S_g$ is village $g$'s contribution to the localized score at $\hat\beta$ and $\widehat H$ is the Hessian; because villages are independent networks, all within-village dependence, including the estimation error in the codegree distances, is contained in $S_g$, so this is the clustered score sandwich $\widehat V_{\mathrm{cl}}$ of Proposition \ref{prop_villages}(ii). The localized standard errors are between roughly 1.7 and 2.7 times the clustered naive-logit ones (for example, 0.096 against 0.040 for rooms and 0.305 against 0.184 for rooms per capita), reflecting that localization discards the concordant and poorly matched pairs and therefore uses far fewer effective comparisons; the comparison remains indicative only, since the estimands differ and the naive logit omits $\lambda(\omega_i)$ and is misspecified under the model. Clustering matters for the pooled columns themselves: the beds coefficient, marginally significant under maximum-likelihood standard errors, has $|t|\approx1.3$ once clustered, and the latrine coefficient in column (1) drops from the 1 percent to the 5 percent level.

\begin{table}[htb]
\centering
\begin{threeparttable}
\makebox[\textwidth][c]{\footnotesize\setlength{\tabcolsep}{2.5pt}%
\begin{tabular}{lcccccc}
\hline\hline
&\multicolumn{2}{c}{Naive Logit}&\multicolumn{2}{c}{Logit with Controls}&\multicolumn{2}{c}{Proposed estimator}\\
 & (1) & (2)& (3) & (4)& (5) $\hat\beta(\hat h)$ & (6) $\hat\beta(2\hat h)$ \\
\hline
Intercept
& -1.9059\sym{***} & -1.8011\sym{***} & -2.1084\sym{***}&  -1.9411\sym{***}& - & -\\
& (0.2323) & (0.2015) & (0.1995)& (0.1948)& & \\
No. of Rooms
& 0.1797\sym{***} &0.1897\sym{***}  &0.2071\sym{***} &0.2002\sym{***}& 0.3802\sym{***} & 0.3826\sym{***} \\
& (0.0396) & (0.0427)&(0.0460) &  (0.0471)& (0.0963)& (0.0788)\\
No. of Beds
& -0.1249 &-0.1076& -0.1155&  -0.1118& -0.2710 & -0.3344\sym{*}\\
& (0.0943) & (0.0937)& (0.0977)&  (0.0985)& (0.1831)& (0.1772)\\
Electricity (category)
& 0.2428\sym{***} & 0.2348\sym{***} &0.1810\sym{***}& 0.1822\sym{***}& -0.2014 & -0.0217 \\
& (0.0550) & (0.0554) & (0.0562)& (0.0556)&  (0.1509)& (0.1062)\\
Latrine (category)
& 0.1572\sym{**} & 0.2137\sym{***}& 0.1788\sym{***}&0.1811\sym{***}& 0.2786\sym{**} & 0.2453\sym{***} \\
& (0.0666) &  (0.0643)&(0.0519) & (0.0541)& (0.1291)& (0.0881)\\
Rooms per capita
& -1.2922\sym{***} &  -1.2621\sym{***}&-1.1874\sym{***} &-1.2177\sym{***}&  -2.0559\sym{***} & -1.7967\sym{***} \\
& (0.1844) & (0.1657)&(0.1922) & (0.1868)& (0.3047) & (0.2842) \\
Beds per capita
& 0.1602& 0.1575	& 0.1346&  0.1764& 0.3538 & 0.7749\\
& (0.4330) &(0.4193) & (0.4379)& (0.4278) & (0.7197)& (0.7228)\\
\hline
Village FE & No & Yes& No & Yes& \multicolumn{2}{c}{differenced out} \\
Matched pairs & -- & -- & -- & -- & 11\,361 & 36\,949 \\
Observations & 9126 & 9126 & 9126 & 9126&9126&9126 \\
\hline
\end{tabular}
}
\begin{tablenotes}
\footnotesize
\item Notes: Standard errors in parentheses, clustered at the village level (43 clusters) in all columns: the score sandwich of Proposition \ref{prop_villages}(ii) in columns (5)--(6) and the standard cluster-robust logit sandwich with the $G/(G-1)$ small-cluster correction in columns (1)--(4). Significance in all columns uses $t_{42}$ critical values. Electricity and Latrine are categorical survey codes entered linearly. Village indicators are constant within matched pairs and difference out of columns (5)--(6). \sym{***}, \sym{**}, and \sym{*} denote significance at the 1\%, 5\%, and 10\% levels.
\end{tablenotes}
\end{threeparttable}
\caption{Empirical Results}\label{emp_results}
\end{table}

Three findings emerge from Table \ref{emp_results}. First, housing size and crowding are robustly associated with adoption once latent network heterogeneity is controlled for: the coefficient on the number of rooms roughly doubles relative to the naive logit (0.380 against 0.180) and rooms per capita strengthens from $-1.29$ to $-2.06$, both significant at the 1 percent level with clustered standard errors, and both stable as the bandwidth doubles. The latrine category is positive and significant at the 5 percent level at $\hat h$ and the 1 percent level at $2\hat h$. These are the effects that survive within-village matching on network behavior.

Second, the electricity coefficient illustrates the bandwidth diagnostic of Section \ref{sec_misspec} on real data. Electricity is strongly positive in the naive logit (0.243, $t\approx4.4$ with clustered standard errors) but sign-reversed and imprecisely estimated under localized within-village matching ($-0.201$ at $\hat h$, $t\approx-1.3$), and it drifts monotonically back toward the pooled value as the matching window widens ($-0.02$ at $2\hat h$, $+0.12$ at $4\hat h$, $+0.23$ with no localization). Estimates that move steadily with the bandwidth indicate a coefficient that is sensitive to which pairs are compared, a pattern consistent with households with electricity occupying systematically different positions in the village network. Absent additional causal assumptions, the localized estimates are conditional associations, and the bandwidth path does not by itself establish that the naive association operates through network position rather than through the amenity. Beds per capita tells a similar story in reverse: it is statistically indistinguishable from zero at every bandwidth ($t=0.49$ at $\hat h$, $1.07$ at $2\hat h$), and village clustering alone already removes the apparent precision of the pooled specifications.

Third, the within-village design makes village fixed effects unnecessary rather than merely optional: any village-level factor entering the index additively is constant within matched pairs and is removed by the same differencing that removes $\lambda(\omega_i)$. The comparison between columns (2), (4) and columns (5), (6) therefore isolates what network-based matching adds beyond village-level controls, namely the household-level component of latent social heterogeneity. The pseudo-true caveat of Proposition \ref{prop_villages}(iii) applies with villages of median size 200, and the reported estimates are best read as adoption gradients net of village effects and of as much latent network heterogeneity as codegree matching at this network size can remove.

\section{Conclusion}\label{section6}
This paper studies identification and estimation in binary choice models when social networks are endogenous. In many empirical applications, unobserved individual traits influence both economic outcomes and the formation of social links, invalidating standard logit specifications and ad hoc network controls. This paper proposes an alternative approach that uses observed network data to control directly for latent individual heterogeneity, without imposing parametric assumptions on network formation.

The argument rests on one observation: individuals who are observationally equivalent in their network formation behavior share the same value of the heterogeneity term $\lambda(\omega_i)$ under Assumption \ref{ass2}, even though the latent trait $\omega_i$ itself is neither observed nor identified. Exploiting this equivalence, the paper establishes point identification of slope parameters in a semiparametric logit model through pairwise comparisons of agents with similar network formation behavior. A feasible estimator based on codegree information is developed and shown to be consistent. Its limiting distribution is derived in two regimes: exact root-$n$ normality centered at the truth when network types have finite support, and normality centered at a pseudo-true value in general, with an explicit bias characterization, a jackknife bias correction, and variance estimators that account for matching on estimated network distances.

Monte Carlo simulations demonstrate that the proposed estimator performs well in finite samples across a range of network environments, including stochastic block, beta, and homophily designs. The estimator exhibits small bias, favorable mean squared error properties, and coverage at the nominal level in the finite-support and homophily designs at empirically relevant sample sizes; in the degree-driven beta design, where codegree matching is least informative, the simulations transparently exhibit the pseudo-true centering that the theory predicts, together with the bandwidth-sensitivity diagnostic that reveals it in practice.

An empirical application to microfinance adoption \citep{banerjee2013diffusion} in rural Indian villages illustrates the practical relevance of the method. Using rich pre-treatment social network data, the proposed estimator delivers covariate effects that are stable across matching bandwidths, with additive village-level factors removed by within-village differencing. Network-based matching captures household-level heterogeneity in latent social traits that village-level controls alone do not absorb: housing-size and crowding effects strengthen, while the naive association between electricity and adoption dissipates under localized matching, moving monotonically back toward the pooled value as the matching window widens.

More broadly, the approach gives a transparent way to use network data in nonlinear models in which latent heterogeneity drives both outcomes and link formation. While the analysis focuses on cross-sectional binary choice models, the underlying ideas extend naturally to other nonlinear settings and to models with repeated observations. Future work may consider extensions to panel data, alternative notions of network similarity, or settings with multiple outcomes and interacting networks.

\section{Appendix}

\subsection*{A.1 Auxiliary lemmas}

The first lemma records the uniform convergence rate of the estimated codegree distances. It corresponds to Lemma B1 in \cite{auerbach2022identification} with an explicit rate; a self-contained proof is given so that the network-side results of this paper do not rest on unpublished material. The lemma is invoked in Step 4 of the proof of Theorem \ref{theo2}, in Step 2 of the proof of Theorem \ref{theo3}, and in Steps 1--2 of the proof of Theorem \ref{theo4}, each time to show that replacing $\delta^2_{ij}$ with $\hat\delta^2_{ij}$ is negligible at the relevant scale.

\begin{lemma}\label{lemB1}
Under Assumption \ref{ass1}, $\max_{i\neq j}\left|\hat{\delta}^2_{ij}-\delta^2_{ij}\right|=O_p\left(n^{-1/2}\log^{1/2}n\right)$.
\end{lemma}

\begin{proofoflemma}
Write $\hat\delta^2_{ij}=\frac1n\sum_t\hat a_t(i,j)^2$ with \\$\hat a_t:=\frac1n\sum_sD_{ts}(D_{is}-D_{js})$, and let $\boldsymbol\omega:=(\omega_1,\ldots,\omega_n)$.

Conditional on $\boldsymbol\omega$, the summands $\{D_{ts}(D_{is}-D_{js})\}_{s\notin\{t,i,j\}}$ are independent across $s$ (they involve the disjoint dyad shocks $\eta_{ts},\eta_{is},\eta_{js}$) and are bounded by one, with conditional means $f(\omega_t,\omega_s)(f(\omega_i,\omega_s)-f(\omega_j,\omega_s))$; the at most three excluded terms contribute $O(1/n)$ deterministically. Hoeffding's inequality gives $\PP\left(|\hat a_t-\bar a_t|\ge x\,|\,\boldsymbol\omega\right)\le2e^{-nx^2/2}$ with $\bar a_t:=\E[\hat a_t|\boldsymbol\omega]$; a union bound over the at most $n^3$ triples $(t,i,j)$ yields $\max_{t,i,j}|\hat a_t-\bar a_t|=O_p(\sqrt{\log n/n})$.

For $t\notin\{i,j\}$, conditional on $(\omega_t,\omega_i,\omega_j)$, $\bar a_t-A_t=\frac1n\sum_{s\notin\{t,i,j\}}\{g(\omega_s)-\E g(\omega_s)\}+O(1/n)$, where $g(u):=f(\omega_t,u)(f(\omega_i,u)-f(\omega_j,u))\in[-1,1]$ and \\$A_t:=\int f(\omega_t,u)(f(\omega_i,u)-f(\omega_j,u))du$. 
Hoeffding's inequality and the same union bound give $\max_{(i,j)}\max_{t\notin\{i,j\}}|\bar a_t-A_t|=O_p(\sqrt{\log n/n})$. For the two coincident indices $t\in\{i,j\}$ this display fails: at $t=i$ the summand $D_{is}(D_{is}-D_{js})$ has conditional mean $f_{is}(1-f_{js})$ rather than $g(\omega_s)=f_{is}(f_{is}-f_{js})$, because $D_{is}^2=D_{is}$, so $\bar a_i-A_i$ is of order one; these two indices are excluded here and their contribution is bounded separately in Step 3.

Since $|\hat a_t|,|A_t|\le1$, $\left|\frac1n\sum_{t\notin\{i,j\}}\hat a_t^2-\frac1n\sum_{t\notin\{i,j\}}A_t^2\right|\le2\max_{t\notin\{i,j\}}|\hat a_t-A_t|=O_p(\sqrt{\log n/n})$ uniformly over pairs, combining Steps 1--2; the two coincident terms contribute $\frac1n(\hat a_i^2+\hat a_j^2)\le2/n$ and $\frac1n(A_i^2+A_j^2)\le2/n$, both $o(\sqrt{\log n/n})$ deterministically.

Conditional on $(\omega_i,\omega_j)$, $\frac1n\sum_{t\ne i,j}A_t^2-\int A(u;\omega_i,\omega_j)^2du$ is a centered average of i.i.d.\ bounded terms; Hoeffding's inequality with a union bound over the $n^2$ pairs gives $O_p(\sqrt{\log n/n})$. 
\\Summing Steps 3--4 and noting $\delta^2_{ij}=\int A(u;\omega_i,\omega_j)^2du$ completes the proof. $\blacksquare$
\end{proofoflemma}

The second lemma sharpens Lemma \ref{lemB1} in the finite-support case: for exactly matched pairs the estimated squared distance concentrates at the faster rate $\log n/n$, and pairs that are not exact matches are eventually never selected by the kernel. It corresponds to Lemma C1 in \cite{auerbach2021inference}. The lemma is invoked in Step 2 of the proof of Theorem \ref{theo3}, where part (i) controls the kernel weights on exact matches and part (ii) removes all other pairs, and it underlies the finite-support branch of the bandwidth rule (\ref{eq_bandwidthrule}).

\begin{lemma}\label{lemC1}
Under Assumptions \ref{ass1}, \ref{ass3} and \ref{ass5}: 
\begin{enumerate}[(i)]
\item  $\max_{i\neq j}\hat{\delta}^2_{ij}\,\mathbbm{1}\{\delta^2_{ij}=0\}=O_p\left(n^{-1}\log n\right)$; 
\item $\PP\left(\exists\,(i,j):\ \delta^2_{ij}\ge\bar{\epsilon} \mbox{ and } \hat{\delta}^2_{ij}<h\right)\to0$.
\end{enumerate}
\end{lemma}

\begin{proofoflemma}
(i) If $\delta^2_{ij}=0$ then $A(\cdot\,;\omega_i,\omega_j)=0$ almost everywhere, so for $t\notin\{i,j\}$ the decomposition $\hat a_t=(\hat a_t-\bar a_t)+(\bar a_t-A_t)$ is a sum of two centered averages; the sub-Gaussian tails of Steps 1--2 of the previous proof give, after the union bound, $\max_{t\notin\{i,j\}}\hat a_t^2=O_p(n^{-1}\log n)$ uniformly over exactly matched pairs. The two coincident indices are not centered: at $t=i$, $\hat a_i=\frac1n\sum_sD_{is}(1-D_{js})$ concentrates at $\int f_{iu}(1-f_{ju})\,du$, which is of order one even on exact matches; but $|\hat a_i|,|\hat a_j|\le1$, so their contribution to $\hat\delta^2_{ij}$ is $\frac1n(\hat a_i^2+\hat a_j^2)\le2/n$. Hence $\hat\delta^2_{ij}\mathbbm{1}\{\delta^2_{ij}=0\}=O_p(n^{-1}\log n)$, the coincident terms entering at the same $n^{-1}$ order as the dyad-noise floor. (ii) On $\{\delta^2_{ij}\ge\bar\epsilon\}$, Lemma \ref{lemB1} gives $\min_{(i,j):\delta^2_{ij}\ge\bar\epsilon}\hat\delta^2_{ij}\ge\bar\epsilon/2$ with probability approaching one, while $h\to0<\bar\epsilon/2$ eventually; hence no such pair enters the kernel window. $\blacksquare$
\end{proofoflemma}

The third lemma has a single purpose: it transfers Theorems \ref{theo2}-\ref{theo5}, which are stated for deterministic bandwidth sequences, to the data-driven rule (\ref{eq_bandwidthrule}) used in Sections \ref{section4} and \ref{section5}.

\begin{lemma}\label{lem_bandwidth}
(a) (Continuous regime.) Suppose Assumption \ref{ass1} holds, \\$U(t):=\PP(\delta^2_{ij}\le t)$ is continuous and strictly increasing on $(0,t_0]$ for some $t_0>0$, the population $c_n$-quantile $q_n$ satisfies $q_n\gg n^{-1/2}\log^{1/2}n$ together with the local moduli
\[U\big(q_n(1+\varepsilon)\big)-U\big(q_n(1-\varepsilon)\big)\;\le\;C\,\varepsilon\,c_n \qquad\text{and}\qquad U\big(q_n(1+\varepsilon)\big)-U(q_n)\;\ge\;c\,\varepsilon\,c_n\]
for all small $\varepsilon>0$ and constants $0<c\le C<\infty$, with $U(q_n\pm\varepsilon_n)-U(q_n\mp\varepsilon_n)=o(c_n)$ for $\varepsilon_n=O(n^{-1/2}\log^{1/2}n)$, and the population median $q_{1/2}$ of $\delta^2_{ij}$ is unique. Let $h_n:=\max(q_n,\,n^{-2/5}q_{1/2})$. Then the continuous-branch bandwidth in (\ref{eq_bandwidthrule}) satisfies $\hat h_n/h_n-1=O_p(\rho_n)$ with $\rho_n:=n^{-1/2}\log^{1/2}n/c_n=o(n^{-1/4})$ for the rule $c_n\asymp n^{-1/16}$.
(b) (Finite-support regime.) Suppose Assumptions \ref{ass1} and \ref{ass5} hold with $\PP(\delta^2_{ij}=0)<1/2$, the noise scale $\hat m$ is computed over a random subsample of $B_n\ge n$ pairs, and the conditional link variance $\nu(\omega_i,\omega_j):=\E\big[f(\omega_t,\omega_s)\{f_{is}+f_{js}-2f_{is}f_{js}\}\,\big|\,\omega_i,\omega_j\big]$, the expectation over independent $\omega_t,\omega_s$, is bounded away from zero and infinity almost surely (which excludes only networks whose links are deterministic given types), and the distribution of $\nu(\omega_i,\omega_j)$ has a positive density in a neighborhood of its median. Then $n\hat m\to_p\nu_{\mathrm{med}}$, the population median of $\nu$, and the finite-support-branch bandwidth satisfies $\hat h_n=C_0\,n^{-1/3}(1+O_p(n^{-1/2}\log^{1/2}n\vee B_n^{-1/2}))$ with $C_0=(3\nu_{\mathrm{med}})^{1/3}q_{1/2}^{2/3}>0$: it eventually lies above the exact-match noise scale of Lemma \ref{lemC1}, below $\bar\epsilon$, and within the range permitted by Assumption \ref{ass3}(\ref{ass3ii}).
(c) (Transfer.) In the respective regimes, the conclusions of Theorems \ref{theo2} and \ref{theo3} hold with $h$ replaced by $\hat h_n$; under the additional smoothness of $h\mapsto\beta_h$ imposed in Assumption \ref{ass_jack}, the conclusions of Theorems \ref{theo4} and \ref{theo5} hold with $h$ replaced by $\hat h_n$ and the pseudo-true value evaluated at the realized bandwidth, $\beta_{\hat h_n}$, and Theorem \ref{theo5}'s centering at $\beta$ is unaffected.
\end{lemma}

\begin{proofoflemma}
\textbf{Part (a).} Let $\widehat U_n(t)=\binom n2^{-1}\sum_{i<j}\mathbbm{1}\{\hat\delta^2_{ij}\le t\}$ and $\widetilde U_n(t)$ its infeasible version with $\delta^2_{ij}$. The class $\{(u,v)\mapsto\mathbbm{1}\{\delta(u,v)^2\le t\}:t\ge0\}$ is linearly ordered, hence VC, and the law of large numbers for U-processes (Theorem 3.1 of \citealp{arcones1993limit}, with the CLT rate from their Theorem 4.9) gives $\sup_t|\widetilde U_n(t)-U(t)|=O_p(n^{-1/2})$. By Lemma \ref{lemB1}, $\varepsilon_n:=\max_{i\ne j}|\hat\delta^2_{ij}-\delta^2_{ij}|=O_p(n^{-1/2}\log^{1/2}n)$, and pointwise
\[|\widehat U_n(t)-\widetilde U_n(t)|\le\binom n2^{-1}\sum_{i<j}\mathbbm{1}\{|\delta^2_{ij}-t|\le\varepsilon_n\}\le \widetilde U_n(t+\varepsilon_n)-\widetilde U_n(t-\varepsilon_n),\]
which at $t$ near $q_n$ is $o_p(c_n)$ by the local moduli in (a). Now fix $\varepsilon>0$. By the lower modulus bound, $U(q_n(1+\varepsilon))\ge c_n+c\varepsilon c_n$, so on the event that all the empirical-distribution errors above are at most $\tfrac12c\varepsilon c_n$, whose probability tends to one because $n^{-1/2}=o(\varepsilon c_n)$ (this is where $q_n\gg n^{-1/2}\log^{1/2}n$ and the displayed moduli are used), we have $\widehat U_n(q_n(1+\varepsilon))\ge c_n$, hence $\hat q_{c_n}\le q_n(1+\varepsilon)$; the symmetric argument gives $\hat q_{c_n}\ge q_n(1-\varepsilon)$. Making $\varepsilon$ proportional to $\rho_n$ and repeating the argument yields the rate $O_p(\rho_n)$, since the error terms are $O_p(n^{-1/2}\log^{1/2}n)$ against a modulus of size $c\varepsilon c_n$. The empirical median satisfies $\hat q_{1/2}\to_pq_{1/2}$ by the same argument at $c=1/2$ (uniqueness of the median), with rate $O_p(n^{-1/2}\log^{1/2}n)$ against a fixed modulus; hence the floor $n^{-2/5}\hat q_{1/2}$ inherits relative rate $O_p(n^{-1/2}\log^{1/2}n)=o(\rho_n)$. The maximum of two quantities each within relative $O_p(\rho_n)$ of its deterministic counterpart is within relative $O_p(\rho_n)$ of the maximum. Finally $\rho_n=n^{-1/2}\log^{1/2}n\cdot n^{1/16}=n^{-7/16}\log^{1/2}n=o(n^{-1/4})$.

\textbf{Part (b).} Fix a pair $(i,j)$ and write the statistic inside the median of (\ref{eq_noisefloor}) as
\[\mathrm{nf}_{ij}=\frac{1}{n^3}\sum_{s}\big(D_{is}-D_{js}\big)^2\,\mathrm{deg}_s,\qquad \mathrm{deg}_s=\sum_tD_{ts}.\]
Conditional on $\boldsymbol\omega=(\omega_1,\dots,\omega_n)$, the summands of $\mathrm{nf}_{ij}$ are \emph{not} mutually independent across $s$: $\mathrm{deg}_s$ and $\mathrm{deg}_{s'}$ share the single dyad $D_{ss'}$, and $\mathrm{deg}_s$ overlaps $(D_{is}-D_{js})^2$ through $t\in\{i,j\}$. Concentration therefore follows from the bounded-differences inequality rather than from Hoeffding's inequality for independent sums. Viewing $\mathrm{nf}_{ij}$ as a function of the independent dyad shocks $\{\eta_{uv}\}_{u<v}$, changing one dyad $\{t,s\}$ with $t,s\notin\{i,j\}$ moves $\mathrm{nf}_{ij}$ by at most $2n^{-3}$ (through $\mathrm{deg}_s$ and $\mathrm{deg}_t$), while changing one of the $O(n)$ dyads involving $i$ or $j$ moves it by at most $2n^{-2}$ (through the factor $(D_{is}-D_{js})^2$). The sum of squared differences is $O(n^2)\cdot n^{-6}+O(n)\cdot n^{-4}=O(n^{-3})$, so McDiarmid's inequality gives $\mathrm{nf}_{ij}-\E[\mathrm{nf}_{ij}\mid\boldsymbol\omega]=O_p(n^{-3/2})=o_p(n^{-1})$, and
\[\E[\mathrm{nf}_{ij}\mid\boldsymbol\omega]=\frac1{n^3}\sum_{s\ne i,j}e_{ij}(\omega_s)\sum_{t}f(\omega_t,\omega_s)+O(n^{-2}),\]
where $e_{ij}(u):=f_{iu}+f_{ju}-2f_{iu}f_{ju}$, with $f_{iu}:=f(\omega_i,u)$. Averaging over the i.i.d.\ draws $\omega_t,\omega_s$ (Hoeffding again, terms bounded by $n^{-2}$) yields $n\,\mathrm{nf}_{ij}\to_p\nu(\omega_i,\omega_j)$ for each pair; the same computation with $D_{ts}$ replaced by its conditional mean shows that the dyad-noise contribution to $\E[\hat\delta^2_{ij}\mid\text{exact match},\boldsymbol\omega]$ has the identical leading term; the two coincident outer indices $t\in\{i,j\}$ add a further systematic $O(n^{-1})$ term (Lemma \ref{lemC1}(i)), so the exact-match scale of $\hat\delta^2_{ij}$ is $\asymp_pn^{-1}$, and only this order, not its constant, enters the construction of the finite-support branch. For the median over the subsample, note that the statistics $\mathrm{nf}_{ij}$ share the adjacency matrix and are dependent across pairs; the argument therefore runs through the latent values. First, $\max_{(i,j)\in\mathcal B_n}|n\,\mathrm{nf}_{ij}-\nu(\omega_i,\omega_j)|=o_p(1)$ by the Hoeffding bounds above with a union over the $B_n\le\binom n2$ subsampled pairs (the exponential tails absorb the union at rate $n^{-1/2}\log^{1/2}n$), so the empirical median of $\{n\,\mathrm{nf}_{ij}\}$ and that of $\{\nu(\omega_i,\omega_j)\}$ differ by $o_p(1)$. Second, the empirical distribution function of $\{\nu(\omega_i,\omega_j)\}$ over pairs is a U-process in the i.i.d.\ types, and $\sup_t$-convergence at rate $O_p(B_n^{-1/2})$ follows from the law of large numbers and central limit theorem for U-processes \citep{arcones1993limit}; provided the distribution of $\nu(\omega_i,\omega_j)$ has a positive density at its median, which is imposed in part (b), the sample median converges to $\nu_{\mathrm{med}}$ at rate $O_p(B_n^{-1/2})$ (if the median is an atom or a flat stretch, the argument delivers convergence to the median interval, which suffices for $\hat m\asymp_pn^{-1}$ and for every use below). The bandwidth limit then follows from the continuous-mapping theorem applied to $(x,y)\mapsto(3x)^{1/3}y^{2/3}$ and $\hat q_{1/2}\to_pq_{1/2}$; under Assumption \ref{ass5} the median of $\delta^2_{ij}$ is an atom of the finite support, and the hypothesis $\PP(\delta^2_{ij}=0)<1/2$ delivers $q_{1/2}\ge\bar\epsilon>0$. This hypothesis is imposed directly because it does not follow from Assumption \ref{ass5} alone: two separated network types with masses $0.9$ and $0.1$ give $\PP(\delta_{ij}=0)=0.9^2+0.1^2=0.82$, so the median of $\delta^2_{ij}$ is zero despite there being two distinct types. In such heavily concentrated designs the finite-support branch should replace the median by a quantile known to exceed the exact-match mass; none of the designs of Section \ref{section4} is of this kind ($\PP(\delta_{ij}=0)=1/3$ in the block design).

\textbf{Part (c).} Write $h_n$ for the deterministic counterpart, equal to $\max(q_n,n^{-2/5}q_{1/2})$ in the continuous regime and to $C_0n^{-1/3}$ in the finite-support regime, and $\rho_n$ for the relative rate from parts (a)--(b); choose a deterministic sequence $\epsilon_n\downarrow0$ with $\epsilon_n/\rho_n\to\infty$ and $\epsilon_n=o(n^{-1/4})$ (possible since $\rho_n=o(n^{-1/4})$), and let $A_n:=\{|\hat h_n/h_n-1|\le\epsilon_n\}$, so $\PP(A_n)\to1$. All statements below are on $A_n$; the construction is the standard sandwich for uniform-in-bandwidth results \citep{einmahl2005uniform}.

\emph{Consistency (Theorem \ref{theo2}).} For any array $Z_{ij}$ with $\sup_{\omega,\omega'}\E[|Z_{ij}|\,|\,\omega_i,\omega_j]<\infty$ and any $h\in[(1-\epsilon_n)h_n,(1+\epsilon_n)h_n]$,
\begin{multline}\label{eq_pert}
\Big|\tbinom n2^{-1}\!\sum_{i<j}\big[K(\hat\delta^2_{ij}/h)-K(\hat\delta^2_{ij}/h_n)\big]Z_{ij}\Big|\\
\le \|K'\|_\infty\frac{2\epsilon_n}{1-\epsilon_n}\, \tbinom n2^{-1}\!\sum_{i<j}\mathbbm{1}\{\hat\delta^2_{ij}\le2h_n\}\,|Z_{ij}|,
\end{multline}
because $|K(u/h)-K(u/h_n)|\le\|K'\|_\infty\,u\,|h^{-1}-h_n^{-1}|\le\|K'\|_\infty\,(u/h_n)\cdot\epsilon_n/(1-\epsilon_n)$ on the union of supports, and $u/h_n\le2$ there. As in Step 4 of the proof of Theorem \ref{theo2} (Lemma \ref{lemB1} plus Assumption \ref{ass4}(\ref{ass4iv})), the localized sum on the right is $O_p(r_n)$, so the perturbation is $O_p(\epsilon_nr_n)=o_p(r_n)$. Applying this with $Z_{ij}=m(v_i,v_j,b)$ and $Z_{ij}=1$ shows the objective and the match count at $\hat h_n$ equal their values at $h_n$ up to $o_p(1)$ after normalization, pointwise in $b$; the convexity argument is unaffected, and $\hat\beta\to_p\beta$ follows. The consistency of $\hat\lambda$ transfers identically.

\emph{Finite support (Theorem \ref{theo3}).} With $\hat h_n\asymp_pn^{-1/3}$: eventually $\hat h_n<\bar\epsilon$, so by Lemma \ref{lemC1}(ii) no separated pair enters the window and the exact localization of the proof of Theorem \ref{theo3} holds along $A_n$; for exact pairs, $\hat\delta^2_{ij}/\hat h_n=O_p(n^{-1}\log n/n^{-1/3})=O_p(n^{-2/3}\log n)\to0$ uniformly by Lemma \ref{lemC1}(i), so $\widehat K_{ij}\to K(0)$ uniformly and the first-stage bound in Step 2 of that proof becomes $O_p(\log n/(\sqrt n\,\hat h_n))=O_p(n^{-1/6}\log n)=o_p(1)$. The remaining steps do not involve the bandwidth, and the conclusion of Theorem \ref{theo3} follows unchanged.

\emph{General regime (Theorems \ref{theo4} and \ref{theo5}).} Write $S(h)$ for the localized score at bandwidth $h$ and expand around $h_n$: for some $\tilde h$ between $\hat h_n$ and $h_n$,
\[S(\hat h_n)-S(h_n)=(\hat h_n-h_n)\,\partial_hS(h_n)+\tfrac12(\hat h_n-h_n)^2\,\partial_h^2S(\tilde h).\]
The derivative $\partial_hS(h)=-\binom n2^{-1}\sum_{i<j}K'(\hat\delta^2_{ij}/h)(\hat\delta^2_{ij}/h^2)\,s_{ij}$ is itself a localized pair statistic; by the argument of Steps 1--3 of the proof of Theorem \ref{theo4}, $\partial_hS(h_n)=\E[\partial_hS(h_n)]+O_p(n^{-1/2}r_n/h_n)$, and $\E[\partial_hS(h_n)]=\partial_h\{r_nB_h\}|_{h_n}$. Therefore, on $A_n$,
\[S(\hat h_n)-S(h_n)=(\hat h_n-h_n)\,\partial_h\{r_nB_h\}\big|_{h_n}+O_p\big(\epsilon_nh_n\cdot n^{-1/2}r_n/h_n\big)+O_p\big(\epsilon_n^2r_n\big).\]
The second term is $O_p(\epsilon_n\, n^{-1/2}r_n)=o_p(n^{-1/2}r_n)$. The third uses $\partial_h^2S(\tilde h)=O_p(r_n/h_n^2)$ (same localization argument, one more derivative of the kernel, Assumption \ref{ass3}(\ref{ass3i})) and is $o_p(n^{-1/2}r_n)$ because $\epsilon_n=o(n^{-1/4})$. The first term is deterministic and equals, after the sandwich, exactly the drift of the pseudo-true value: $\beta_{\hat h_n}-\beta_{h_n}=(\hat h_n-h_n)\,\partial_h\beta_h|_{h_n}+o(\cdot)$, using the differentiability of $h\mapsto\beta_h$ imposed in Assumption \ref{ass_jack} and now listed among part (c)'s hypotheses; the second-order remainder is $(\epsilon_nh_n)^2\,O(h_n^{1/\theta-2})=\epsilon_n^2\,O(h_n^{1/\theta})=o(n^{-1/2})$ by $\epsilon_n=o(n^{-1/4})$ and the second-derivative bound in the same assumption. Hence
\[\sqrt n\,\big\{ \big(\hat\beta(\hat h_n)-\beta_{\hat h_n}\big)-\big(\hat\beta(h_n)-\beta_{h_n}\big)\big\}=o_p(1),\]
at the $\Omega_h^{1/2}$ scale, and Theorem \ref{theo4}'s conclusion holds at the realized bandwidth: $V_n^{-1/2}(\hat\beta-\beta_{\hat h_n})\to_d\mathcal N(0,I_k)$. For Theorem \ref{theo5}, note that the three estimation bandwidths are the common multiples $(c_1\hat h_n,c_2\hat h_n,c_3\hat h_n)$ of the same realized $\hat h_n$, and the expansion of Assumption \ref{ass_jack} holds for every $h$ in a fixed neighborhood of $0$; conditional on any realization of $\hat h_n$ in that neighborhood, the generalized Vandermonde weights annihilate the powers $(\,\hat h_n)^{m/\theta}$, $m=1,\dots,L$, exactly as in the deterministic case, so $\sum_la_l\beta_{c_l\hat h_n}=\beta+O(\hat h_n^{(L+1)/\theta})=\beta+o_p((nr_n)^{-1/2})$. Combining with the transferred Theorem \ref{theo4} at each bandwidth (Cram\'er--Wold as in the single-bandwidth case) gives Theorem \ref{theo5}'s conclusion centered at $\beta$, unchanged. $\blacksquare$
\end{proofoflemma}

\begin{lemma}\label{lem_sigma}
Under Assumptions \ref{ass1}--\ref{ass4}, the limit $\Sigma_0=\lim_{h\downarrow0}\Sigma_h$
exists and is positive definite.
\end{lemma}

\begin{proofoflemma}
Write $a_i:=X_i'\beta+\lambda(\omega_i)$ and $u_{ij}:=(X_i-X_j)'\beta$, and set
$\Lambda:=\sup_{u\in[0,1]}|\lambda(u)|<\infty$ by
Assumption \ref{ass1}(vi). For $M>0$ let $A_M:=\{\|X_i\|\le M\}\cap\{\|X_j\|\le M\}$.
On $A_M$ we have $|u_{ij}|\le2M\|\beta\|$ and $|a_i|\vee|a_j|\le M\|\beta\|+\Lambda=:B_M$,
so that
\[
F(u_{ij})F(-u_{ij})\ \ge\ F(-2M\|\beta\|)^2=:c_M>0 ,
\]
and, conditioning on $(X_i,X_j,\omega_i,\omega_j)$ and using the independence in
Assumption \ref{ass1}(iv),
\[
\PP(y_i\ne y_j\mid X_i,X_j,\omega_i,\omega_j)
=F(a_i)\big(1-F(a_j)\big)+F(a_j)\big(1-F(a_i)\big)\ \ge\ 2F(-B_M)^2=:\eta_M>0 .
\]
Both bounds are deterministic and depend only on $M$.

Fix $v\in\R^k$ with $\|v\|=1$. The integrand defining $\Sigma_h$ is non-negative,
so restricting it to $A_M$ and applying the two bounds gives
\[
v'\Sigma_h v\ \ge\ c_M\,\eta_M\; r_n^{-1}\,\E\!\left[(v'\Delta X_{ij})^2\,
\mathbbm{1}\{A_M\}\,K\big(\delta^2_{ij}/h\big)\right].
\]
By Assumption \ref{ass4}(3)--(4) the kernel-weighted localization has the same
limit as conditioning on the shrinking codegree balls, so as $h\downarrow0$ the
right-hand side converges to
$c_M\eta_M\,\E[(v'\Delta X_{ij})^2\mathbbm{1}\{A_M\}\mid\delta_{ij}=0]$.
By Assumption \ref{ass_rank}(iii) the family $\{\|\Delta X_{ij}\|^2\}$ is
uniformly integrable over these events, so
$\E[\|\Delta X_{ij}\|^2\mathbbm{1}\{A_M^c\}\mid\delta_{ij}=0]\to0$ as
$M\to\infty$; choose $M$ so that this quantity is at most
$\lambda_{\min}(\Xi_0)/2$, which is possible uniformly over $\|v\|=1$. Then
\[
\E\big[(v'\Delta X_{ij})^2\mathbbm{1}\{A_M\}\mid\delta_{ij}=0\big]
\ \ge\ v'\Xi_0v-\tfrac12\lambda_{\min}(\Xi_0)\ \ge\ \tfrac12\lambda_{\min}(\Xi_0)>0
\]
by Assumption \ref{ass_rank}(ii), and hence
$\liminf_{h\downarrow0}v'\Sigma_hv\ge c_M\eta_M\lambda_{\min}(\Xi_0)/2>0$
uniformly in $v$. Existence of the limit follows from the same uniform
integrability, since the integrand is dominated by $\|\Delta X_{ij}\|^2$ and the
conditional expectations over the shrinking events converge by
Assumption \ref{ass_rank}(iii). $\blacksquare$
\end{proofoflemma}

\subsection*{A.2 Verification of the bandwidth rule}
\hypertarget{app_bandwidth}{}
This subsection verifies that the continuous branch of the rule
\eqref{eq_bandwidthrule} satisfies Assumptions \ref{ass3}(\ref{ass3ii}) and
\ref{ass_holder}(\ref{ass_holder_iv}), and that the finite-support branch
satisfies Assumption \ref{ass3}(\ref{ass3ii}), which is what Theorem \ref{theo3}
requires, and records how $h$ should be chosen more generally.
 Throughout,
suppose $\PP(\delta^2_{ij}\le t)\asymp t^{\beta_0}$ near zero for a small-ball
exponent $\beta_0>0$, so that $r_n(u)\asymp h^{\beta_0}$ up to constants uniform
in $u$, and write $h\asymp n^{-\varrho}$.

\medskip
\noindent\textit{The admissible window.} Assumption \ref{ass3}(\ref{ass3ii})
requires $\gamma<1-2\varrho$ and $\gamma>4\varrho\beta_0$ simultaneously, so its
admissible range of $\gamma$ is nonempty if and only if \\ $\varrho<1/(2+4\beta_0)$.
The window of Assumption \ref{ass_holder}(\ref{ass_holder_iv}), expressed through
the type-mass exponent $\alpha=1/(2\beta_0)$ that corresponds to $\delta$
comparable to the type distance, is
\[
\varrho<\tfrac{1}{4(\beta_0+1)},
\]
which is the binding restriction.

\medskip
\noindent\textit{The continuous branch.} The quantile rule with $c_n\asymp n^{-a}$
delivers $q_n\asymp n^{-a/\beta_0}$, hence an effective exponent
$\varrho=a/\beta_0$, and
is therefore compatible with both conditions if and only if
$a<\beta_0/(4(\beta_0+1))$ and, with the $1/6$ cap of
Assumption \ref{ass_holder}(\ref{ass_holder_iv}), $a<\beta_0/6$. The second
constraint binds, and the implemented value $a=1/16$ satisfies it for every
$\beta_0>3/8$, with no upper restriction on $\beta_0$. In the leading case
$\beta_0=1/2$ the effective bandwidth rate is $\varrho=a/\beta_0=1/8<1/6$.
The condition $\beta_0>3/8$ is a property of the design, not of the rule, and it
should be checked rather than assumed. A check is available from the match
counts: if $\PP(\delta^2_{ij}\le t)\asymp t^{\beta_0}$ near zero, the number of
pairs inside the window grows by a factor $2^{\beta_0}$ when the bandwidth is
doubled, so $\log_2$ of the ratio of match counts at $\hat h_n$ and $2\hat h_n$
estimates the exponent directly.

The value $\beta_0=1/2$ is not merely illustrative. Computing $r_n(u)$ directly
in the two continuous designs of Section \ref{section4} gives $r_n(u)\asymp
h^{1/2}$ uniformly in $u$, so $\beta_0=1/2$ is the exponent those designs
actually exhibit.

The derivation above assumes that the empirical quantile of $\hat\delta^2_{ij}$
inherits the small-ball behaviour of $\delta^2_{ij}$. This holds when the window
dominates the first-stage error and fails otherwise: if $\hat h_n$ is of the
order of $\hat m$, the local behaviour of the distribution of
$\hat\delta^2_{ij}$ near $\hat h_n$ reflects estimation error rather than the
design, and the exponent recovered from match counts is not $\beta_0$. A count
exponent far above the leading case should therefore be read as evidence that
the window lies in the noise, not as a favourable value of $\beta_0$.
Outside this
range of $\beta_0$ the rule remains a valid localization device, but the
distribution theory then runs through the pseudo-true centering of Theorem
\ref{theo4} evaluated at the realized bandwidth, as delivered by Lemma
\ref{lem_bandwidth}(c).

Two features of this branch are finite-sample safeguards rather than rate
choices. First, the fraction $c_n$ shrinks with $n$, since a fixed-fraction
quantile rule converges to a positive constant and violates $h\to0$. Second, the
floor $n^{-2/5}\hat q_{1/2}$ is stated relative to the median distance rather
than in absolute terms, because an absolute floor can exceed the entire range of
the codegree distances. Under the homophily design of Section \ref{section4},
$\max_{u,v}\delta^2(u,v)\approx0.046$, attained at a pair with one extreme and
one interior type, approximately $(0.49,1)$, while the two-extreme pair gives the
smaller value $\delta^2(0,1)=1/108$. An absolute floor of $n^{-2/5}$ therefore
exceeds every codegree distance for $n\lesssim2\times10^3$, the range of the
simulations, and would there assign positive weight to every pair and disable
localization. The relative floor binds asymptotically only when
$\beta_0<5a/2=5/32$, which lies outside the compatible range in any case.

\medskip
\noindent\textit{The finite-support branch.} Since $\hat m\asymp_p n^{-1}$ while
$\hat q_{1/2}$ converges to a positive constant, $\hat h_n=(3\hat
m)^{1/3}\hat q_{1/2}^{\,2/3}$ has order $n^{-1/3}$. This satisfies Assumption
\ref{ass3}(\ref{ass3ii}) with any $\gamma\in(0,1/3)$ and keeps $\sqrt n\,\hat
h_n\gg\log n$, as the proof of Theorem \ref{theo3} requires. A placement closer
to the noise scale, such as the geometric midpoint $(3\hat m\cdot\hat
q_{1/2})^{1/2}$ of order $n^{-1/2}$, would violate both requirements. Finally,
the exponent $-1/16$ in $c_n$ keeps the quantile above the uniform first-stage
rate $n^{-1/2}\log^{1/2}n$ of Lemma \ref{lemB1} whenever $\beta_0>1/8$, which
holds throughout the compatible range and in particular at $\beta_0=1/2$, the
value obtained when $\delta$ scales linearly in the latent distance.

\phantomsection
\addcontentsline{toc}{subsection}{Proof of Theorem  \ref{theo1}}
\subsection*{A.3 Proof of Theorem \ref{theo1}}

\begin{proofoftheorem}\hypertarget{app_theo1}{}
\textbf{Step 1 (conditional logit representation at the latent level).} By equation (\ref{eq1}) and Assumption \ref{ass1}(iv),
\begin{equation}\label{e1}
    \PP(y_i=1|X_i,\omega_i)=F\left(X_i'\beta+\lambda(\omega_i)\right).
\end{equation}
Throughout this proof, conditioning is on the latent quadruple $Z_{ij}:=(X_i,X_j,\omega_i,\omega_j)$; the values $\lambda(\omega_i)$ and $\lambda(\omega_j)$ are random variables, and no equality between them is imposed at this stage. By Assumption \ref{ass1}(i) and (iv), $y_i$ and $y_j$ are conditionally independent given $Z_{ij}$. Writing $\Delta X_{ij}=X_i-X_j$, $\Delta\lambda_{ij}=\lambda(\omega_i)-\lambda(\omega_j)$, and $F_k:=F(X_k'\beta+\lambda(\omega_k))$ for $k\in\{i,j\}$, the same algebra as in Chamberlain's conditional logit gives
\begin{align*}
    \PP(y_i=1|y_i+y_j=1,Z_{ij})
    &=\frac{F_i(1-F_j)}{F_i(1-F_j)+F_j(1-F_i)}\\
    &=\frac{\exp\left(X_i'\beta+\lambda(\omega_i)\right)}{\exp\left(X_i'\beta+\lambda(\omega_i)\right)+\exp\left(X_j'\beta+\lambda(\omega_j)\right)}\\
    &=F\left(\Delta X_{ij}'\beta+\Delta\lambda_{ij}\right),
\end{align*}
using $F(x)=\exp(x)/(1+\exp(x))$. Define $\pi_{ij}:=F(\Delta X_{ij}'\beta+\Delta\lambda_{ij})$ and $c_{ij}:=\PP(y_i\neq y_j|Z_{ij})\in(0,1)$.

\textbf{Step 2 (localized objective and its limit).} For $h>0$ define
\begin{align*}
\Omega_h(b):={}&-\E\Big[y_{i}\log F[\Delta X_{ij}'b]+y_j\log F[-\Delta X_{ij}'b]\,\Big|\\
&\qquad\ \rho_{ij}\le h,\ y_i+y_j=1\Big],
\end{align*}
which is well defined by Assumption \ref{ass1}(vii) and integrable by Assumption \ref{ass_rank} since $|\log F(u)|\le|u|+\log2$. By the law of iterated expectations, conditioning first on $Z_{ij}$ and using Step 1,
\[\Omega_h(b)=\E\left[w_{ij}\,\ell\left(\pi_{ij},\Delta X_{ij}'b\right)\,\Big|\,\rho_{ij}\le h\right],\]
where $\ell(\pi,u):=-\pi\log F(u)-(1-\pi)\log F(-u)$ and $w_{ij}:=c_{ij}/\E[c_{ij}|\rho_{ij}\le h]>0$ almost surely. Since $\ell(\pi,u)-\ell(\pi,u^*)=\mathrm{KL}\big(\pi\,||\,F(u)\big)-\mathrm{KL}\big(\pi\,||\,F(u^*)\big)$ for any $u,u^*$, where $\mathrm{KL}(p||q)=p\log(p/q)+(1-p)\log((1-p)/(1-q))\ge0$ is the Kullback--Leibler divergence between Bernoulli distributions,
\begin{equation}\label{eq_KLdecomp}
\Omega_h(b)-\Omega_h(\beta)=\E\Big[w_{ij}\big\{\mathrm{KL}\big(\pi_{ij}||F(\Delta X_{ij}'b)\big)-\mathrm{KL}\big(\pi_{ij}||F(\Delta X_{ij}'\beta)\big)\big\}\,\Big|\,\rho_{ij}\le h\Big].
\end{equation}
By Assumption \ref{ass2}, $\nu_h:=\sup_{u,v:||f_u-f_v||_2\le h}|\lambda(u)-\lambda(v)|\to0$ as $h\downarrow0$. On the conditioning event, the logistic structure delivers a Kullback--Leibler bound that is uniform in the covariates: for the logit cdf, $\mathrm{KL}\big(F(u+d)\,||\,F(u)\big)=\int_0^dF'(u+t)\,t\,dt\le d^2/8$ for every $u\in\R$, since $F'\le1/4$; applying this with $u=\Delta X_{ij}'\beta$ and $d=\Delta\lambda_{ij}$ gives $\mathrm{KL}\big(\pi_{ij}\,||\,F(\Delta X_{ij}'\beta)\big)\le\nu_h^2/8$ uniformly. Taking $h\downarrow0$ in (\ref{eq_KLdecomp}), where the limits exist by the convention of equation (\ref{eq_condexp}) and the interchange is justified by dominated convergence (the envelope $|\log F(u)|\le|u|+\log2$ together with the uniform integrability over shrinking conditioning events in Assumption \ref{ass_rank}(iii)), the second divergence vanishes ($\mathrm{KL}(\pi_{ij}||F(\Delta X_{ij}'\beta))\le\nu_h^2/8\to0$ uniformly) and
\[\Omega_0(b)-\Omega_0(\beta)=\E\left[w_{ij}\,\mathrm{KL}\big(F(\Delta X_{ij}'\beta)\,||\,F(\Delta X_{ij}'b)\big)\,\Big|\,\rho_{ij}=0\right]\ \ge\ 0.\]

\textbf{Step 3 (uniqueness).} Suppose $\Omega_0(b)=\Omega_0(\beta)$ for some $b\neq\beta$. The argument stays at positive $h$ throughout, so no conditional law at $h=0$ is ever invoked. Combining (\ref{eq_KLdecomp}) with the uniform bound $\mathrm{KL}(\pi_{ij}||F(\Delta X_{ij}'\beta))\le\nu_h^2/8$ and $\E[w_{ij}|\rho_{ij}\le h]=1$ gives
\begin{equation}\label{eq_KLnull}
\E\left[w_{ij}\,\mathrm{KL}\big(\pi_{ij}||F(\Delta X_{ij}'b)\big)\,\Big|\,\rho_{ij}\le h\right]\ \le\ \Omega_h(b)-\Omega_h(\beta)+\nu_h^2/8\ \longrightarrow\ 0.
\end{equation}
Fix $M>0$ and let $A_M:=\{\|\Delta X_{ij}\|\le M\}\cap\{c_{ij}\ge1/M\}$. Since $c_{ij}\le1$ implies $\E[c_{ij}|\rho_{ij}\le h]\le1$, we have $w_{ij}\ge c_{ij}\ge1/M$ on $A_M$. On $A_M$, once $\nu_h\le1$, the indices $\Delta X_{ij}'\beta+\Delta\lambda_{ij}$ and $\Delta X_{ij}'b$ lie in the compact interval $[-M(\|\beta\|\vee\|b\|)-1,\,M(\|\beta\|\vee\|b\|)+1]$, on which $F'\ge\eta_M>0$. Pinsker's inequality for Bernoulli distributions, $\mathrm{KL}(p||q)\ge2(p-q)^2$, the mean-value bound $|F(s)-F(t)|\ge\eta_M|s-t|$ on that interval, and $(x-y)^2\ge x^2/2-y^2$ then yield, on $A_M$,
\[\mathrm{KL}\big(\pi_{ij}||F(\Delta X_{ij}'b)\big)\ \ge\ 2\big(\pi_{ij}-F(\Delta X_{ij}'b)\big)^2\ \ge\ \eta_M^2\big(\Delta X_{ij}'(b-\beta)\big)^2-2\nu_h^2.\]
With $w_{ij}\ge1/M$ on $A_M$, (\ref{eq_KLnull}) therefore forces, for every fixed $M$,
\[\E\left[\big(\Delta X_{ij}'(b-\beta)\big)^2\,\mathbbm{1}_{A_M}\,\Big|\,\rho_{ij}\le h\right]\longrightarrow0.\]
On the complement of $A_M$, $(\Delta X_{ij}'(b-\beta))^2\le\|b-\beta\|^2\|\Delta X_{ij}\|^2$, and $\E[\|\Delta X\|^2\{\mathbbm{1}\{\|\Delta X\|>M\}+\mathbbm{1}\{c_{ij}<1/M\}\}|\rho_{ij}\le h]$ is small uniformly in $h$ for $M$ large: the first indicator by the uniform integrability in Assumption \ref{ass_rank}(iii); for the second, the logistic bounds $F(t)\ge e^{-|t|}/2$ and $1-F(t)\ge e^{-|t|}/2$ give $c_{ij}\ge\tfrac14\exp\{-|X_i'\beta+\lambda(\omega_i)|-|X_j'\beta+\lambda(\omega_j)|\}$, so by Markov's inequality $\PP(c_{ij}<1/M\,|\,\rho_{ij}\le h)\le\sup_{h'\le h_0}\E[|X_i'\beta+\lambda(\omega_i)|+|X_j'\beta+\lambda(\omega_j)|\,|\,\rho_{ij}\le h']/\log(M/4)\to0$ as $M\to\infty$ uniformly in $h$ by the moment bound in Assumption \ref{ass_rank}(iii), and uniform integrability converts a uniformly small probability into a uniformly small expectation. Combining the three displays, $\lim_{h\downarrow0}\E[(\Delta X_{ij}'(b-\beta))^2|\rho_{ij}\le h]=(b-\beta)'\Xi_0(b-\beta)=0$, contradicting the positive definiteness of $\Xi_0$ in Assumption \ref{ass_rank}(ii). Note that Jensen's inequality applied to (\ref{eq_KLdecomp}) would deliver only $\Omega_0(\beta)\le\Omega_0(b)$; it is the strict positivity of the Kullback--Leibler divergence combined with Assumption \ref{ass_rank} that rules out other minimizers. The same argument applies when the conditioning is on shrinking codegree balls $\{\delta^2_{ij}\le h\}$ instead of link-function balls, with Assumption \ref{ass_rank}(iii) invoked for the codegree-ball events, for which it is also imposed: by Lemma 1 of \cite{auerbach2022identification}, for every $\varepsilon>0$ there is $\bar{h}>0$ with $\rho_{ij}\mathbbm{1}\{\delta_{ij}\le\bar h\}\le\varepsilon$, so $\Delta\lambda_{ij}\to0$ uniformly on codegree balls as well, and the limiting objective $Q_0$ of Section \ref{section_Consistency} has the same unique minimizer $\beta$.

\textbf{Step 4 (identification of $\lambda(\omega_i)$).} Fix a network type $\varphi$ in the support of $f_\omega$ and write $B_h(\varphi):=\{u\in[0,1]:||f_u-\varphi||_2\le h\}$; fix a covariate value $x$ in the support of the conditional distribution of $X_j$ given $\omega_j\in B_h(\varphi)$ for all small $h>0$ (almost every $x$ under the conditional distribution of $X_j$ given $f_{\omega_j}=\varphi$ has this property). For $h>0$, Assumption \ref{ass1}(vii) gives the event $\{\omega_j\in B_h(\varphi)\}$ positive probability, so $\E[y_j\,|\,X_j=x,\ \omega_j\in B_h(\varphi)]$ is well defined, and by (\ref{e1}), which conditions on $(X_j,\omega_j)$ jointly, it equals $\E[F(x'\beta+\lambda(\omega_j))\,|\,X_j=x,\ \omega_j\in B_h(\varphi)]$; no independence between $X_j$ and $\omega_j$ is used. Evaluating at $\varphi=f_{\omega_i}$, Assumption \ref{ass2} bounds $|\lambda(\omega_j)-\lambda(\omega_i)|\le\nu_h$ on the conditioning event, and $F'\le1/4$ places the display within $\nu_h/4$ of $F(x'\beta+\lambda(\omega_i))$. Letting $h\downarrow0$ along the convention stated before Assumption \ref{ass2}, $\PP(y_j=1\,|\,X_j=x,\ f_{\omega_j}=f_{\omega_i})=F(x'\beta+\lambda(\omega_i))$. Since $F$ is strictly increasing and $\beta$ is unique by Step 3, inverting at any such $x$ yields (\ref{eq5}), the derivation shows the right-hand side is the same for every such $x$, and $\lambda(\omega_i)$ is thereby determined uniquely as a functional of the network-type class of agent $i$; under Assumption \ref{ass2} the function $\lambda$ is constant on these classes, so this is identification of the value $\lambda(\omega_i)$ itself. $\blacksquare$
\end{proofoftheorem}

\phantomsection
\addcontentsline{toc}{subsection}{Proof of Theorem  \ref{theo2}}
\subsection*{A.4 Proof of Theorem \ref{theo2}}

\begin{proofoftheorem}\hypertarget{app_theo2}{}
Write $K_{ij}:=K(\delta^2_{ij}/h)$ and $\widehat{K}_{ij}:=K(\hat{\delta}^2_{ij}/h)$, and let
\[A_n(\delta,b):=\binom n2^{-1}\sum_{i<j}K_{ij}\,m(v_i,v_j,b), \quad r_n=\E[K_{ij}], \quad \hat{r}_n=\binom n2^{-1}\sum_{i<j}\widehat{K}_{ij},\]
so that $Q_n(\hat{\delta},b)=A_n(\hat{\delta},b)/\hat{r}_n$. The limiting objective function is $Q_0(b)=\lim_{h\downarrow0}Q_h(b)$ with $Q_h(b)=r_n^{-1}\E[K_{ij}m(v_i,v_j,b)]$; the limit exists by Assumption \ref{ass4}, and $\beta$ is its unique minimizer by Theorem \ref{theo1} (whose Step 3 applies to codegree-ball localizations, as noted there). Throughout, fix $b$ in the parameter space; all bounds below are at fixed $b$.

\textbf{Step 1 (convexity reduction).} For each pair, $b\mapsto m(v_i,v_j,b)$ is convex, because $u\mapsto-\log F(u)$ is convex ($(-\log F)''=F(1-F)>0$) and $m$ is a non-negative combination of convex functions composed with linear maps. Hence $Q_n(\hat{\delta},\cdot)$ is a random convex function on the convex parameter space. By the convexity lemma for extremum estimators \citep[Theorem 2.7]{newey1994large}, pointwise convergence $Q_n(\hat{\delta},b)\to_pQ_0(b)$ for each $b$, together with uniqueness of the minimizer of $Q_0$, implies $\hat{\beta}\to_p\beta$. It therefore suffices to establish pointwise convergence; no uniform convergence or stochastic equicontinuity argument is required.

\textbf{Step 2 (expected localized objective).} Let $g(\omega_i,\omega_j,b)$ denote $\E[m(v_i,v_j,b)\,|\,\omega_i,\omega_j]$, which exists by the square integrability in Assumption \ref{ass4}(1) and is bounded by Assumption \ref{ass4}(4). Conditioning on $(\omega_i,\omega_j)$,
\[r_n^{-1}\E\left[K_{ij}\,m(v_i,v_j,b)\right]=r_n^{-1}\E\left[K_{ij}\,g(\omega_i,\omega_j,b)\right]=Q_h(b)\longrightarrow Q_0(b).\]
The convergence deserves a bridge, because $Q_h$ is a kernel-weighted average while $Q_0$ is defined as a hard-ball conditional limit. Because $K$ is nonincreasing (Assumption \ref{ass3}(\ref{ass3i})), the layer-cake representation $K(x)=\int_0^{K(0)}\mathbbm{1}\{x\le\kappa(u)\}\,du$ with $\kappa(u):=\sup\{x\ge0:K(x)\ge u\}\le1$ gives
\[Q_h(b)=\frac{\int_0^{K(0)}\E\big[m(v_i,v_j,b)\,\big|\,\delta^2_{ij}\le h\kappa(u)\big]\,\PP\big(\delta^2_{ij}\le h\kappa(u)\big)\,du}{\int_0^{K(0)}\PP\big(\delta^2_{ij}\le h\kappa(u)\big)\,du},\]
a weighted average of hard-ball conditional means over radii $h\kappa(u)\le h$; since each of these converges to $Q_0(b)$ as $h\downarrow0$ (the limit exists by Assumption \ref{ass4}(3)) and the weights integrate to one, $Q_h(b)\to Q_0(b)$. Note that $K_{ij}=K(\delta^2_{ij}/h)$ is a deterministic function of $(\omega_i,\omega_j)$, so no substitution of its argument is involved: the kernel localizes in the codegree distance, and the connection to the differences $\lambda(\omega_i)-\lambda(\omega_j)$ is made only through Lemma 1 of \cite{auerbach2022identification} and Assumption \ref{ass2}, which guarantee that $\Delta\lambda_{ij}\to0$ uniformly on the kernel support $\{\delta^2_{ij}\le h\}$ as $h\downarrow0$.

\textbf{Step 3 (variance of the localized objective).} $A_n(\delta,b)$ is a second-order U-statistic. By the standard variance decomposition, with $\zeta_2:=\Var(K_{ij}m_{ij})\le\E[K_{ij}^2m_{ij}^2]\le\|K\|_\infty M\,r_n$ and $\zeta_1:=\mathrm{Cov}(K_{ij}m_{ij},K_{ik}m_{ik})\le M\,\E[K_{ij}K_{ik}]\le M\|K\|_\infty\E[r_n(\omega_i)]\cdot\sup_ur_n(u)/\E[r_n(\omega_i)]\cdot r_n\le CM\sup_ur_n(u)\,r_n$, where $M:=\sup_{\omega,\omega'}\E[m^2|\omega,\omega']<\infty$ by Assumption \ref{ass4}(\ref{ass4iv}),
\[\Var\left(\frac{A_n(\delta,b)}{r_n}\right)\le \frac{C}{r_n^2}\left[\frac{\zeta_2}{n^2}+\frac{\zeta_1}{n}\right]\le C'\left[\frac{1}{n^2r_n}+\frac{\sup_ur_n(u)}{n\,r_n}\right]\longrightarrow0,\]
since $\sup_ur_n(u)\le\|K\|_\infty$ and $nr_n\ge n\inf_ur_n(u)\ge n^{1-\gamma/4}\cdot(n^{\gamma/4}\inf_ur_n(u))\to\infty$ by Assumption \ref{ass3}(\ref{ass3ii}). By Chebyshev's inequality, $A_n(\delta,b)/r_n=Q_h(b)+o_p(1)$. The same computation with $m\equiv1$ gives $\binom n2^{-1}\sum_{i<j}K_{ij}/r_n\to_p1$.

\textbf{Step 4 (first-stage error).} By the mean value theorem, with $a^*_{ij}$ between $\hat{\delta}^2_{ij}/h$ and $\delta^2_{ij}/h$,
\begin{align*}
\left|A_n(\hat{\delta},b)-A_n(\delta,b)\right|\le\frac{\|K'\|_\infty}{h}\,\max_{i\neq j}\left|\hat{\delta}^2_{ij}-\delta_{ij}^2\right|\;&\binom n2^{-1}\sum_{i<j}\left|m(v_i,v_j,b)\right|\\
&\times\mathbbm{1}\left\{\min(\hat\delta^2_{ij},\delta^2_{ij})< h\right\},
\end{align*}
where the maximum over pairs, rather than a single generic pair, appears because the mean-value point $a^*_{ij}$ differs across pairs, and the indicator reflects the support of $K'$. By Lemma \ref{lemB1}, $\max_{i\neq j}|\hat{\delta}^2_{ij}-\delta^2_{ij}|=O_p(n^{-1/2}\log^{1/2}n)$, so on an event of probability approaching one the indicator is bounded by $\mathbbm{1}\{\delta^2_{ij}\le2h\}$, and $\E[|m_{ij}|\mathbbm{1}\{\delta^2_{ij}\le2h\}]\le M^{1/2}\PP(\delta^2_{ij}\le2h)$. Under the kernel-mass regularity in Assumption \ref{ass4} (i.e., $\limsup_n\PP(\delta^2_{ij}\le2h)/r_n<\infty$, imposed in Assumption \ref{ass4}(4) and automatic in the finite-support case),
\[\frac{\left|A_n(\hat{\delta},b)-A_n(\delta,b)\right|}{r_n}=O_p\left(\frac{\log^{1/2}n}{n^{1/2}\,h}\right)=o_p(1),\]
where the last equality holds because $n^{1-\gamma}h^2\to\infty$ implies $n^{1/2}h\ge n^{\gamma/2}\to\infty$ faster than $\log^{1/2}n$. Note that under the bandwidth conditions of Assumption \ref{ass3}(\ref{ass3ii}), an unnormalized bound of the form $O_p(h^{-2}n^{-1/2})$ would not be $o_p(1)$ in general; the normalization by $r_n$, the uniform rate of Lemma \ref{lemB1}, and the localization of the sum are all needed. The same bound with $m\equiv1$ gives $\hat{r}_n/r_n\to_p1$.

\textbf{Step 5 (conclusion for $\hat\beta$).} Combining Steps 2--4,
\[Q_n(\hat{\delta},b)=\frac{A_n(\hat{\delta},b)}{\hat r_n}=\frac{A_n(\delta,b)/r_n+o_p(1)}{\hat r_n/r_n}=Q_h(b)+o_p(1)\longrightarrow_pQ_0(b)\]
for every fixed $b$, and Step 1 delivers $\hat{\beta}\to_p\beta$.

\textbf{Step 6 (consistency of $\hat{\lambda}(\omega_i)$).} Fix $i$; each part of Assumption \ref{ass_lambda} is used where named. \emph{(a) Uniform consistency of $\widehat\PP$ on the trimmed set.} Within each cell of $X^d$ (Assumption \ref{ass_lambda}(i)) and each network ball, $\widehat\PP$ is a Nadaraya--Watson estimator whose effective sample size is $\tilde n:=n\,r_n\,h_X^{k_c}\to\infty$ (Assumption \ref{ass_lambda}(ii)); by the density bounds and the uniform continuity of $x^c\mapsto\PP(y=1|X=x,f_\omega)$ on $\X_0$ (Assumption \ref{ass_lambda}(iii)), a standard kernel-regression argument gives
\[\sup_{x\in\X_0}\big|\widehat\PP(y=1|x,f_{\omega_j})-\PP(y=1|x,f_{\omega_j})\big|=O_p\big(\tilde n^{-1/2}\log^{1/2}\tilde n\big)+o(1),\]
the $o(1)$ collecting the $h_X$-smoothing and $\nu_h$-matching biases, both vanishing by Assumptions \ref{ass_lambda}(ii) and \ref{ass2}. \emph{(b) The clipping is eventually inactive and $F^{-1}$ is Lipschitz where evaluated.} Conditional on $\omega_i$, matched agents $j$ have $\lambda(\omega_j)\to\lambda(\omega_i)$, so on the compact $\X_0$ the target probabilities lie in $[c,1-c]$ for a constant $c>0$ uniform over $i$, by the boundedness of $\lambda$ in Assumption \ref{ass1}(vi) and compactness of $\X_0$; since $\tau_n\to0$ and the estimation error in (a) is $o_p(\tau_n)$ by the trimming rate $\tau_n\gg(\log\tilde n/\tilde n)^{1/2}$ (Assumption \ref{ass_lambda}(iv)), the clip binds with probability tending to zero, and $F^{-1}$ is Lipschitz on $[c_i/2,1-c_i/2]$. Hence, uniformly over matched $j$ with $X_j\in\X_0$ (the restriction imposed by Assumption \ref{ass_lambda}(iv)),
\begin{multline*}F^{-1}\big(\widehat{\PP}(y_j=1|X_j,f_{\omega_j})\big)-X_j'\hat{\beta}\\=\lambda(\omega_j)+O_p\big(\tilde n^{-1/2}\log^{1/2}\tilde n\big)+O_p\big(\|\hat\beta-\beta\|\big)+o_p(1)=\lambda(\omega_j)+o_p(1),\end{multline*}
using $\hat\beta\to_p\beta$ from the first part of the theorem and boundedness of $\X_0$. \emph{(c) Matched heterogeneity converges.} For pairs receiving positive weight, $\hat{\delta}^2_{ij}<h$, so $\delta^2_{ij}<2h$ with probability approaching one by Lemma \ref{lemB1}; Lemma 1 of \cite{auerbach2022identification} and Assumption \ref{ass2} then give $\max_{j:\hat\delta^2_{ij}<h}|\lambda(\omega_j)-\lambda(\omega_i)|\to_p0$. Since $\hat{\lambda}(\omega_i)$ is a weighted average of the summands in (b) with weights summing to one, $\hat{\lambda}(\omega_i)\to_p\lambda(\omega_i)$. $\blacksquare$
\end{proofoftheorem}

\phantomsection
\addcontentsline{toc}{subsection}{Proof of Theorem  \ref{theo3}}
\subsection*{A.5 Proof of Theorem \ref{theo3} (finite support)}

\begin{proofoftheorem}\hypertarget{app_theo3}{}
Recall $\nabla_bm(v_i,v_j,b)=-s(v_i,v_j,b)$ and
\[\nabla^2_bm(v_i,v_j,b)=\mathbbm{1}(y_i\ne y_j)F[u_{ij}(b)]F[u_{ji}(b)]\Delta X_{ij}\Delta X_{ij}',\qquad u_{ij}(b)=\Delta X_{ij}'b,\]
obtained by direct differentiation of the logit log-likelihood. Since $\hat\beta$ minimizes $Q_n(\hat\delta,b)$ and lies in the interior of the parameter space with probability approaching one (Theorem \ref{theo2}), the first-order condition and a mean value expansion give
\begin{equation}\label{eq_expansion}
\sqrt{n}\left(\hat\beta-\beta\right)=\widehat{H}_n(\tilde\beta)^{-1}\,\hat{r}_n^{-1}\,\sqrt{n}\,S_n, \qquad S_n:=\binom n2^{-1}\sum_{i<j}\widehat{K}_{ij}\,s(v_i,v_j,\beta),
\end{equation}
where $\widehat{H}_n(b):=\hat{r}_n^{-1}\binom n2^{-1}\sum_{i<j}\widehat{K}_{ij}\nabla^2_bm(v_i,v_j,b)$ and $\tilde\beta$ lies between $\hat\beta$ and $\beta$ componentwise.

\textbf{Step 1 (exact localization).} Under Assumption \ref{ass5}, $\delta^2_{ij}\in\{0\}\cup[\bar{\epsilon},\infty)$ almost surely. For $h<\bar{\epsilon}$ (which holds eventually), $K_{ij}=K(0)\mathbbm{1}\{\delta^2_{ij}=0\}$ exactly, so $r_n=K(0)\pi_0>0$ is constant in $h$.

\textbf{Step 2 (first-stage error is negligible at the $\sqrt{n}$ scale).} Decompose
\[S_n=\binom n2^{-1}\sum_{i<j}K_{ij}s_{ij}+\binom n2^{-1}\sum_{i<j}\left(\widehat{K}_{ij}-K_{ij}\right)s_{ij}=:U_n+T_n.\]
Split $T_n$ by the support of $\delta^2_{ij}$. On pairs with $\delta^2_{ij}\ge\bar{\epsilon}$, $K_{ij}=0$, and by Lemma \ref{lemC1}(ii), $\widehat{K}_{ij}=0$ for all such pairs with probability approaching one; their contribution to $\sqrt n\,T_n$ is $o_p(1)$. On pairs with $\delta^2_{ij}=0$, a mean value expansion and Lemma \ref{lemC1}(i) give
\[\big|\widehat{K}_{ij}-K(0)\big|\le\|K'\|_\infty\frac{\hat{\delta}^2_{ij}}{h}\le\|K'\|_\infty\,\frac{O_p(n^{-1}\log n)}{h} \quad\text{uniformly over such pairs},\]
so the corresponding portion of $\sqrt{n}|T_n|$ is bounded by $\sqrt{n}\cdot O_p(n^{-1}\log n/h)\cdot\binom n2^{-1}\sum_{i<j}\|s_{ij}\|=O_p\left(\log n/(\sqrt{n}h)\right)=o_p(1)$, since $\binom n2^{-1}\sum\|s_{ij}\|=O_p(1)$ by $\E\|X_i\|^2<\infty$ and $\sqrt{n}h\gg n^{\gamma/2}$ by Assumption \ref{ass3}(\ref{ass3ii}). Hence $\sqrt{n}\,S_n=\sqrt{n}\,U_n+o_p(1)$. The same argument applied with $s_{ij}$ replaced by $1$ and by $\nabla^2_bm$ gives $\hat{r}_n\to_pK(0)\pi_0$ and $\widehat{H}_n(\tilde\beta)\to_p\Sigma_0$, the latter using also $\tilde\beta\to_p\beta$ (Theorem \ref{theo2}), continuity of $\nabla^2_bm$ in $b$, and the dominance $\|\nabla^2_bm\|\le\|\Delta X_{ij}\|^2$; $\Sigma_0$ is positive definite by Assumption \ref{ass_rank}, since $F(u)F(-u)$ is bounded below on compact index sets and the conditional distribution of $\Delta X_{ij}$ given $\delta_{ij}=0$ is nondegenerate.

\textbf{Step 3 (the population score is exactly zero).} We have $U_n=K(0)\tbinom n2^{-1}\sum_{i<j}\phi_{ij}$, where $\phi_{ij}:=\mathbbm{1}\{\delta_{ij}=0\}\,s_{ij}$. This is a genuine second-order U-statistic: $\phi_{ij}$ depends only on $(v_i,v_j,\omega_i,\omega_j)$ because $\delta_{ij}=\delta(\omega_i,\omega_j)$ is a deterministic function of the latent types, the estimated distances having been removed in Step 2. Its mean is exactly zero:
conditional on $Z_{ij}=(X_i,X_j,\omega_i,\omega_j)$ with $\delta_{ij}=0$, Lemma 1 of \cite{auerbach2022identification} gives $\rho_{ij}=0$, Assumption \ref{ass2} gives $\Delta\lambda_{ij}=0$, and by Step 1 of the proof of Theorem \ref{theo1},
\[\E\left[s_{ij}|Z_{ij},\delta_{ij}=0\right]=c_{ij}\left\{F\left(\Delta X_{ij}'\beta+\Delta\lambda_{ij}\right)-F\left(\Delta X_{ij}'\beta\right)\right\}\Delta X_{ij}=0.\]
This is the step that fails when types vary continuously, since $\Delta\lambda_{ij}\neq0$ on inexact matches, and that generates the bias term $B_h$ of Theorem \ref{theo4}; under Assumption \ref{ass5} no such bias arises, and no condition of the form $\sqrt{n}\,\E[U_n]\to0$ needs to be assumed: it holds exactly, with $\E[U_n]=0$.

\textbf{Step 4 (Hoeffding decomposition and CLT).} Since $\E[\phi_{ij}]=0$, the Hoeffding decomposition gives $U_n/K(0)=\frac2n\sum_{i=1}^nq(Z_i)+R_n$ with $q(Z_i)=\E[\phi_{ij}|Z_i]=\pi(\omega_i)\,t(y_i,X_i,\omega_i)$, $Z_i=(y_i,X_i,\omega_i)$, and $R_n$ the degenerate remainder, which satisfies the standard bound $\Var(R_n)\le C\,\E\|\phi_{ij}\|^2/n^2\le C\,\E\|\Delta X_{ij}\|^2/n^2$, so $\sqrt{n}R_n\to_p0$. The vectors $q(Z_i)$ are i.i.d., mean zero, with finite variance $V_0=\Var(\pi(\omega_i)t_i/\pi_0)$, so by the multivariate CLT and Slutsky's theorem,
\[\sqrt{n}\left(\hat\beta-\beta\right)=\Sigma_0^{-1}\,\frac{K(0)}{K(0)\pi_0}\cdot\frac{2}{\sqrt n}\sum_{i=1}^n q(Z_i)+o_p(1)\xrightarrow{d}\mathcal{N}\left(0,4\Sigma_0^{-1}V_0\Sigma_0^{-1}\right),\]
with $V_0=\Var\left(\pi(\omega_i)t_i/\pi_0\right)$ as stated. $\blacksquare$
\end{proofoftheorem}

\phantomsection
\addcontentsline{toc}{subsection}{Proof of Theorem  \ref{theo4}}
\subsection*{A.6 Proof of Theorem \ref{theo4} (general network types)}

\begin{proofoftheorem}\hypertarget{app_theo4}{}
The expansion (\ref{eq_expansion}) continues to hold with $\beta$ replaced by the pseudo-true value $\beta_h$ of (\ref{eq_pseudotrue}): $\hat\beta$ is interior with probability approaching one, $\beta_h\to\beta$ is interior for all small $h$, and the mean value expansion around $\beta_h$ gives $\sqrt n(\hat\beta-\beta_h)=\widehat H_n(\tilde\beta)^{-1}\hat r_n^{-1}\sqrt n\,S_n$ with $S_n:=\binom n2^{-1}\sum_{i<j}\widehat K_{ij}\,s(v_i,v_j,\beta_h)$ and $\tilde\beta$ between $\hat\beta$ and $\beta_h$. Throughout this proof $s_{ij}:=s(v_i,v_j,\beta_h)$, matching the definition of the influence function and of $\Omega_h$ in Section \ref{section332}; since $\|\beta_h-\beta\|=O(h^\kappa)\to0$ and the score is Lipschitz in $b$ with square-integrable envelope, every moment bound stated in the assumptions transfers to $\beta_h$ with adjusted constants. (Evaluation at $\beta$ instead would change the first-stage channels $\psi_2,\psi_3$ by $O(h^{\kappa-1})$, the same order as the channels themselves; this is why $\Omega_h$ is defined at $\beta_h$.) The Hessian analysis is as in Step 2 of the proof of Theorem \ref{theo3}, now using Lemma \ref{lemB1} and the bandwidth window of Assumption \ref{ass_holder}(\ref{ass_holder_iv}), and yields $\widehat{H}_n(\tilde\beta)\to_p\Sigma_h$ (with $\Sigma_h\to\Sigma_0$) and $\hat r_n/r_n\to_p1$. The substance of the proof is the analysis of the score $S_n$, which parallels the proof of Proposition C2 in \cite{auerbach2021inference}, with the linear pairwise difference $\Delta_{ij}=(x_i-x_j)'(u_i-u_j)$ of that paper replaced throughout by the logit score $s_{ij}$; I reproduce the structure and indicate where the argument is identical.

\textbf{Step 1 (second-order Taylor expansion of the kernel).} By Assumption \ref{ass3}(\ref{ass3i}), with $\iota_{ij}$ a mean value between $\hat{\delta}^2_{ij}$ and $\delta^2_{ij}$,
\[\widehat{K}_{ij}=K_{ij}+K'\big(\delta^2_{ij}/h\big)\,\frac{\hat{\delta}^2_{ij}-\delta^2_{ij}}{h}+\frac12K''\big(\iota_{ij}/h\big)\left(\frac{\hat{\delta}^2_{ij}-\delta^2_{ij}}{h}\right)^2.\]
The quadratic term is $o_p(n^{-1/2}r_n)$ after averaging, by a direct bound: its absolute contribution to $S_n$ is at most $\tfrac12\|K''\|_\infty\max_{i\ne j}\big(\hat\delta^2_{ij}-\delta^2_{ij}\big)^2h^{-2}$ times $\binom n2^{-1}\sum_{i<j}\|s_{ij}\|\mathbbm{1}\{\min(\hat\delta^2_{ij},\delta^2_{ij})\le h\}$. By Lemma \ref{lemB1} the maximum is $O_p(n^{-1}\log n)$; by the kernel-mass condition of Assumption \ref{ass4}(4) and Cauchy--Schwarz the localized sum is $O_p(r_n)$. The contribution to $\sqrt n\,S_n/r_n$ is therefore $O_p\big(\log n/(\sqrt n\,h^2)\big)=o_p(1)$, since $\varrho<\alpha/(2+4\alpha)\le1/4$ under Assumption \ref{ass_holder}(\ref{ass_holder_iv}) gives $\sqrt n\,h^2\gg n^{1/2-2\varrho}\cdot n^{-\epsilon}\gg\log n$. This step uses only the network objects and the moment bound $\E\|s_{ij}\|^2<\infty$.

\textbf{Step 2 (U-statistic representation of the linear term).} Write $\hat{\delta}^2_{ij}$ as an average over triples of the products $D_{ts_1}D_{ts_2}(D_{is_1}-D_{js_1})(D_{is_2}-D_{js_2})$ and replace these products by their conditional means $F_{ijts_1s_2}$, defined in equation (\ref{eq_influence}), given the latent types. The substitution error has conditional mean zero given the types (the $\eta$'s are independent across dyads). Its variance is controlled by a dyad-sharing count: write $e_{ijts_1s_2}:=D_{ts_1}D_{ts_2}(D_{is_1}-D_{js_1})(D_{is_2}-D_{js_2})-F_{ijts_1s_2}$, so that the substitution error is $(M_nr_n)^{-1}$ times a sum of terms $s_{ij}\,h^{-1}K'(\delta^2_{ij}/h)\,e_{ijts_1s_2}$ over distinct tuples, in which the factor $s_{ij}h^{-1}K'(\delta^2_{ij}/h)$ is measurable with respect to the types and outcomes. Conditional on these, $\E[e\,e'\,|\,\boldsymbol\omega]=0$ unless the two tuples share at least one dyad among $\{t,s_1\},\{t,s_2\},\{i,s_1\},\{j,s_1\},\{i,s_2\},\{j,s_2\}$, because distinct dyads are independent given types. Ordered pairs of tuples sharing at least one dyad and having distinct scored pairs number $O(n^8)$ out of $O(n^{10})$, with each covariance term bounded by $h^{-2}\|K'\|_\infty^2\,\E[|s_{ij}||s_{i'j'}|\mathbbm{1}\{\delta^2_{ij}\le h\}\mathbbm{1}\{\delta^2_{i'j'}\le h\}]=O(r_n^2/h^2)$, using the score moments of Assumption \ref{ass4}(2), the window mass of Assumption \ref{ass4}(\ref{ass4iv}), and the balanced matching intensity of Assumption \ref{ass4}(3) for windows sharing an agent; tuple pairs with the same scored pair or with more than one shared dyad are fewer by powers of $n$ and are dominated because $nr_n\to\infty$. Hence the variance of the substitution error is $O\big((M_nr_n)^{-2}\,n^8\,r_n^2h^{-2}\big)=O\big(1/(n^2h^2)\big)$, so the error is $O_p(1/(nh))=o_p(n^{-1/2})$ on the $S_n/r_n$ scale under Assumption \ref{ass3}(\ref{ass3ii}), matching the corresponding count in the proof of Proposition C2 of \cite{auerbach2021inference} with the linear pairwise difference replaced by the bounded-moment logit score. The second-order kernel remainder, which the first-order expansion leaves behind, is of the same $1/(nh)$ order once the first-stage error is measured pair by pair rather than uniformly. By a second-order Taylor expansion of $K$ (Assumption \ref{ass3}(\ref{ass3i}) provides the bounded $K''$), the residual $\widehat K_{ij}-K_{ij}-h^{-1}K'(\delta^2_{ij}/h)(\hat\delta^2_{ij}-\delta^2_{ij})$ is bounded by $\tfrac12\|K''\|_\infty h^{-2}\Delta_{ij}^2\,\mathbbm{1}\{\min(\hat\delta^2_{ij},\delta^2_{ij})<h\}$ with $\Delta_{ij}:=\hat\delta^2_{ij}-\delta^2_{ij}$, and the per-pair mean square of $\Delta_{ij}$ scales with the distance being estimated:
\begin{equation}\label{eq_pairMSE}
\E\big[\Delta_{ij}^2\,\big|\,\omega_i,\omega_j\big]\ \le\ C\,\frac{\delta^2_{ij}\vee n^{-1}}{n}.
\end{equation}
To see (\ref{eq_pairMSE}), write $\Delta_{ij}=\frac1n\sum_t\{2A_t(\hat a_t-A_t)+(\hat a_t-A_t)^2\}$, with the two coincident indices $t\in\{i,j\}$ contributing at most $2/n$ deterministically as in Lemma \ref{lemB1}. The squared-error part has conditional mean $O(1/n)$ (the noise-floor scale) and smaller fluctuations. The cross part is $\frac2n\sum_tA_t\xi_t$ with $\xi_t:=\hat a_t-A_t$, $\E[\xi_t^2\,|\,\boldsymbol\omega\text{-marginal}]\le C/n$, and $|\mathrm{Cov}(\xi_t,\xi_{t'})|\le C/n$ for $t\ne t'$ (the covariance runs through the shared dyads $(i,s),(j,s)$ and the shared type draws $\omega_s$); hence its variance is at most $\frac4{n^2}\big[\sum_tA_t^2\cdot\tfrac Cn+\big(\sum_t|A_t|\big)^2\tfrac Cn\big]\le\frac{4C}{n^3}\big[n\delta^2_{ij}+n\cdot n\delta^2_{ij}\big]\le\frac{8C\,\delta^2_{ij}}{n}$, using $\frac1n\sum_tA_t^2\to\delta^2_{ij}$ and $(\sum_t|A_t|)^2\le n\sum_tA_t^2$. On the event $\min(\hat\delta^2_{ij},\delta^2_{ij})<h$, which is contained in $\{\delta^2_{ij}\le2h\}$ with probability approaching one by Lemma \ref{lemB1}, the bound (\ref{eq_pairMSE}) is $\le C(2h\vee n^{-1})/n$; combining with $\E[|s_{ij}|\mathbbm{1}\{\delta^2_{ij}\le2h\}]\le Cr_n$ (Assumption \ref{ass4}(\ref{ass4iv})) and the Cauchy--Schwarz inequality, the second-order remainder contributes $O_p\big(h^{-2}\cdot h/n\big)=O_p\big(1/(nh)\big)$ to $S_n/r_n$, which is $o_p(n^{-1/2})$ under Assumption \ref{ass3}(\ref{ass3ii}) exactly as the two terms above. (The cruder route through the uniform rate of Lemma \ref{lemB1}, $\max_{ij}\Delta_{ij}^2=O_p(n^{-1}\log n)$, gives only $O_p(\log n/(nh^2))$ and would appear to require $nh^4/\log^2n\to\infty$; the per-pair bound (\ref{eq_pairMSE}) shows that requirement is an artifact of the uniform bound. The window $\varrho<1/4$ of Assumption \ref{ass_holder}(\ref{ass_holder_iv}) is consumed elsewhere: by the Hessian and studentization rates at the end of this proof, and $\varrho<1/6$ by the variance estimation of Appendix \hyperlink{app_var}{A.8}.) Note that $\E[F_{ijts_1s_2}|\omega_i,\omega_j]=\delta^2_{ij}$ by definition of the codegree distance. Consequently,
\begin{align*}
\frac{S_n}{r_n}&=U_n'+o_p\left(n^{-1/2}\right),\\
U_n'&:=\frac{1}{M_n\, r_n}\sum\Delta\!\!\!\!\sum_{i<j,\,t,s_1,s_2}s_{ij}\Big[K_{ij}+\tfrac1hK'\big(\delta^2_{ij}/h\big)\big(F_{ijts_1s_2}-\delta^2_{ij}\big)\Big],
\end{align*}
a fifth-order U-statistic over distinct indices, normalized by the number of index tuples in the sum, $M_n:=\tbinom n2(n-2)(n-3)(n-4)$ whose kernel depends on $n$ through $h$. Restricting the sum to distinct indices excludes every repeated-index class, and each class must be accounted for, not only the collision $s_1=s_2$. The excluded tuples fall into four classes: (a) $s_1=s_2$; (b) $t\in\{i,j\}$; (c) $s_k\in\{i,j\}$ for $k\in\{1,2\}$; and (d) $s_k=t$. Each class fixes at least one index coincidence, so for a given scored pair $(i,j)$ each class contains $O(n^2)$ of the $n^3$ triples $(t,s_1,s_2)$; class (d) in fact vanishes identically, since $s_k=t$ produces the diagonal entry $D_{tt}=0$. Each surviving term is bounded ($|D|\le1$, $|F_{ijts_1s_2}|\le1$) and carries the factor $h^{-1}K'(\delta^2_{ij}/h)$, bounded by $\|K'\|_\infty/h$ and supported on $\{\delta^2_{ij}\le h\}$. The conditional means differ by class and none is assumed to vanish: class (a) carries the noise-floor scale of equation (\ref{eq_noisefloor}); classes (b)--(c) carry the coincident-index means made explicit in Lemma \ref{lemB1} (for example, at $t=i$ the product $D_{is_1}D_{is_2}$ has conditional mean $f_{is_1}f_{is_2}$ only when $s_1\ne s_2$, and $D_{is}^2=D_{is}$ otherwise). Summing: the excluded classes contribute to $|S_n|$ at most $\binom n2^{-1}\sum_{i<j}|s_{ij}|\,\tfrac{\|K'\|_\infty}{h}\,\mathbbm{1}\{\delta^2_{ij}\le h\}\cdot n^{-3}\cdot O(n^2)$, and since $\E\big[\binom n2^{-1}\sum|s_{ij}|\mathbbm{1}\{\delta^2_{ij}\le h\}\big]\le Cr_n$ by Assumptions \ref{ass4}(2) and \ref{ass4}(\ref{ass4iv}), the total is $O_p\big(r_n/(nh)\big)$, that is, $O_p(1/(nh))$ on the $S_n/r_n$ scale, which is $o_p(n^{-1/2})$ because $\sqrt n/(nh)=1/(\sqrt n\,h)\to0$ under Assumption \ref{ass3}(\ref{ass3ii}). This is the same accounting that the many-network result of Proposition \ref{prop_villages} must redo at the pooled scale, where the factor $\sqrt G$ makes it binding.

\textbf{Step 3 (the population score vanishes exactly at $\beta_h$).} By the conditional mean-zero property of $F_{ijts_1s_2}-\delta^2_{ij}$ given $(\omega_i,\omega_j)$ and its independence from $(v_i,v_j)$ given the types, the $K'$ component has expectation zero, so
\[\E[U_n']=r_n^{-1}\E\left[s(v_i,v_j,\beta_h)K_{ij}\right]=0\]
\emph{exactly}, by the first-order condition of the minimization (\ref{eq_pseudotrue}) defining $\beta_h$. This is the payoff of centering at the exact minimizer: no bias term survives in the expansion, and no Slutsky product with a diverging drift ever arises. The magnitude of the centering drift itself is controlled as in the text: $\beta_h-\beta=\Sigma_h^{-1}B_h+O(h^{2\kappa})$ with $B_h=r_n^{-1}\E[c_{ij}\{F(\Delta X_{ij}'\beta+\Delta\lambda_{ij})-F(\Delta X_{ij}'\beta)\}\Delta X_{ij}K_{ij}]$, computed from Step 1 of the proof of Theorem \ref{theo1}. On the kernel support $\delta^2_{ij}\le h$, Lemma A1 of \cite{auerbach2022identification} (under Assumption \ref{ass_holder}(\ref{ass_holder_ii})) and Assumption \ref{ass_holder}(\ref{ass_holder_iii}) give $|\Delta\lambda_{ij}|\le C_\lambda^{1/2}\left(2C^{\frac1{2+4\alpha}}h^{\frac{\alpha}{2(1+2\alpha)}}\right)^{\varphi}=O(h^\kappa)$, and since $|F(a+d)-F(a)|\le|d|/4$ and $r_n^{-1}\E[\|\Delta X_{ij}\|K_{ij}]=O(1)$ by the conditional moment bound of Assumption \ref{ass4}(2), $\|B_h\|=O(h^\kappa)$ as claimed in the text.

\textbf{Step 4 (projection and CLT).} $U_n'$ is a fifth-order U-statistic with an $n$-varying kernel. The moment condition of the projection lemma \citep[Lemma 3.1]{powell1989semiparametric}, extended to higher-order U-statistics as in Lemma A.3 of \cite{ahn1993semiparametric}, is verified explicitly rather than assumed: the second moment of the (normalized) kernel of $U_n'$ is bounded by
\[\frac{C}{r_n^2}\left(\E\left[\|s_{ij}\|^2K_{ij}^2\right]+\frac1{h^2}\E\left[\|s_{ij}\|^2K'\!\left(\tfrac{\delta^2_{ij}}{h}\right)^2\right]\right)\le\frac{C'}{r_nh^2},\]
using $|F_{ijts_1s_2}-\delta^2_{ij}|\le1$, $\E[\|s_{ij}\|^2|\omega_i,\omega_j]\le C$ (Assumption \ref{ass4}(2), since $\|s_{ij}\|\le\|\Delta X_{ij}\|$), and $\PP(\delta^2_{ij}\le h)\le Cr_n$ (Assumption \ref{ass4}(\ref{ass4iv})). 
Under Assumption \ref{ass_holder}(\ref{ass_holder_ii}), $r_n\ge\underline
K\,C^{-1/\alpha}h^{1/\alpha}$, so with $h=C_7n^{-\varrho}$,
\[
n\,r_n\,h^2\ \ge\ \underline K\,C^{-1/\alpha}C_7^{1/\alpha+2}\;
n^{\,1-\varrho\left(1/\alpha+2\right)} ,
\]
which diverges precisely when $\varrho\left(1/\alpha+2\right)<1$, that is when
$\varrho<\alpha/(1+2\alpha)$. Assumption
\ref{ass_holder}(\ref{ass_holder_iv}) imposes
$\varrho<\alpha/(2+4\alpha)=\tfrac12\cdot\alpha/(1+2\alpha)$, so the requirement
holds with a factor of two to spare in the exponent and $n\,r_n\,h^2\gg n^{1/2}$.
Hence the kernel's second moment is $o(n)$ and the projection lemma applies:

\[U_n'=\frac1n\sum_{\tau=1}^n\varphi_\tau+R_n, \qquad \sqrt nR_n\to_p0,\]
where $\varphi_\tau=\psi_1(Z_\tau)+\psi_2(Z_\tau)+\psi_3(Z_\tau)$ collects the three projection channels displayed in equation (\ref{eq_influence}): an index $\tau$ contributes as a member of the scored pair ($i$ or $j$; factor $2$ in $\psi_1$), as the bridging agent $t$ (in $\psi_2$), and as one of the two $s$-agents (factor $2$ in $\psi_3$). No cancellation among the channels is claimed or used: all cross-covariances are retained inside $\Omega_h=\Var(\varphi_\tau)$. (Indeed they do not vanish in general: conditional on $\omega_\tau$, $\E[F_{ij\tau s_1s_2}-\delta^2_{ij}|\omega_i,\omega_j,\omega_\tau]=(p_{\omega_i}(\omega_\tau)-p_{\omega_j}(\omega_\tau))^2-\delta^2_{ij}\neq0$, and with the logit score $\E[s_{ij}|Z_{ij}]$ is the nonzero matching bias on inexact matches, so the exact cancellations available in the linear model are unavailable here.)

For the central limit theorem, the summands $\varphi_\tau$ are i.i.d.\ across $\tau$, mean zero, with variance $\Omega_h$, and the two families of channels are bounded differently. The first-stage channels satisfy $\|\psi_2\|,\|\psi_3\|\le C/h$ pointwise: the scored pair $(i,j)$ excludes $\tau$, so the conditional expectations localize through $K'$, whose window has unconditional probability $\PP(\delta^2_{ij}\le h)\le Cr_n$ by Assumption \ref{ass4}(4), cancelling the $r_n^{-1}$ factor. The score channel $\psi_1$ is not bounded, but has fourth moments: $\|\psi_1(Z_\tau)\|\le C(1+\|X_\tau\|)\big(r_n(\omega_\tau)/r_n+1\big)\le C'(1+\|X_\tau\|)$, again by Assumption \ref{ass4}(3), so $\E\|\psi_1\|^4\le C''$ by the conditional moment bound of Assumption \ref{ass4}(2). Hence
\[\E\|\varphi_\tau\|^4\le C\Big(1+h^{-2}\,\E\|\varphi_\tau\|^2\Big),\]
and, since $\mathrm{tr}\,\Omega_h\le Ch^{-2}$, the Lyapunov ratio obeys
\[\frac{\E\|\varphi_\tau\|^4}{n\,\lambda_{\min}(\Omega_h)^2}\ \le\ \frac{C}{n\,\lambda_{\min}(\Omega_h)^2}+\frac{C}{n h^4\,\lambda_{\min}(\Omega_h)^2}\ \longrightarrow\ 0,\]
using $nh^4\to\infty$ (from $\varrho<1/4$, Assumption \ref{ass_holder}(\ref{ass_holder_iv})) and $\liminf_n\lambda_{\min}(\Omega_h)>0$ (Assumption \ref{ass_holder}(\ref{ass_holder_v})); no upper bound on $\Omega_h$ is used, consistent with the remark after Theorem \ref{theo4} that $V_n$ need not be of order $1/n$. The Lyapunov condition holds for the triangular array and
\[\Omega_h^{-1/2}\sqrt{n}\,U_n'\xrightarrow{d}\mathcal{N}(0,I_k).\]
Combining with the expansion around $\beta_h$: because $\Omega_h$ need not be bounded, rates rather than plain consistency are used for the outer factors. The kernel and score substitution bounds (Lemma \ref{lemB1} for the estimated distances, Lipschitz continuity of the score, and the localized moment bounds of Assumption \ref{ass4}, the same ingredients later reused in Appendix \hyperlink{app_var}{A.8}) give $\|\widehat H_n(\tilde\beta)-\Sigma_h\|=O_p(n^{-1/2}\log^{1/2}n\,h^{-1}+\|\tilde\beta-\beta_h\|)$, with $\|\tilde\beta-\beta_h\|\to_p0$ from Theorem \ref{theo2}'s argument applied to the fixed-$h$ localized objective and $|\hat r_n/r_n-1|=O_p(n^{-1/2}\log^{1/2}n\,h^{-1})$, and $\lambda_{\max}(\Omega_h)\le\mathrm{tr}\,\Omega_h\le Ch^{-2}$, so the relative error these factors contribute to the studentized statistic is $O_p(h^{-1}\cdot n^{-1/2}\log^{1/2}n\,h^{-1})=O_p(n^{-1/2}\log^{1/2}n\,h^{-2})=o_p(1)$ under $\varrho<1/4$ (Assumption \ref{ass_holder}(\ref{ass_holder_iv})). Hence,
\[V_n^{-1/2}\left(\hat\beta-\beta_h\right)\xrightarrow{d}\mathcal{N}(0,I_k), \qquad V_n=\Sigma_h^{-1}\Omega_h\Sigma_h^{-1}/n,\]
which is the claim. $\blacksquare$
\end{proofoftheorem}

\phantomsection
\addcontentsline{toc}{subsection}{Proof of Theorem  \ref{theo5}}
\subsection*{A.7 Proof of Theorem \ref{theo5} (jackknife bias correction)}

\begin{proofoftheorem}
Applying the projection argument of the proof of Theorem \ref{theo4} jointly to the finite collection $\{\hat{\beta}_{c_lh}\}_{l=1}^{L+1}$ (the joint asymptotic normality follows from the Cram\'er--Wold device, since each estimator admits an asymptotically linear representation in the same i.i.d.\ observations, with cross-bandwidth covariance blocks $\Omega_{h,l_1l_2}=\mathrm{Cov}(\varphi^{(l_1)}_\tau,\varphi^{(l_2)}_\tau)$) gives
\[V_{5,n}^{-1/2}\left(\bar{\beta}_L-\bar{\beta}_{L,h}\right)\xrightarrow{d}\mathcal{N}(0,I_k), \qquad \bar{\beta}_{L,h}:=\sum_{l=1}^{L+1}a_l\,\beta_{c_lh},\]
with $V_{5,n}=\bar V_n$ as in the statement. By Assumption \ref{ass_jack},
\begin{align*}
\bar{\beta}_{L,h}&=\beta+\sum_{m=1}^{L}C_{m}\underbrace{\left[\sum_{l=1}^{L+1}a_{l}c_{l}^{m/\theta}\right]}_{=\,0\ \text{for}\ m=1,\ldots,L}h^{m/\theta}+O\left(h^{(L+1)/\theta}\right)\\
&=\beta+O\left(h^{(L+1)/\theta}\right)=\beta+o\left(n^{-1/2}\right),
\end{align*}
because the $L+1$ weights satisfy $\sum_la_l=1$ and annihilate every retained power $h^{m/\theta}$, $m=1,\ldots, L$, by construction of the generalized Vandermonde system, and the remainder is negligible relative to the standard error, which is of order at least $n^{-1/2}$ times the root of the combined-variance lower bound imposed in Assumption \ref{ass_jack}, by that assumption's $o(n^{-1/2})$ rate condition; the same lower bound makes the Cram\'er--Wold limit nonsingular. (With only $L$ weights, as in an earlier version of this construction, the bracket at $m=L$ is nonzero (for instance $\sum_la_lc_l^2=3-2(1.5)^2=-1.5$ for weights $(3,-2)$ at $c=(1,1.5)$, $\theta=1$), and the surviving $C_Lh^{L/\theta}$ term is not controlled by the rate condition; the extra bandwidth is essential.) The claim follows. $\blacksquare$
\end{proofoftheorem}

\subsection*{A.8 Variance estimation}\hypertarget{app_var}{}

The variance estimator is the influence-function object (\ref{eq_varest}) of Section \ref{section332}; this subsection proves its consistency, with rates. Write $\varphi_\tau$ for the population influence function (\ref{eq_influence}) evaluated at $\beta_h$ and $\hat\varphi_\tau$ for its estimate, and set $\varepsilon_n:=\max_{i\ne j}|\hat\delta^2_{ij}-\delta^2_{ij}|=O_p(n^{-1/2}\log^{1/2}n)$ (Lemma \ref{lemB1}) and $b_n:=\|\hat\beta-\beta_h\|=O_p(n^{-1/2}h^{-1})$ (a crude bound from Theorem \ref{theo4}, using $\mathrm{tr}\,\Omega_h\le Ch^{-2}$). Each channel of $\hat\varphi_\tau$ is a V-statistic in the sense of \cite{ahn1993semiparametric}, and its deviation from $\varphi_\tau$ decomposes into four substitution errors, bounded uniformly over $\tau$:
kernel arguments, $|K(\hat\delta^2/h)-K(\delta^2/h)|\le\|K'\|_\infty\varepsilon_n/h$ and $|K'(\hat\delta^2/h)-K'(\delta^2/h)|\le\|K''\|_\infty\varepsilon_n/h$, contributing $O_p(\varepsilon_n/h)$ to $\hat\varphi_{1,\tau}$ and, through the $1/h$ prefactor, $O_p(\varepsilon_n/h^2)$ to $\hat\psi_{2,\tau},\hat\psi_{3,\tau}$;
scores, $\|\hat s_{ij}-s_{ij}\|\le C(1+\|\Delta X_{ij}\|^2)b_n$, contributing $O_p(b_n)$ and $O_p(b_n/h)$ respectively after the localized averages;
network products, $\widehat F_{ij\tau},\widehat F'_{ij\tau}$ concentrating on their conditional means at rate $O_p(n^{-1/2}\log^{1/2}n)$ by the Hoeffding-plus-union-bound argument of Lemma \ref{lemB1}, contributing $O_p(n^{-1/2}\log^{1/2}n/h)$;
and V-statistic sampling error, $O_p((nr_n)^{-1/2}/h)$ by the second-moment bound of Step 4 of the proof of Theorem \ref{theo4}.
Collecting terms, $\max_\tau\|\hat\varphi_\tau-\varphi_\tau\|=O_p\big(n^{-1/2}\log^{1/2}n\,h^{-2}\big)$. Since $\frac1n\sum_\tau\|\varphi_\tau\|^2=O_p(\mathrm{tr}\,\Omega_h)=O_p(h^{-2})$,
\begin{multline*}\big\|\widehat\Omega_h-\Omega_h\big\|\ \le\ \max_\tau\|\hat\varphi_\tau-\varphi_\tau\|\Big(2\Big(\tfrac1n\sum_\tau\|\varphi_\tau\|^2\Big)^{1/2}+\max_\tau\|\hat\varphi_\tau-\varphi_\tau\|\Big)\\+O_p\big(n^{-1/2}(\E\|\varphi_\tau\|^4)^{1/2}\big)=O_p\big(n^{-1/2}\log^{1/2}n\,h^{-3}\big),\end{multline*}
which is $o_p(1)$ because Assumption \ref{ass_holder}(\ref{ass_holder_iv}) imposes $\varrho<1/6$ strictly, so $n^{1/2}h^3=n^{1/2-3\varrho}$ diverges polynomially and absorbs the logarithm. (The V-statistic term can dominate the collected rate when $\alpha<1/2$; it is $O_p(n^{-1/2}h^{-1/(2\alpha)-1})$ and remains $o_p(h)$-negligible in the aggregation since $\varrho(2+\tfrac1{2\alpha})<\tfrac{1+4\alpha}{4+8\alpha}<\tfrac12$ for every $\alpha$.) With $\lambda_{\min}(\Omega_h)$ bounded below (Assumption \ref{ass_holder}(\ref{ass_holder_v})) and $\widehat\Sigma_h\to_p\Sigma_h$ (by the same substitution bounds, valid in both regimes; see the Hessian step of Appendix \hyperlink{app_theo4}{A.6}), $\widehat\Sigma_h^{-1}\widehat\Omega_h\widehat\Sigma_h^{-1}/n\cdot V_n^{-1}\to_pI_k$. The cross-bandwidth blocks $\widehat\Omega_{h,l_1l_2}$ are handled identically. Under the hypotheses of Theorem \ref{theo3} (finite support), $\hat\varphi_{2,\tau}$ and $\hat\varphi_{3,\tau}$ are asymptotically negligible and may be omitted. Two implementation identities are useful as unit tests: with $C=DD$ and $E=DC$, $\widehat F_{ij\tau}=((C_{\tau i}-C_{\tau j})/n)^2$ and $\widehat F'_{ij\tau}=(D_{i\tau}-D_{j\tau})(E_{\tau i}-E_{\tau j})/n^2$, and both average exactly to $\hat\delta^2_{ij}$ over $\tau=1,\ldots,n$.

For grouped data the reported clustered variance is the score sandwich $\widehat V_{\mathrm{cl}}$ of Proposition \ref{prop_villages}(ii); its consistency is proved in the cluster-variance step of Appendix \hyperlink{app_villages}{A.9}, which requires no delta-method correction because all within-village estimation, including the codegree distances, is a function of village data alone.

\subsection*{A.9 Proof of Proposition \ref{prop_villages}}\hypertarget{app_villages}{}

Throughout, villages $g=1,\dots,G$ are independent, each generated by the single-network model with common primitives $(\beta,\lambda,f)$, sizes $n_g\in[\underline n,C\underline n]$, and all assumptions are imposed within each network with constants uniform over $g$. Superscript $(g)$ denotes village-$g$ objects: $\hat\delta^{2,(g)}_{ij}$, $r^{(g)}_n$, $\widehat K^{(g)}_{ij}=K(\hat\delta^{2,(g)}_{ij}/h_g)$, and so on. The pooled estimator minimizes $\sum_g\sum_{i<j\in g}\widehat K^{(g)}_{ij}\,m(v_i,v_j,b)$, and $N=\sum_gn_g$.

\emph{Part (i).} Write the pooled objective, normalized, as
\begin{align*}Q_N(b)&=\sum_{g=1}^G\pi_g\,Q^{(g)}(b),\\ \pi_g:=\frac{\sum_{i<j\in g}\widehat K^{(g)}_{ij}}{\sum_{g'}\sum_{i<j\in g'}\widehat K^{(g')}_{ij}},&\qquad Q^{(g)}(b):=\frac{\sum_{i<j\in g}\widehat K^{(g)}_{ij}m_{ij}(b)}{\sum_{i<j\in g}\widehat K^{(g)}_{ij}},\end{align*}
a convex combination of village objectives, each convex in $b$. Steps 2--4 of the proof of Theorem \ref{theo2} apply within each village with $n$ replaced by $n_g\ge\underline n$. The only step requiring care for growing $G$ is the union bound in Lemma \ref{lemB1}: applied simultaneously to all villages, the bound runs over at most $G\cdot(C\underline n)^3\le\underline n^{C'+3}$ index triples, so the uniform rate becomes $\max_g\max_{i\ne j\in g}|\hat\delta^{2,(g)}_{ij}-\delta^{2,(g)}_{ij}|=O_p(\underline n^{-1/2}\log^{1/2}\underline n)$, the same rate up to the constant in the logarithm; likewise for Lemma \ref{lemC1}. Consequently $Q^{(g)}(b)=Q_0(b)+o_p(1)$ uniformly over $g$ (the $o_p$ bounds in Steps 3--4 depend on $g$ only through $n_g$ and the uniform constants), hence $Q_N(b)=Q_0(b)+o_p(1)$ pointwise in $b$. The population objective and its unique minimizer $\beta$ are village-invariant because the primitives are common. The convexity lemma \citep[Theorem 2.7]{newey1994large} yields $\hat\beta\to_p\beta$. $\square$

\emph{Part (ii).} Let $S_N(b)=\sum_gS_g(b)$ with $S_g(b)=\sum_{i<j\in g}\widehat K^{(g)}_{ij}s(v_i,v_j,b)$ the village scores, and let $H_N$ be the pooled Hessian. The expansion (\ref{eq_expansion}) holds for the pooled criterion by the same mean-value argument, with $H_N/\!\sum_g\binom{n_g}2r^{(g)}_n\to_p\Sigma_h$ by the village-level Hessian analysis applied per village and averaged (uniformity over $g$ as in part (i)).

For the score, apply Steps 1--3 of the proof of Theorem \ref{theo4} within each village, with all scores evaluated at the pooled pseudo-true value $\beta_h$ (with a common bandwidth, $\beta_h$ minimizes each village's localized population objective, so each village's localized population score at $\beta_h$ is exactly zero, as in Step 3 of that proof): on the common bandwidth window,
\[S_g(\beta_h)=a_g\Big(\frac1{n_g}\sum_{\tau\in g}\varphi^{(g)}_\tau+R_g\Big), \qquad \sqrt{n_g}\,R_g\to_p0 \text{ uniformly over } g,\]
with $a_g:=\binom{n_g}2r^{(g)}_n$ as in the statement,
where $\varphi^{(g)}_\tau$ is the influence function of (\ref{eq_influence}) computed in village $g$ at $\beta_h$. With $\Phi_g:=a_g\,n_g^{-1}\sum_{\tau\in g}\varphi^{(g)}_\tau$ as in the statement, the $\Phi_g$ are independent across $g$ (Assumption \ref{ass1} across villages), mean zero, with $\Var(\Phi_g)=a_g^2n_g^{-1}\Omega_h(1+o(1))$, and the expansion of the pooled first-order condition gives $\Sigma_h(\sum_ga_g)(\hat\beta-\beta_h)=\sum_g\Phi_g+o_p\big((\sum_g\Var\Phi_g)^{1/2}\big)$, matching the statement's normalization. Two remainders in this expansion are not mean zero and therefore do not average away across villages; they require the joint rate condition rather than the within-village bounds. Both are of the same order: the repeated-index tuples dropped in Step 2 of the proof of Theorem \ref{theo4} contribute $a_g\,O(1/(n_gh))$ per village to the pooled score, and the second-order kernel remainder of that proof, controlled by the per-pair mean-square bound (\ref{eq_pairMSE}), contributes $a_g\,O(1/(n_gh))$ as well. Summed over villages the total is $O\big(G\,a/(\underline n\,h)\big)$ with $a:=\max_ga_g$, while the pooled score scale is $(\sum_g\Var\Phi_g)^{1/2}\asymp\sqrt G\,a\,\underline n^{-1/2}\lambda_{\min}(\Omega_h)^{1/2}$; the ratio is $\sqrt G\big/(\sqrt{\underline n}\,h\,\lambda_{\min}(\Omega_h)^{1/2})$, which the joint rate condition sends to zero (for fixed $G$ it reduces to $1/(\sqrt{\underline n}\,h)\to0$, implied by Assumption \ref{ass3}(\ref{ass3ii})). All remaining terms are mean zero and controlled by the variance bounds below.

\emph{Case $G$ fixed, $\underline n\to\infty$.} Each village satisfies the central limit theorem of the proof of Theorem \ref{theo4}; the villages are independent; the sum of $G$ independent asymptotically normal vectors is asymptotically normal with summed variances. \emph{Case $G\to\infty$.} By the Cram\'er--Wold device it suffices to verify Lyapunov's condition for the scalar arrays $X_g:=u'\big(\sum_{g'}\Var\Phi_{g'}\big)^{-1/2}\Phi_g$, $\|u\|=1$. Write $w_\tau:=u'(\sum\Var\Phi)^{-1/2}\varphi^{(g)}_\tau$ and $s_g^2:=\E w_\tau^2$. Then $\Var(X_g)=a_g^2n_g^{-1}s_g^2$ and $\sum_g\Var(X_g)=1$, so with comparable villages $s_g^2\asymp n_g/(a_g^2G)$. Rosenthal's inequality for the i.i.d.\ within-village sum gives
\[\E X_g^4\ \le\ C\,a_g^4n_g^{-4}\big(n_g^2s_g^4+n_g\,\E w_\tau^4\big),\]
and summing over $g$: the first (Gaussian) part contributes $\sum_gC\,a_g^4n_g^{-2}s_g^4\asymp C/G\to0$ regardless of the size of $\Omega_h$; for the second, note
\[\E w_\tau^4\le\Big(\frac{s_g^2}{\lambda_{\min}(\Omega_h)}\Big)^2\,\E\|\varphi\|^4, \qquad \E\|\varphi\|^4\le C\big(1+h^{-2}\,\mathrm{tr}\,\Omega_h\big)\le Ch^{-4},\]
by the split bound of the proof of Theorem \ref{theo4} and $\mathrm{tr}\,\Omega_h\le Ch^{-2}$, so it contributes at most $C/(G\,\underline n\,h^4\lambda_{\min}(\Omega_h)^2)\to0$, using $\underline n\,h^4\to\infty$ (implied by the within-village window $\varrho<\alpha/(2+4\alpha)\le1/4$ of Assumption \ref{ass_holder}(\ref{ass_holder_iv})) and Assumption \ref{ass_holder}(\ref{ass_holder_v}); no condition linking $G$ to the bandwidth is needed \emph{for this variance step}; the joint rate condition of the statement is used only for the centering remainder above. Lyapunov's condition holds and $\big(\sum_g\Var(\Phi_g)\big)^{-1/2}\sum_g\Phi_g\to_d\mathcal N(0,I_k)$. Combining with the expansion yields the displayed convergence of Proposition \ref{prop_villages}(ii).

\emph{Cluster variance.} $\widehat V_{\mathrm{cl}}$ is built from the realized village scores, and its consistency follows from the standard cluster argument, with no delta-method correction needed: every object estimated within village $g$, including the codegree distances entering $\widehat K^{(g)}_{ij}$, is a function of village-$g$ data alone, so the $S_g(\beta_h)$ are independent across $g$ with mean exactly zero (the pooled first-order condition at $\beta_h$) and their cross-village variation already carries the first-stage estimation noise. Throughout this step $G\to\infty$ and $\mathrm{tr}\,\Omega_h\le\bar C\lambda_{\min}(\Omega_h)$, as in the statement. Three steps. (1) $\sum_gS_g(\beta_h)S_g(\beta_h)'\big/\sum_g\Var(S_g(\beta_h))\to_pI$ by the law of large numbers for independent outer products with fourth moments bounded as in the Lyapunov step; well-conditioning makes the elementwise convergence uniform over directions. (2) $\Var(S_g(\beta_h))=\Var(\Phi_g)(1+o(1))$: the non-projection remainder $a_gR_g$ has mean zero (the H\'ajek projection preserves the mean and $\E U'_g=0$ by the first-order condition) and $\Var(\sqrt{n_g}R_g)\to0$ by the second-moment projection bound of Step 4 of the proof of Theorem \ref{theo4}, so it moves the variance by a vanishing factor and $\sum_ga_gR_g=o_p\big((\sum_g\Var\Phi_g)^{1/2}\big)$. (3) Substituting $\hat\beta$ for $\beta_h$: $S_g(\hat\beta)=S_g(\beta_h)-H_g(\bar\beta_g)(\hat\beta-\beta_h)$ with $\|H_g\|=O_p(a_g)$ uniformly, and $\|\hat\beta-\beta_h\|=O_p\big((\sum_ga_g)^{-1}(\sum_g\Var\Phi_g)^{1/2}\big)$ from part (ii), so the substitution changes $\sum_gS_gS_g'$ by a factor $1+o_p(1)$. Combining with $\widehat H_N/\sum_ga_g\to_p\Sigma_h$ gives $\widehat V_{\mathrm{cl}}^{-1/2}(\hat\beta-\beta_h)\to_d\mathcal N(0,I_k)$, the operational statement of part (ii). $\square$

\emph{Part (iii).} Fix $\bar n$ and let villages be i.i.d.\ draws of a network of size at most $\bar n$ (sizes drawn from a fixed distribution on $[\underline n,\bar n]$; the argument with deterministic sizes is identical along subsequences). Each village contributes an i.i.d.\ copy of the random function
\[b\;\mapsto\;M(b):=\sum_{i<j}\widehat K_{ij}\,m(v_i,v_j,b),\]
a bounded-degree polynomial in village data with $\E\sup_{b\in\mathcal B}|M(b)|<\infty$ on the compact parameter space by Assumption \ref{ass4}. By the uniform law of large numbers for i.i.d.\ random convex functions, $G^{-1}\sum_gM_g(b)\to_p\overline M(b):=\E[M(b)]$ pointwise, $\overline M$ is convex, and it is strictly convex at its minimizer (the expected Hessian is $\E[\widehat K_{ij}F(1-F)\Delta X\Delta X']\succ0$ whenever kernel weights are positive with positive probability). Hence $\hat\beta\to_p\beta_{h,\bar n}:=\arg\min_b\overline M(b)$ as $G\to\infty$, by the convexity lemma. It remains to show $\beta_{h,\bar n}\ne\beta$ in general. The first-order condition at $\beta$ reads $\E\big[\widehat K_{ij}\,s(v_i,v_j,\beta)\big]=0$; conditioning on the latent types and on the network,
\[\E\big[\widehat K_{ij}s_{ij}\big]=\E\Big[\widehat K_{ij}\,c_{ij}\big\{F(\Delta X_{ij}'\beta+\Delta\lambda_{ij})-F(\Delta X_{ij}'\beta)\big\}\Delta X_{ij}\Big],\]
where $\widehat K_{ij}$ is measurable with respect to the network and independent of $(y_i,y_j)$ given the types. At fixed $\bar n$ the estimated distances do not concentrate: $\Var(\hat\delta^2_{ij}\mid\boldsymbol\omega)\ge c/\bar n>0$ under the nondegeneracy hypothesis of Lemma \ref{lem_bandwidth}(b), so $\widehat K_{ij}$ assigns positive probability to pairs with $\Delta\lambda_{ij}\ne0$ whenever types with distinct $\lambda$ have positive mass, and the display is a fixed nonzero vector except in degenerate cases (exact finite-support designs in which within-village exact matching is achievable at size $\bar n$ and the kernel weight concentrates on exact matches, or designs in which $\Delta\lambda_{ij}$-weighted terms cancel by symmetry). A one-dimensional example with strictly monotone $\lambda$ and homophilous $f$ verifies non-cancellation directly: there $\Delta\lambda_{ij}$ has the sign of $\omega_i-\omega_j$ and the weighted term has a strictly positive projection on $\E[\Delta X_{ij}\,\mathrm{sign}(\omega_i-\omega_j)]\ne0$ when $X$ is correlated with $\omega$. Hence $\beta_{h,\bar n}\ne\beta$ in general, and no growth in $G$ repairs the discrepancy, since the limit depends on $G$ not at all. $\square$

\subsection*{A.10 Implementation details}
This subsection records every implementation choice a replicator needs; the simulation and estimation code implements exactly these steps.
\begin{enumerate}[(1)]
\item  \emph{Distances.} $C=DD$, $\hat\delta^2_{ij}=n^{-3}\|C_{\cdot i}-C_{\cdot j}\|^2$; in grouped data all objects are computed within network $g$, with $n$ replaced by $n_g$ throughout (this is the ``village-specific normalization'').
\item  \emph{Quantiles.} Empirical quantiles $\hat q_c$ use linear interpolation between order statistics (the default of standard software); ties are handled by the interpolation and require no separate rule.
\item  \emph{Noise scale.} $\hat m$ is the median of $\mathrm{nf}_{ij}$ over a uniform random subsample of $\min\{2000,\binom n2\}$ pairs, drawn once per dataset ($B_n\ge n$ at every size used; in a village, over that village's pairs, with $n$ replaced by $n_g$).
\item  \emph{Regime rule.} Finite-support branch if and only if the share of pairs with $\hat\delta^2_{ij}\le3\hat m$ is at least $0.15$ and the share in $(3\hat m,9\hat m]$ is at most one third of it; otherwise continuous branch. Equation (\ref{eq_bandwidthrule}) then gives $\hat h$.
\item  \emph{Estimation.} Newton iterations on the localized conditional-logit objective over discordant pairs with positive kernel weight, ridge $10^{-10}$, convergence $10^{-10}$ in the max-norm of the step.
\item  \emph{Jackknife.} Fits at $(\hat h,1.5\hat h,2\hat h)$ with weights $a(\hat\kappa)$ solving the generalized Vandermonde system of Section \ref{section332} at the estimated exponent, equal to $(6,-8,3)$ at $\hat\kappa=1$; exponent from the grid $(2\hat h,4\hat h,8\hat h)$ as in Section \ref{section332}, clamped to $[1/2,2]$, applied only if the raw $\hat\kappa\ge0.4$; when the correction is refused, the jackknife column reports the uncorrected estimator \emph{and its uncorrected variance} (\ref{eq_varest}).
\item  \emph{Variance.} Single network: (\ref{eq_varest}), with the sums over $\tau$ including $\tau\in\{i,j\}$ (the diagonal terms are $O(1/n)$ of the total and immaterial); the identities $\widehat F_{ij\tau}=((C_{\tau i}-C_{\tau j})/n)^2$ and $\widehat F'_{ij\tau}=(D_{i\tau}-D_{j\tau})(E_{\tau i}-E_{\tau j})/n^2$, $E=DC$, are the recommended unit test. Grouped data: the score sandwich $\widehat V_{\mathrm{cl}}$ of Proposition \ref{prop_villages}(ii).
\item \emph{Heterogeneity estimator.} $\hat\lambda$ uses $K_X(z)\propto(1-z^4)^3\,\mathbbm{1}\{|z|<1\}$ (the triweight-type kernel applied to the squared standardized difference) for $K_X$, $h_X=1.06\,\hat\sigma_Xn^{-1/5}$, and trimming $\tau_n=1/(1+\log^2n)$. In the simulation designs the operational $\X_0$ is the full sample, the target probabilities being interior by construction; in applications a central-range restriction on the covariates implements $\X_0$.
\end{enumerate}


\section*{Declaration of generative AI and AI-assisted technologies in the manuscript preparation process}
During the preparation of this work, the author used Claude Anthropic to assist with language editing, improving the clarity of the
exposition, and checking the internal consistency of the theoretical arguments. All content
was reviewed, verified, and edited by the author, who takes full responsibility for the final
version of the manuscript and its accuracy.

\bibliographystyle{elsarticle-harv}
 \bibliography{mybib}



\end{document}